\documentclass[twocolumn,prd,aps,superscriptaddress,preprintnumbers,tightenlines,showpacs,nofootinbib,eqsecnum,amsfonts,amsmath]{revtex4}
\pdfoutput=1

\usepackage{epsfig}
\usepackage{graphics}
\usepackage{graphicx}
\usepackage{amsmath,amssymb,mathrsfs}
\usepackage{amsfonts}
\usepackage{color}
\usepackage{wasysym}
\usepackage{times}
\usepackage{mathptmx}
\usepackage{gensymb}

\usepackage[caption=false]{subfig} 

\DeclareMathAlphabet{\mathcalstd}{OMS}{cmsy}{m}{n}
\DeclareMathAlphabet{\mathpzc}{OT1}{pzc}{m}{it}

\DeclareMathOperator{\atan2}{atan2}

\newcommand{\bigO}{\mathcal O}
\newcommand{\pd}[2]{\frac{\partial #1}{\partial #2}}

\newcommand{\be}{\begin{equation}}
\newcommand{\ee}{\end{equation}}
\newcommand{\ber}{\begin{eqnarray}}
\newcommand{\eer}{\end{eqnarray}}
\newcommand{\bea}{\begin{eqnarray}}
\newcommand{\eea}{\end{eqnarray}}

\newcommand{\RM}{\mathcal{R}_\text{M}}
\newcommand{\Apr}{A_r}
\newcommand{\Apt}{A_t}
\newcommand{\lpr}{\lambda_r}
\newcommand{\lpt}{\lambda_t}

\newcommand{\R}[1]{\tilde{\mathcal{R}}_\text{#1}}

\newcommand{\RMd}{\tilde{\mathcal{R}}_\text{M}}

\newcommand{\phiGW}{\phi_\text{GW}}
\newcommand{\omGW}{\omega_\text{GW}}
\newcommand{\omGWcl}{\omega_\text{GW,cleaned}}
\newcommand{\omOrb}{\Omega}

\newcommand{\ephiGW}{e_{\phi,\text{GW}}}
\newcommand{\eomOrb}{e_\Omega}

\newcommand{\UIB}{Departament de F\'isica, Universitat de les Illes Balears, 
Crta. Valldemossa km 7.5, E-07122 Palma, Spain}
\newcommand{\Vienna}{Gravitational Physics, Faculty of Physics, University of Vienna, Boltzmanngasse
5, A-1090 Vienna, Austria}
\newcommand{\Cardiff}{School of Physics and Astronomy, Cardiff University, Queens Building, CF24 3AA, Cardiff, United Kingdom}

\begin{document}



\title{An efficient iterative method to reduce eccentricity in numerical-relativity simulations of \\ 
compact binary inspiral}

\author{Michael P\"urrer}
\affiliation{\Cardiff}
\affiliation{\Vienna}

\author{Sascha Husa}
\affiliation{\UIB}

\author{Mark Hannam}
\affiliation{\Cardiff}

\begin{abstract}
We present a new iterative method to reduce eccentricity in black-hole-binary simulations. 
Given a good first estimate of low-eccentricity starting momenta, we evolve puncture initial data for 
$\sim4$ orbits and construct improved initial parameters by comparing the inspiral with 
post-Newtonian calculations. Our method is the first to be applied directly to 
the \emph{gravitational-wave (GW) signal}, rather than the orbital motion. 
The GW signal is in general less contaminated by gauge effects, which, 
in moving-puncture simulations, limit orbital-motion-based measurements of the eccentricity to an 
uncertainty of $\Delta e \sim 0.002$, making it difficult to reduce the eccentricity below this value. 
Our new method can reach eccentricities below $10^{-3}$ in one or two iteration steps; we find
that this is \emph{well below} the requirements for GW astronomy in the advanced detector
era. Our method can be readily adapted to any compact-binary simulation with GW emission, including 
black-hole-binary simulations that use alternative approaches, 
and neutron-star-binary simulations. We also comment on the differences in eccentricity estimates 
based on the strain $h$, and the Newman-Penrose scalar $\Psi_4$. 
\end{abstract}


\maketitle
 

\section{Introduction} 
\label{sec:introduction}

There is a large-scale effort underway to produce models of the gravitational-wave (GW)
signal from the late inspiral, merger and ringdown of binary systems of black holes,
calibrated against large numbers of numerical 
simulations~\cite{Ohme:2011rm}. These models will be essential to 
locate and interpret black-hole-binary GW signals in the data from second-generation 
laser-interferometric detectors, the first of which, Advanced LIGO, may be 
commissioned as early as 2014~\cite{Abbott:2007kv,Shoemaker:aLIGO,2010CQGra..27h4006H}. 
The most pressing need is for models of binaries 
that undergo {\it non-eccentric} inspiral. It is non-trivial to prescribe initial conditions 
that lead to non-eccentric inspiral in numerical simulations, and to date there is no 
systematic procedure to do this for simulations performed using the 
``moving-puncture method''~\cite{Campanelli:2005dd,Baker:2005vv}, which is the most common 
in the field. 

In fully general relativistic numerical simulations the binary's eccentricity cannot be prescribed. 
The best we can do is to use some model to guess initial parameters that may lead to low-eccentricity
inspiral and, if necessary, adjust those parameters until the eccentricity falls below some 
acceptable tolerance. The problem is further complicated by the difficulty of eccentricity 
measurement. There is no rigorous or unique definition of eccentricity for binary systems in 
general relativity. We can define a number of quantities that all reduce to the Newtonian limit,
and all agree at zero eccentricity~\cite{Mroue:2010re}, but many of these depend on the 
motion of the black holes, which is gauge dependent. 
(It should be emphasized that \emph{all} coordinate 
black-hole motion in these simulations is \emph{entirely due to the gauge variables}.) 
It would be preferable to use the GW signal, which is far less gauge-dependent.

Various techniques have been proposed to obtain momenta leading to low eccentricity by employing Newtonian or post-Newtonian information and short NR simulations~\cite{Husa:2007rh,Hannam:2007ik,Hannam:2010ec,Tichy:2010qa,Pfeiffer:2007yz,Boyle:2007ft,Buonanno:2010yk}.
In previous work we estimated the initial parameters from solutions of the post-Newtonian 
(PN) equations of motion~\cite{Husa:2007rh,Hannam:2007ik}. For an equal-mass nonspinning 
binary these resulted in an eccentricity of $e \sim 0.0025$ (from the NR eccentricity estimator that
we use in this paper).
For larger mass ratios, and for binaries made up of spinning black holes,
the eccentricity was larger, even when higher-order PN spin contributions were 
included~\cite{Hannam:2010ec}. 
In some cases we further used PN solutions to estimate the overall magnitude of the 
perturbation in the initial momenta necessary to correct for the eccentricity~\cite{Hannam:2010ec}. 
This procedure worked well, but in providing only the \emph{magnitude} of the momentum 
adjustment, it was not possible to 
independently refine both the radial and tangential momenta. The method 
also relied on the gauge-dependent coordinate motion of the black holes, which further 
limited its potential; a second iteration of the method was usually not possible, and 
eccentricities could not be reduced below $e \sim 0.004$. 

A powerful iteration method was proposed in Ref.~\cite{Pfeiffer:2007yz}, 
in which eccentricities below $e \sim 10^{-5}$ could be achieved in two iteration steps. 
This method was further extended to precessing binaries in Ref.~\cite{Buonanno:2010yk}.
This method also relies on the coordinate motion of the black holes. 
It has been applied to simulations that use
initial data in particular quasi-equilibrium coordinates that are adapted to the motion of the 
black holes, and this means that (a) large non-physical gauge effects in the coordinate motion 
are not apparent, and (b) the phase of the black holes' motion can be used from $t=0$, which is
necessary for the implementation of the method presented 
in~\cite{Pfeiffer:2007yz,Boyle:2007ft,Buonanno:2010yk}. These features make it 
difficult to apply that method to moving-puncture simulations. (We discuss this in more
detail in Sec.~~\ref{sub:comparison_between_eccentricity_reduction_methods}.) 
In moving-puncture simulations the orbital phase cannot be 
used from $t=0$, and we must instead wait until the gauge has settled down (at least one 
orbit into the simulation); and even then it contains additional non-physical oscillations due to 
gauge effects. 
Even in cases where the coordinate motion appears to closely reflect the underlying physics,
it would be preferable to have an eccentricity-reduction method that can be applied to the 
gravitational-wave signal, which is far less gauge dependent and is, ultimately, the 
physically measurable quantity that we are interested in modeling. 

In this paper we present a robust iterative method that overcomes these issues. 
The idea is as follows. Start with a short NR simulation that exhibits eccentricity, and a non-eccentric 
PN/EOB (effective one body, see e.g.~\cite{Buonanno:1998gg,Damour:2001tu}) 
evolution of the same system. Adjust the initial momenta in the
PN/EOB evolution until it exhibits eccentricity oscillations that agree with those in the NR waveforms, 
in both amplitude and phase. The inverse adjustment is then applied to the NR initial momenta, and 
a new NR simulation performed, and the process repeated. The use of the amplitude {\it and} phase 
of the eccentricity oscillations makes it possible to independently determine the required adjustment 
in both the tangential and radial momenta of the black holes. The problem of matching the 
amplitude and phase of the eccentricity oscillations can be cast as a minimization problem, and its
solution semi-automated. 

We describe our approach in more detail in Sec.~\ref{sec:description_of_ER_method},
and illustrate it with 
simple PN examples (which avoid troublesome gauge and noise issues). 
In Sec.~\ref{sec:NR} we turn to full NR simulations. We first describe a method
to filter the GW signal, and make the crucial observation that the 
eccentricities measured from the phase of the GW strain $h$ and the Newman-Penrose
scalar $\Psi_4$ are not the same; this point is elaborated further with a 1PN calculation
in Appendix~\ref{sub:eccentricity_estimators_for_gw}. We then apply our method to three NR 
configurations. 

In Sec.~\ref{sec:implementation} we develop a systematic procedure to determine 
the momentum adjustment factors, which makes it possible to semi-automate our method. 
We also make some estimates of the computational overhead of applying our method in 
large parameter studies. 

The method could in principle use the orbital phase rather than the GW phase. We show
in Sec.~\ref{sub:gauge_dependence_of_the_orbital_motion}, however, that the orbital
frequency contains additional oscillations due to gauge effects, which make it difficult
to use it for eccentricity reduction below $e \sim 0.002$.

Our method can reach eccentricities below $10^{-3}$ in one or two steps. In 
Sec.~\ref{sub:errors} we  
demonstrate that, perhaps surprisingly, NR simulations with eccentricities 
even as high as $e \sim 0.01$ are unlikely to introduce noticeable errors into GW
searches or parameter estimation in the advanced-detector era, and that a target 
eccentricity of $e \sim 10^{-3}$ reduces phase oscillations to well below our most 
stringent current requirements on NR phase accuracy. 


\section{Description of the eccentricity reduction method} 
\label{sec:description_of_ER_method}

\subsection{Sketch of the eccentricity algorithm algorithm} 
\label{sec:ER-sketch}

In our numerical simulations we start with two black holes with masses $m_1$ and $m_2$ and 
spins $\mathbf{S}_1$ and $\mathbf{S}_2$, separated by a coordinate distance $D$. 
In this work the spins are both parallel or anti-parallel to the orbital angular momentum 
of the binary, so there is no precession and the orbital plane is fixed.
Given such a configuration, our goal is to estimate values of the radial and tangential
momenta $(p_r, p_t)$ that lead to inspiral of (in principle) arbitrarily
low eccentricity, i.e. \emph{quasi-circular} (QC) inspiral.
In this work we will formulate our method for moving-puncture evolutions of 
Bowen-York-puncture initial data, but it can be generalized to other approaches.

Our method is based on having a sufficiently accurate approximate model of the
frequency evolution of the GW signal as a function of the initial momenta, $\omega_\text{M}(p_t, p_r)$, 
for the same initial configuration as used in a numerical simulation.
We choose initial momenta $(p_r^0, p_t^0)$ for a first numerical simulation, such 
that the eccentricity in our model is zero, $e_M(p_r^0, p_t^0) = 0$. 
Since the model is approximate, the eccentricity in the waveform that results from 
a numerical simulation using these parameters, $e_\text{NR}^0$, will be in general non-zero.
However, we assume that the model, although not precisely faithful to a
full numerical simulation, does capture much of the dependence of the GW signal
(and its eccentricity) on the initial parameters. 

We then try to remove the eccentricity in our simulations by 
adjusting the initial parameters by the same amount as is required to {\it produce} the same eccentricity
in the model solution. This basic idea was already presented in previous 
work~\cite{Husa:2007rh,Hannam:2010ec} and we will justify that this is a valid assumption in 
Sec.~\ref{sub:requirements_on_the_model}.
In those applications we adjusted the tangential and radial momenta by the same factor, 
i.e., we found a factor $\lambda$ such that $e_M(\lambda p_r^0, \lambda p_t^0) = e_\text{NR}^0$,
and then updated the parameters by $(p_r^1, p_t^1) = (p_r^0, p_t^0)/\lambda$. 
This procedure does not allow for the separate identification of $p_r$ and $p_t$, placing a 
lower limit on the eccentricity that can be obtained. 
In addition, we measured the eccentricity using the puncture motion of the two black holes. 
This motion is gauge dependent (and indeed the motion is {\it entirely} due to the gauge choice), and 
our experiments show (see Sec.~\ref{sub:gauge_dependence_of_the_orbital_motion}) that this means 
that we cannot reduce the eccentricity to lower than about $e \sim 0.002$. 

Eccentricity is only uniquely defined for conservative Newtonian dynamics. Based on an expansion of 
analytic solutions to the Kepler problem for small eccentricities one can define eccentricity estimators 
(see appendix~\ref{sub:newtonian_eccentricity_estimators} or \cite{Mroue:2010re} for definitions).
Due to the lack of a unique eccentricity for BH-binary evolution it is important to understand how different 
eccentricity estimators are related. In appendix~\ref{sec:1pn_eccentricity_estimators} we make such a 
comparison for a 1PN binary and the GW signal obtained from the quadrupole formula.
To estimate the eccentricity from NR data, we employ the eccentricity estimators
\begin{equation}
  \label{eq:e_phi_GW}
  \ephiGW(t) := \frac{\phi_\text{GW}(t) - \phi_\text{GW,fit}(t)}{4}
\end{equation}
and
\begin{equation}
  \label{eq:e_omega_orb}
  \eomOrb(t) := \frac{\Omega(t) - \Omega_\text{fit}(t)}{2\Omega_\text{fit}(t)},
\end{equation}
where $\phi_\text{GW,fit}(t)$ and $\Omega_\text{fit}(t)$ are approximations to the non-eccentric phase and
frequency, obtained via a fit over several orbital periods. $\Omega$ is the orbital frequency and 
$\phi_\text{GW}$ is the GW phase obtained from the Newman-Penrose scalar $\Psi_4$ at some 
fixed extraction radius $r_\text{ex}$.
The eccentricity is estimated from the amplitude of the eccentricity oscillations.

To determine separately the radial and tangential momenta, we consider the eccentricity function Eqn.~(\ref{eq:e_omega_orb}), which can be written with reference to the GW frequencies from the NR simulation 
and the approximate model,
\begin{equation}
  \label{eq:e_NR}
  e_\text{NR}(t) = \frac{\omega_\text{NR}(t) - \omega_\text{M}(t)}{2 \omega_\text{M}(t)}.
\end{equation} 
In practice the frequency from the NR waveform, $\omega_\text{NR}$, will be very noisy, and it is more 
convenient to consider the \emph{residual}
\begin{equation}
  \label{eq:R_NR}
  \mathcal{R}(t) = \omega_\text{NR}(t) - \omega_\text{M}(t).
\end{equation} 
If we perturb the momenta by the factors $(\lambda_r, \lambda_t)$, then we
can also calculate a residual between the perturbed model and the non-eccentric model,
\begin{equation}
  \RM^{\lambda}(t) := \RM(\lambda_r, \lambda_t; t) = \omega_\text{M}(\lambda_r p_r^0, \lambda_t p_t^0; t) - \omega_\text{M}(t).
\end{equation} 
Our modified method then consists of choosing $(\lambda_r, \lambda_t)$ such that 
\begin{equation}
  \label{eq:match_residuals_vague}
  \RM^{\lambda}(t) \approx \mathcal{R}(t),
\end{equation}
with agreement both in the
\emph{amplitude} and \emph{phase} of the residuals. Having determined these factors 
$(\lambda_r^0, \lambda_t^0)$ for the first NR residual $\mathcal{R}^0(t)$, we produce 
updated parameters for the next numerical simulation, 
\begin{eqnarray}
  p_r^1 & = & p_r^0 / \lambda_r^0,\label{eq:p_update_1_r}\\
  p_t^1 & = & p_t^0 /  \lambda_t^0.\label{eq:p_update_1_t}
\end{eqnarray} 

We then perform a second numerical simulation using $(p_r^1, p_t^1)$. If our technique works, then
the waveform produced in this simulation will contain less eccentricity, $e_\text{NR}^1 < e_\text{NR}^0$. 
The entire process is then repeated, and successive updates are made, 
\begin{eqnarray}
  p_r^{i+1} & = & p_r^i /  \lambda_r^i,   \label{eq:p_update_i_r}\\
  p_t^{i+1} & = & p_t^i /  \lambda_t^i ,  \label{eq:p_update_i_t}
\end{eqnarray} until the eccentricity has fallen below some desired threshold. 

This is our eccentricity reduction procedure. What remains is to specify the model of the GW phase and
frequency evolution, procedures filter the NR GW signal for analysis, and a method
to locate the optimal $(\lambda_r, \lambda_t)$ parameters. We now focus on these issues.

We expect that the efficiency of any iterative procedure will depend on the fidelity of the 
model $\omega_\text{M}$ to the results of a numerical simulation. In our procedure, we use 
as our model solutions of the EOB equations of motion for nonspinning point particles, 
augmented by the highest known PN spin effects. Details are given in 
Appendix~\ref{sec:eob_pn_equations_of_motion}.
We produce the initial guess for $(p_r^0, p_t^0)$ using
the same procedure as in our past work~\cite{Husa:2007rh,Hannam:2010ec}: we solve the 
PN/EOB equations of
motion with a large initial separation (typically $\sim 40M$), using initial momenta from
PN circular-orbit expressions (where $p_r = 0$), so that the eccentricity has essentially
reduced to zero through radiation reaction by the time the solution reaches $D \sim 10M$, 
where we are interested in starting a full numerical simulation, and at this point we
read off the parameters $(p_r^0, p_t^0)$. 

Since we base our model for the GW signal on the EOB orbital frequency $\Omega_\text{EOB}$
we take into account the retardation of the GW and the relation between the orbital and GW frequencies.
We define the EOB model 
$\omega_\text{M}$ and the NR frequency $\omega_\text{NR}$ for the GW frequency at a finite extraction radius $r_\text{ex}$ as 
\begin{align}
  \omega_\text{NR}(t) &:= \omega_\text{GW}(t + r_\text{ex})\\
  \omega_\text{M}(p_r, p_t; t) &:= 2\Omega_\text{EOB}(p_r, p_t; t).
\end{align}

It is also in principle possible to apply the same method to the orbital frequency of the puncture motion,
or to the separation of the two punctures. We find that this works adequately well if we place only moderate 
requirements on the final eccentricity. In these cases we use 
$\Omega_\text{M}(p_r, p_t; t) = \Omega_\text{EOB}(p_r, p_t; t)$, or $r_{\text{M},\text{orb}}(p_r, p_t; t) = r_\text{EOB,orb}(p_r, p_t; t)$, respectively, instead of $\omega_\text{M}(p_r, p_t)$. An orbital NR frequency residual can then be defined as
\begin{equation}
  \label{eq:R_orb}
  \mathcal{R}_\text{orb}(t) = 2 \left(\Omega(t) - \Omega_\text{M}(t)\right),
\end{equation}
to replace $\mathcal{R}$ in Eqn. \eqref{eq:match_residuals_vague}.
This residual will be used in the PN/EOB example \ref{sec:PN-EOB-example}, where we use only orbital quantities, and in section~\ref{sec:NR_example} as a comparison to $\mathcal{R}$. In full NR examples, however, we will
see in Sec.~\ref{sec:NR_example} that this method does not allow us to achieve eccentricities below about 
$e \sim 0.002$.


\subsection{Requirements on the model} 
\label{sub:requirements_on_the_model}

We return to the main idea that the eccentricity can be reduced by finding scale factors $\lambda$ that fulfill 
\begin{equation}
  e_M(\lambda_r p_r^0, \lambda_t p_t^0) = e_\text{NR}^0 > 0
\end{equation}
and then using $(p_r^0 / \lambda_r, p_t^0 / \lambda_t)$ as new initial parameters.
We would like to make two points related to this assumption.

Given two approximants $A$ and $B$, which could be EOB, PN or NR, let one of the two, say $A$, play the role 
of the model and the other the role of `NR', i.e., we want to find $\lambda$ such that 
$e_A(\lambda_r p_r^0, \lambda_t p_t^0) = e_B^0$.
Define QC parameters $p_A, p_B$ for each model that satisfy
\begin{equation}
  e_A(p_A) = e_B(p_B) = 0.
\end{equation}
Assuming that there is only a single eccentricity minimum and $p_A \neq p_B$, it then follows that $e_A(p_B) > 0$ 
and $e_B(p_A) > 0$. The values $e_A(p_B)$ and $e_B(p_A)$ can be interpreted as a distance 
between the two QC parameter sets $p_A, p_B$ for the two approximants $A$ and $B$. If these approximants 
model the binary evolution up to a similar order of accuracy, then it is natural to assume the symmetry
\begin{equation}
  e_A(p_B) = e_B(p_A).
\end{equation}
PN and EOB evolutions are symmetric in this sense up to numerical accuracy in determining the eccentricity. 
For equal-mass non-spinning inspiral PN/EOB QC data at a separation of $D = 12M$ we find that 
$e_\text{PN}(p_\text{EOB}) = 0.003192$, 
$e_\text{EOB}(p_\text{PN}) = 0.003179$, an asymmetry of merely $0.4\%$. This distance will vary depending 
on the location in the BBH parameter space, i.e., on the mass ratio and spins of the black holes, and on the 
initial separation. 

The symmetry is weaker between NR and PN/EOB. This is to be expected, for two main reasons. The first is
physical: PN and EOB evolutions differ from each other only in higher-order PN contributions, while the NR 
waveforms capture the full general relativistic physics. The second reason is related to gauge: the PN/EOB
parameters formally map to the Bowen-York-puncture parameters only up to 2PN 
order~\cite{Jaranowski:1997ky}, and that is
only true for the initial data. From the construction of the initial data, through the gauge changes that the 
wormholes undergo at early times in a moving-puncture 
simulation~\cite{Hannam:2006vv,Hannam:2006xw,Hannam:2008sg}, up to the point where the gauge
has settled down after approximately one orbit, there do not exist any quantitative predictions of the relationship
between the momenta in a PN/EOB calculation and the physical momenta during the inspiral in the 
moving-puncture simulation. All we have are observations that suggest that there is a close relationship 
between the PN/EOB and NR momenta (see Sec. V D of Ref.~\cite{Hannam:2007ik}). 
However, the condition $e_A(p_B) = e_B(p_A)$ is not strictly required by our method.

Rather, the crucial assumption is that the \emph{behavior} of the model and its eccentricity $e_M(p)$ in the 
vicinity of the model QC parameters $p_M$ is close to that of $e_\text{NR}(p)$ near its QC parameters $p_\text{NR}$. 
That is, the gradient $\nabla_p e_M$ should be close to $\nabla_p e_\text{NR}$ in a region around the 
respective QC solutions extending to the highest required eccentricity, say $10^{-2}$. We have checked 
that this is indeed the case by comparing $\pd{e}{p_t}$ for NR and EOB data and the explicit Newtonian 
eccentricity formula in the sensitivity analysis given in section \ref{sec:e_sensitivity}; details are given 
in Sec.~\ref{sub:errors}

In addition, the model must be sufficiently faithful to the 
real physics to produce reasonable starting momenta for our procedure. This rules out Newtonian or 1PN 
models for our method. Obviously, the better the starting momenta, the less work is required in reducing 
eccentricity. Therefore it makes sense to use the highest order PN/EOB equations of motion available.


\subsection{An example with two post-Newtonian approximants} 
\label{sec:PN-EOB-example}

We will illustrate the procedure with a simplified example, where a PN 
solution plays the role of the NR simulation. In this way issues of numerical noise 
and gauge effects are removed, and we can focus only on the eccentricity reduction 
algorithm. The model remains the EOB solution described previously. Note that for
this illustration, we could equally well swap the roles of the EOB and PN solutions due to the symmetry 
discussed in Sec.~\ref{sub:requirements_on_the_model}.
The configuration is an 
equal-mass non-spinning binary with an initial separation of $D=12M$.

\begin{figure}[htbp]
 \centering
 \includegraphics[width=0.48\textwidth]{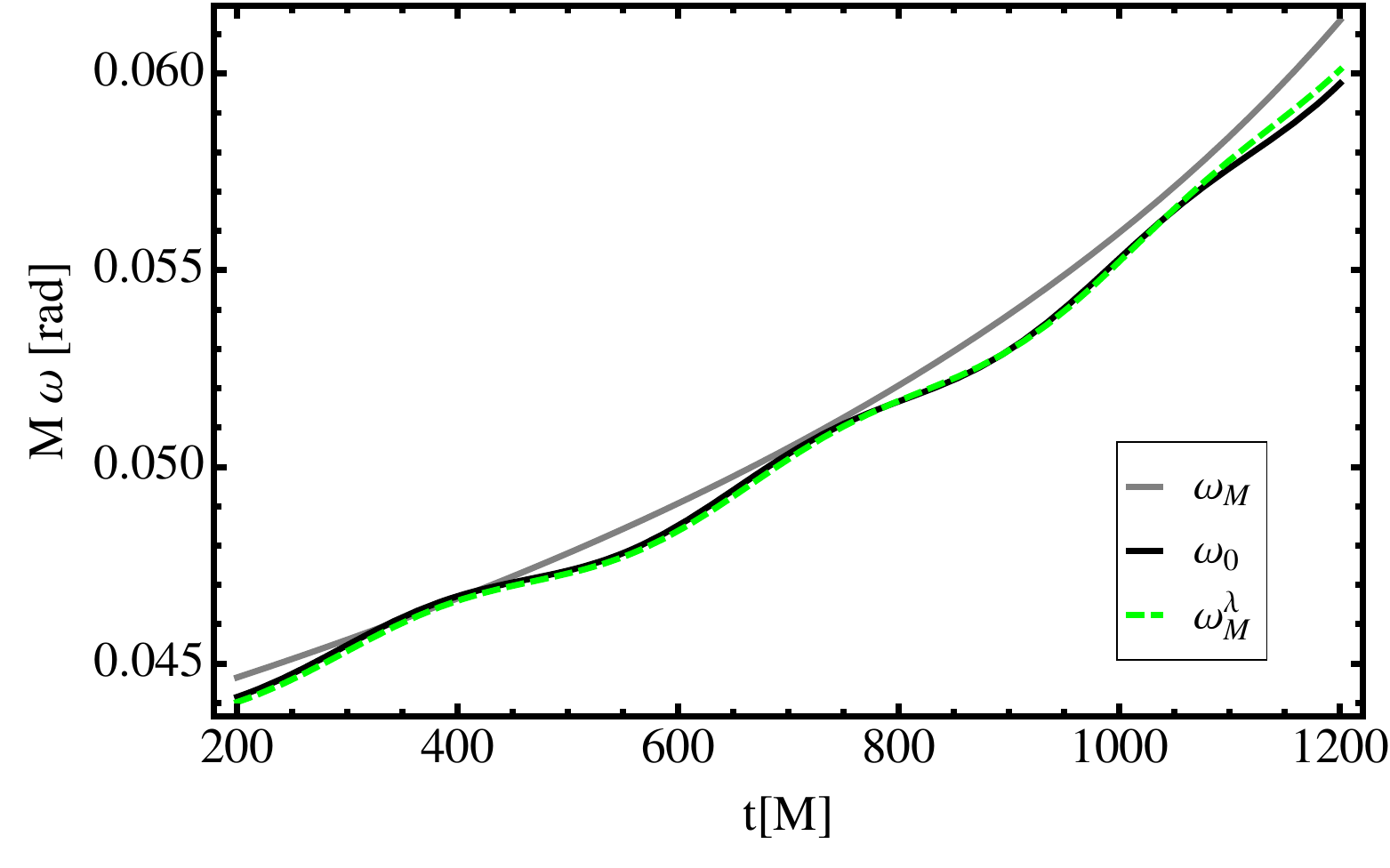}
 \caption{
    GW frequencies in the eccentricity reduction example. The grey line indicates the frequency of the 
    non-eccentric reference EOB solution, $\omega_\text{M}$. The frequency of the surrogate eccentric 
    ``NR'' simulation, $\omega_0$, is in black. The results of perturbing the initial momenta of the reference 
    EOB solution are also shown for $\lpt = 1.0015$ (green dashed: $\omega_\text{M}^\lambda$).
  }
 \label{fig:purePNFrequencies}
\end{figure}

\begin{figure}[htbp]
 \centering
 \includegraphics[width=0.48\textwidth]{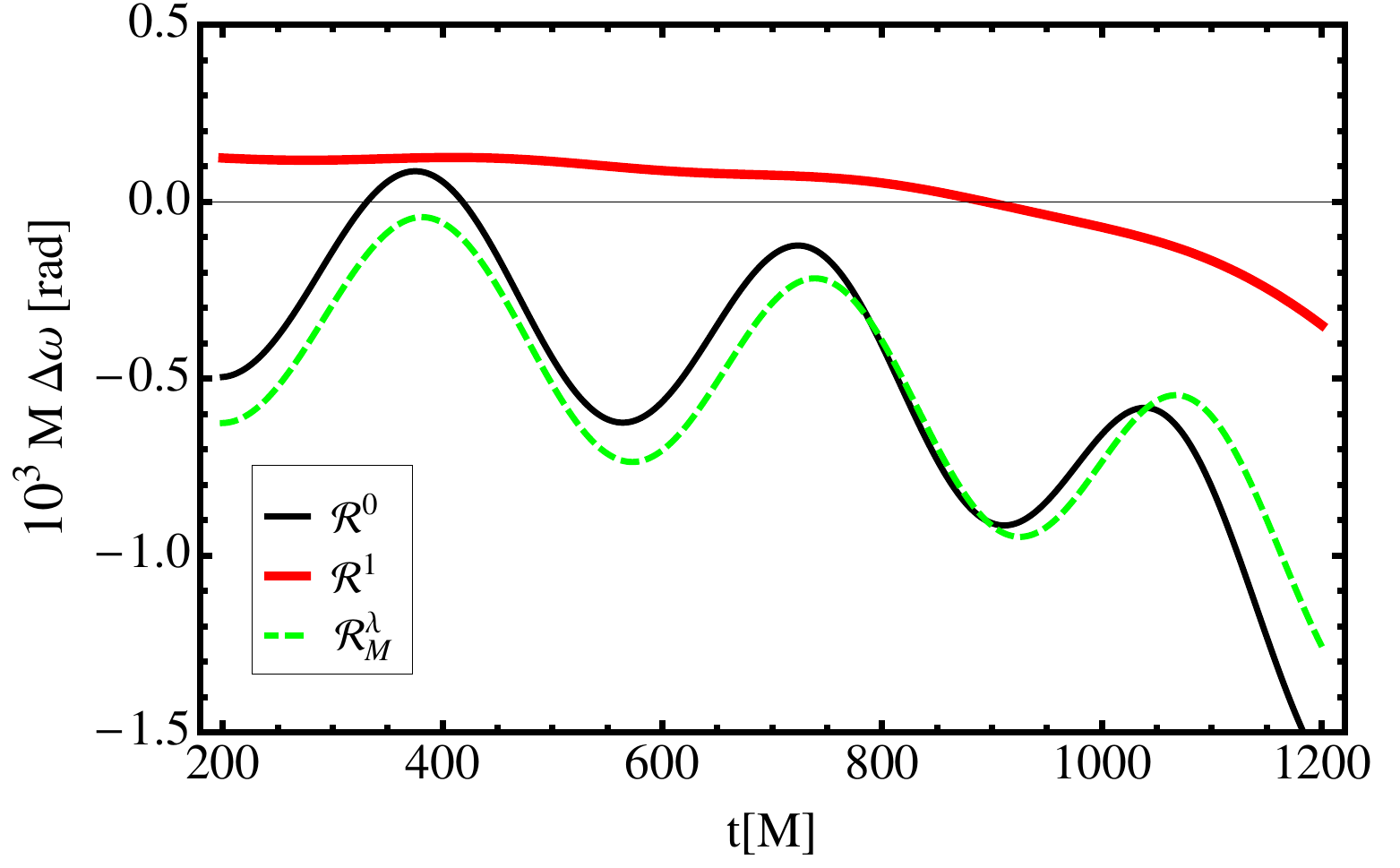}
 \caption{
    Same as Fig.~\ref{fig:purePNFrequencies}, but now showing the frequency residuals (\ref{eq:R_orb}). 
    The additional red curve shows the ``NR'' evolution with 
    the improved initial parameters. In the first iteration the eccentricity has been reduced by a factor of 40,
    $e_1 \sim 8 \times 10^{-5}$. 
    }
 \label{fig:purePN01}
\end{figure}

For the ``NR'' simulation using $(p_r^0, p_t^0)$, the eccentricity is $e_\text{NR}^0 \sim 0.003$. 
The ``NR'' frequency $\omega_0$ and its residual $\mathcal{R}^0$ are shown by the black solid line in 
Figs.~\ref{fig:purePNFrequencies} and~\ref{fig:purePN01}. For the first eccentricity reduction step 
we choose $\lpr = 1$ (i.e., we do not alter the radial momenta). We make a guess for the 
perturbation factor $\lpt$ and calculate $\RM^\lambda$ using the perturbed initial parameters.
The perturbation factor $\lpt$ is then adjusted until good agreement is achieved.
In Section \ref{sec:full_er_algorithm} we will describe an algorithm to automatize this adjustment; here we will
find a good agreement between $\RM^\lambda$ and $\mathcal{R}^i$ ``by eye''. 
Figs.~\ref{fig:purePNFrequencies} and \ref{fig:purePN01} illustrate the effect on 
$\Omega$ and $\RM^\lambda$ of choosing $\lpt = 1.0015$ (green dashed lines). 
Having obtained acceptable scale factors $\lambda^* = (1, 1.0015)$, the new initial parameters 
$(p_r^1, p_t^1)$ are calculated according to Eqns.~\eqref{eq:p_update_1_r} and \eqref{eq:p_update_1_t}. 
This first step results in a reduction of the eccentricity by a factor of 40, and the corresponding 
``NR'' residual $\mathcal{R}^1$ is shown by the thick red line in Fig.~\ref{fig:purePN01}.

In the second step of the example, we have to deal with the problem of `dephasing' between 
the ``NR'' and model residuals. In Fig.~\ref{fig:purePN01} we can already see that the EOB residual 
(the green dashed line) is slightly out of phase with $\mathcal{R}^0$. This dephasing becomes more pronounced
as the eccentricity is reduced (see the residual $\mathcal{R}^1$), and has to be removed if we are to continue with 
the procedure.

A generic (frequency) residual 
\begin{equation}
  \mathcal{R}(t) = A\cos(\Omega_r t + \Phi)  + \mathcal{D}(t),
\end{equation}
is composed of sinusoidal oscillations of frequency $\Omega_r$ and amplitude A (which is directly related to 
eccentricity via equations \eqref{eq:e_NR}) and \eqref{eq:R_NR}, plus a non-oscillatory contribution, 
which stems from the different phase evolutions between the simulation or perturbed model 
and the model waveform. It is this non-oscillatory contribution that gives rise to the dephasing in the residuals. 

To resolve this problem, we note that our eccentricity reduction method aims to capture only the eccentricity-related 
oscillations in the residual,  and the non-oscillatory part contains no useful information. To remove the 
non-oscillatory part we perform a fit in time to each residual and subtract it to obtain the `residual modulo dephasing'
\begin{equation}
\label{eq:RminusFit}
  \R{}(t) := \mathcal{R}(t) - {\mathcal{R}}_\text{fit}(t).
\end{equation}
As the fitting model we choose either a polynomial of order $n$
\begin{equation}
\label{eq:Rfit}
  {\mathcal{R}}_\text{fit}(t) = \sum_{i=0}^n C_i \, t^i,
\end{equation}
or the rational model given in equation \eqref{eq:model_dephasing}. 
A polynomial fit is more robust, but its order (usually 4 or 5) needs to be adjusted according to the length of the
time interval (the `fitting window') used for the fit, to avoid picking up parts of the eccentricity oscillations. 
In addition it is advantageous to discard data in the resulting function $\R{}(t)$ near the boundaries of the fitting 
window to reduce artifacts.

\begin{figure}[htbp]
 \centering
 \includegraphics[width=0.5\textwidth]{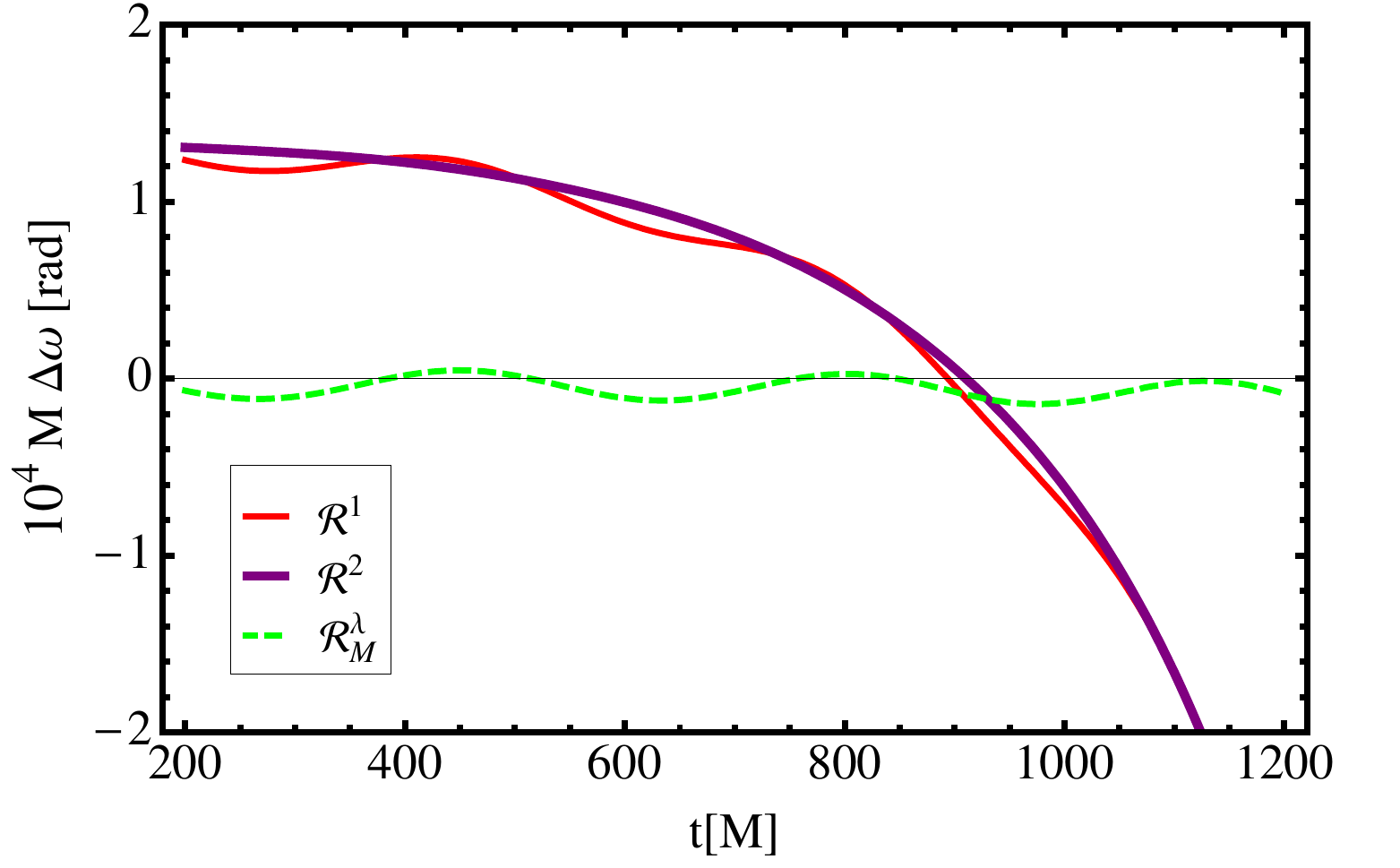}
 \hspace*{-0.25cm}\includegraphics[width=0.48\textwidth]{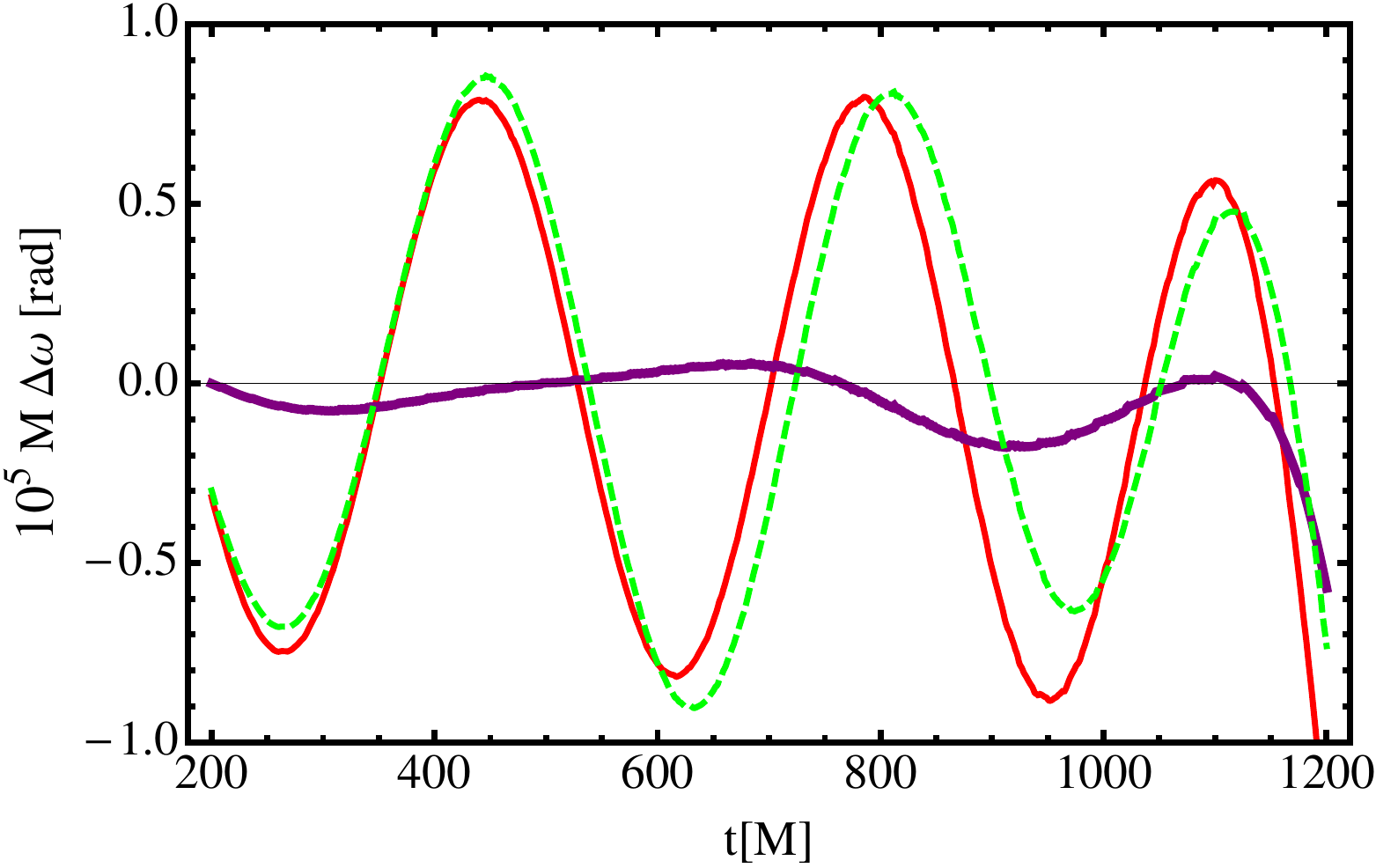}
 \caption{Frequency residuals in the second step of the eccentricity reduction example. 
 Top panel: Raw frequency residuals $\mathcal{R}^1$ (as in Fig.~\ref{fig:purePN01}), and $\mathcal{R}^2$,
 which is the result of this eccentricity reduction step. Also shown are the perturbed-EOB residuals, which cannot
 be easily compared to $\mathcal{R}^1$. Lower panel: Now the dephasing has been removed from $\mathcal{R}^1$,
 making it possible to determine the appropriate perturbation factors. See text for more details.
}
 \label{fig:purePN23}
\end{figure}

We now return to the second reduction step in our example. In the top panel of Fig.~\ref{fig:purePN23}, the 
solid red line indicates the same $\mathcal{R}^1$ as in Fig.~\ref{fig:purePN01}. When we remove the 
dephasing, we recover the solid red line in the lower panel. The eccentricity oscillations are now clearly 
visible, and we can again search for appropriate perturbations $(\lpr,\lpt)$ to the reference EOB solution to
model this residual. We find by trial and error that the optimal perturbation parameters are given by 
$\vec\lambda^* = (\lpr^*, \lpt^*) = (1.013, 1.000015)$.
Fig.~\ref{fig:purePN23} shows the results of applying $\vec\lambda^*$ to the initial parameters of the 
reference EOB solution. 
The perturbation $\RM^\lambda$ matches very well with the PN residual. (This perturbation is also shown in the
top panel of Fig.~\ref{fig:purePN23}, where we see that the matching cannot be performed without first
removing the dephasing in the residual $\mathcal{R}^1$.) Improved initial parameters are then 
obtained by adjusting the momenta: $p_r \rightarrow p_r / 1.013$, $p_t \rightarrow p_t / 1.000015$. 
These momenta result in a further reduction of an order of magnitude of the eccentricity, which leaves us 
with $e_2 \sim 8 \times 10^{-6}$. The momenta, eccentricities and scale factors for each iteration are
given in Tab.~\ref{tab:PN_EOB_q1_results}.

\begin{table}[h]
  \catcode`#=\active\def#{\phantom0} 
  \begin{tabular}{cccc@{\hspace{1.5mm}}||@{\hspace{1.5mm}}ccc}
    \hline
    Iteration & $p_\text{r}$ & $p_\text{t}$ & $\eomOrb$ & $\lpr$ & $\lpt$\\
    \hline
     0 & 0.000541  & 0.0851657   & $0.003##$          & 1     & 1.0015\\
     1 & 0.000541  & 0.0850382   & $8\times 10^{-5}$  & 1.013 & 1.000015\\
     2 & 0.000534  & 0.0850369   & $8\times 10^{-6}$  &       &        \\
    \hline
  \end{tabular}
  \caption{Initial momenta, eccentricity estimates $\eomOrb$ and results for the equal-mass non-spinning PN/EOB example eccentricity reduction case discussed in the text.
  }
  \label{tab:PN_EOB_q1_results}
\end{table}
  

\section{Application to NR simulations} 
\label{sec:NR}

We now apply our eccentricity reduction procedure to full NR simulations. 
In Sec.~\ref{sec:nr_setup} we summarize our numerical methods, and in Sec.~\ref{sec:filtering} 
we summarize our procedure to produce a clean GW signal, which is the key ingredient in
our procedure. The procedure itself is then applied to three non-precessing black-hole-binary
configurations in Sec.~\ref{sec:NR_example}.

\subsection{NR Setup} 
\label{sec:nr_setup}

Our numerical setup is similar to that used in Ref.~\cite{Hannam:2010ec}, but for completeness 
we repeat the details here. We performed numerical simulations with the BAM code
\cite{Brugmann:2008zz,Husa:2007hp}. 
The code starts with black-hole-binary puncture initial data 
\cite{Brandt:1997tf,Bowen:1980yu} generated using a pseudo-spectral 
elliptic solver~\cite{Ansorg:2004ds}, and evolves them with the $\chi$-variant of the
moving-puncture \cite{Campanelli:2005dd,Baker:2005vv} version of the BSSN
\cite{Shibata:1995we,Baumgarte:1998te} formulation of the 3+1 Einstein 
evolution equations. Spatial finite-difference derivatives are
sixth-order accurate in the bulk \cite{Husa:2007hp}, Kreiss-Oliger
dissipation terms converge at fifth order, and a fourth-order Runge-Kutta
algorithm is used for time evolution. 
The gravitational waves emitted by the binary are calculated from the
Newman-Penrose scalar $\Psi_4$, and the details of our implementation of
this procedure are given in \cite{Brugmann:2008zz}.

In each simulation, the black-hole punctures are initially a coordinate distance
$D$ apart, and are placed on the $y$-axis at $y_1 = -qD/(1+q)$ and $y_2 = D/(1+q)$,
where $q = M_2/M_1$ is the ratio of the black-hole masses in the binary, and we always choose 
$M_1 < M_2$. The masses $M_i$ are estimated from the Arnowitt-Deser-Misner
(ADM) mass at each puncture, according to the method described 
in~\cite{Brandt:1997tf}. (This measure becomes inaccurate for high spins~\cite{Hannam:2009ib},
but this does not preclude the application of the eccentricity reduction procedure presented
in this work.) The Bowen-York punctures are given 
momenta $p_x = \mp p_t$ tangential to their separation vector, and $p_y = \pm p_r$
towards each other. The spin parameter of a BH is defined as $\chi_i = S_i/M_i^2$.

All simulations used the ``$\chi$ variant'' of the moving-puncture method and six mesh-refinement 
buffer points. The base configuration consists of $l_1$ nested mesh-refinement 
boxes with a base value of $N^3$ points, which surround each black hole, and $l_2$ nested boxes with 
$(2N)^3$ points, which surround the entire binary system. The value of $l_1$ differs for each black
hole; in equal-mass simulations $l_1$ is the same for each black hole, but when the mass ratio is
1:2, the smaller black hole is given one extra refinement level, so that both black holes are equally
well resolved. In addition, we use $(4N)^3$ points on the level where wave extraction is performed,
to allow accurate wave extraction to larger radii. The levels immediately above and below this are given
an intermediate number of points (typically $(3N)^3$), so that no two levels are of the same size. 
The choices of $N$, $l_1$, $l_2$ and the resolutions are given in Tab.~\ref{tab:grids}. 
The resolution around the puncture is denoted by $M_1/h_\text{min}$.  

Far from the sources, the meaningful length scale is the total mass of the binary, 
$M = M_1+M_2$, and 
so the resolution on the coarsest level is given by $h_\text{max}/M$. We also give the resolution 
on the wave extraction level(s), $h_\text{ex}/M$. 

In section \ref{sec:NR_example} we will present results for eccentricity reduction for the 
configurations listed in table \ref{tab:grids}, with the exception of the equal-mass non-spinning configuration. The latter was used to 
study the dependence of gauge oscillations in the orbital quantities on the parameter $\eta$ that appears in 
the $\Gamma$-driver shift condition. These are discussed in Sec.~\ref{sub:gauge_dependence_of_the_orbital_motion}. 
In all other equal-mass cases, $\eta = 2/M$, and for unequal-mass configurations we 
used a spatially varying $\eta$~\cite{Muller:2009jx,Muller:2010zze,Alic:2010wu} with the  
functional form
\begin{equation}
  \eta(\vec x) = \eta_A + \frac{\eta_p - \eta_A}{1 + \left( \frac{(\vec x_p - \vec x)^2}{w^2} \right)^\delta},
\end{equation}
where we have chosen the asymptotic value $\eta_A = 2/M$, we have set 
$\eta_p = 3/M$ near the location $\vec x_p$ of the small BH and fixed the width 
$w = 2.67M$ and power $\delta = 2$, so that the modification falls off like $1/r^4$.

Since the focus of this paper is on eccentricity reduction, inspiral runs are sufficient. 
Where available, we also give the location of the amplitude maximum in time, $t_{peak}$, 
of the GW signal and the number of GW cycles, $N_\text{GW}$ for each  simulation. A full 
convergence series has been 
performed for the $q=2, \chi_1 = 0, \chi_2 = 0.25$ configuration only, which is used in 
Sec.~\ref{sub:phase_errors_mismatches_between_eccentric_hybrids} to compare 
phase errors and mismatches due to eccentricity with errors due to numerical resolution.

\begin{table*}
\caption{\label{tab:grids}
Summary of grid setup for numerical simulations. The grid parameters follow the
notation introduced in \cite{Brugmann:2008zz}; see text. $M_1/h_\text{min}$ denotes the resolution on the 
finest level with respect to the \emph{smallest black hole}, while $h_\text{max}/M$ is the resolution on the 
coarsest level with respect to the \emph{total mass}, $M = M_1+M_2$. The outer boundary of the 
computational domain is at $x_{i,\text{max}}/M$, where $x_i = \{x,y,z\}$. In general $l_1$ indicates the 
number of moving refinement levels around each puncture, and $l_2$ the number of large refinement
levels that encompass both punctures. In the $q=2$ configurations we use
three refinement levels around the puncture of the large black hole, and four around the other.
$h_\text{ex}/M$ is the resolution on the (main) wave extraction level. 
The simulations were started at an initial coordinate separation is $D$ and include 
$N_\text{GW}$ GW cycles before reaching the amplitude maximum in the GW signal $t_\text{peak}$. 
The starting momenta are given in section~\ref{sec:NR_example}.
}
\begin{ruledtabular}
\begin{tabular}{|| c | c | c | c | c | c | c | c | c | c | c | c ||}
$q$ & $\chi_1$ & $\chi_2$ & $D/M$ & $N$ & $(l_1,l_2)$ & $M_1/h_\text{min}$  & $h_\text{max}/M$ & $h_\text{ex}/M$ & $x_{i,\text{max}}/M$ & $t_\text{peak}/M$ & $N_\text{GW}$ \\[10pt]
\hline
\multicolumn{12}{||l||}{Aligned-spin simulations}\\
\hline
$2$ & $0$     & $0.25$  & 11.3 &80, 88, 96 &(3, 4; 9) & $32, 34.91, 40.73$ & $42.67, 39.11, 33.52$ & $0.67, 0.61, 0.52$ &$2048$ & $1915$ & $21.5$\\
$2$ & $-0.75$ & $-0.75$ & 12.6 &88    &(3, 4; 9) &$32$       &$42.67$     &$0.67$     &$2389$ & $N/A$  & $N/A$\\
$1$ & $0$     & $0$     & 12   &72, 80 &(5; 5)   &$24, 26.67$ &$10.67, 9.6$ &$1.33, 1.2$ &$768$  & $1972$ & $20.5$\\
$1$ & $0.5$   & $0.5$   & 11   &88    &(5; 5)   &$29.33$    &$8.73$      &$1.09$     &$768$  & $1732$ & $21$\\
\end{tabular}
\end{ruledtabular}
\end{table*}



\subsection{Filtering the numerical GW signal} 
\label{sec:filtering}

\begin{figure}[htbp]
  \centering
  \includegraphics[width=0.5\textwidth]{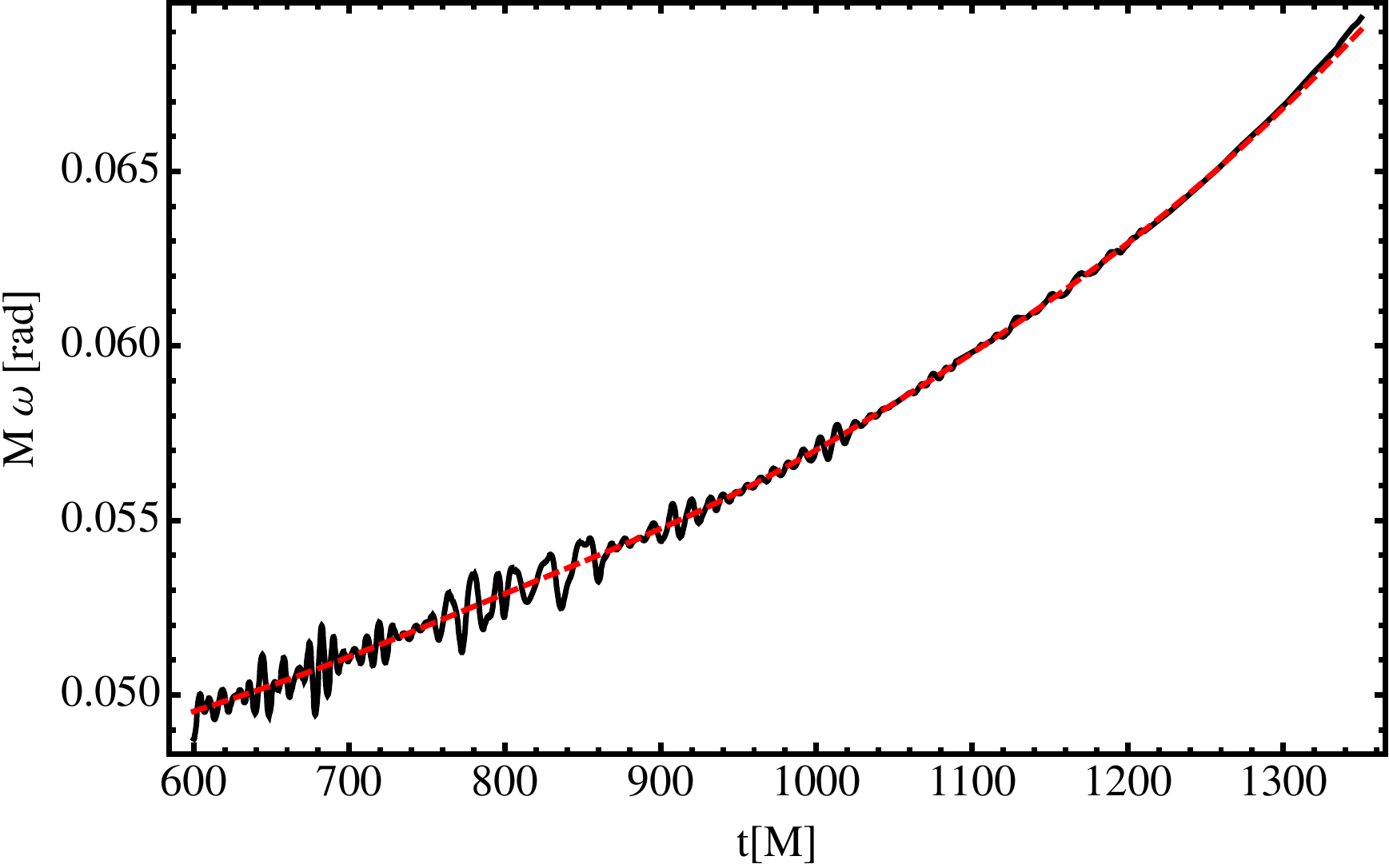}
  \caption{The GW frequency from an equal-mass non-spinning simulation.
  The black line shows the unfiltered frequency obtained from differentiating the GW phase.
  The red dashed line shows
  The filtered GW frequency Eqn.~\eqref{eq:omega_GW_cleaned}.
  }
  \label{fig:omega_cleaning}
\end{figure}

We use the gravitational wave frequency extracted at some finite extraction radius as a 
more gauge-invariant input quantity. We extract the $(\ell=2,m=2)$ mode of the 
Newman-Penrose scalar $\Psi_4$ and define the wave phase $\phiGW$ as 
$r \Psi_4^{22}(t) = A(t) e^{i \phiGW(t)}$. 
The GW frequency $\omGW$ is then obtained as the time derivative of the GW phase. 
The numerical noise in $\Psi_4$ prevents this definition from being directly useful for our setup. 
This is clear in Fig.~\ref{fig:omega_cleaning},
where the raw GW frequency from $\Psi_4$, from a simulation of an equal-mass nonspinning 
binary is shown in grey.

\begin{figure}[htbp]
  \centering
  \includegraphics[width=0.5\textwidth]{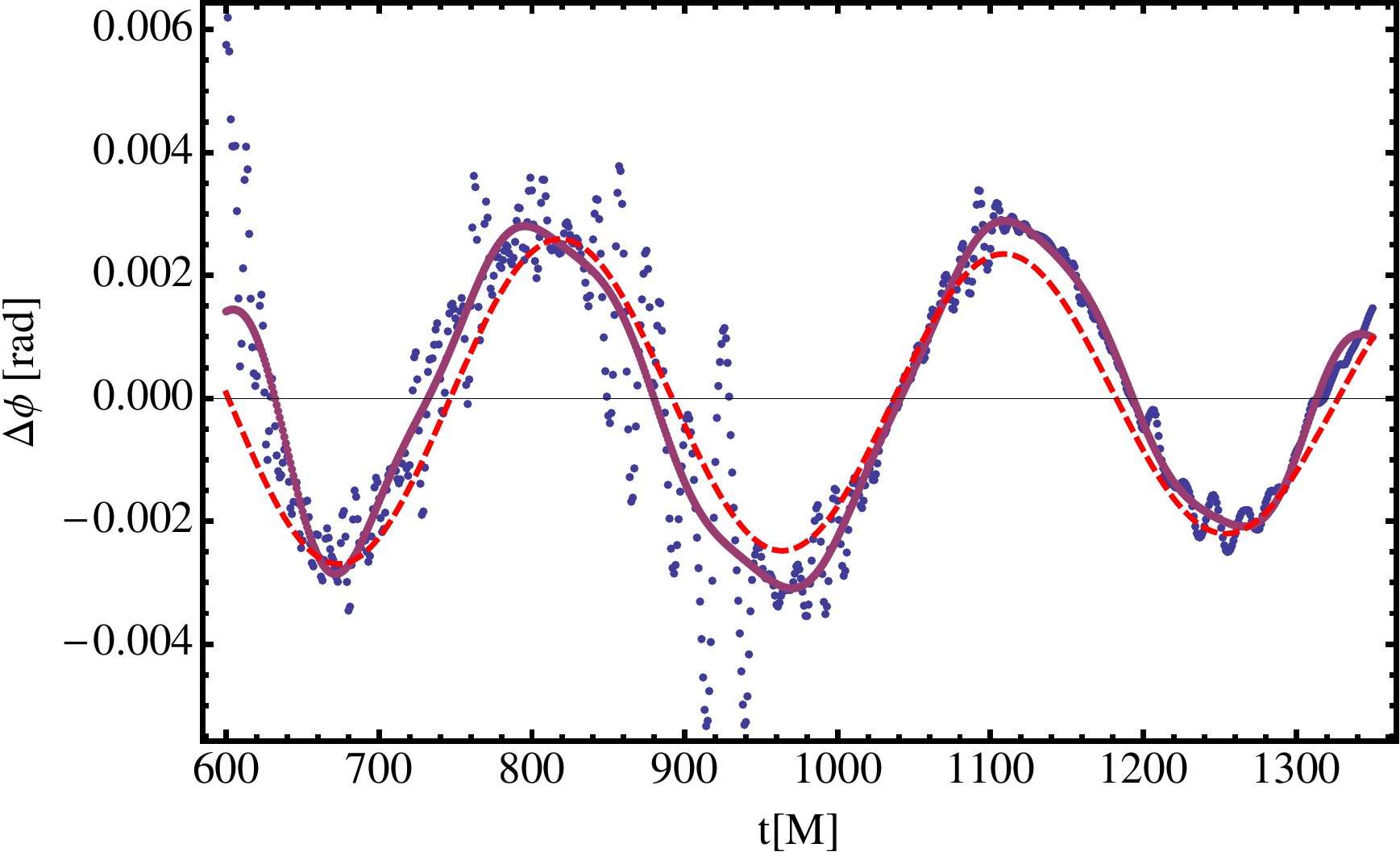}
  \caption{The GW phase from the same simulation as in Fig.~\ref{fig:omega_cleaning}, with the fit 
  (\ref{eq:GW-phase-fit-overall}) removed.  The blue points show the residual obtained after subtracting the 
  fit from the raw NR data. The thick magenta line shows the filtered residual. The red dashed line shows the 
  cosine fit $\mathcal{F}_L [\phi_\text{GW,res}](t)$.}
  \label{fig:phi_cleaning}
\end{figure}

The following fitting and filtering technique has allowed us to construct a cleaned residual eccentricity 
oscillation that can be used with our method. We extract the eccentricity oscillations from the phase $\phiGW$ 
by first removing the overall behavior of the phase via a polynomial fit of order $n$,
\begin{equation}
  \label{eq:GW-phase-fit-overall}
  \phi_\text{GW,fit}(t) = \sum_{i=0}^n C_i \, t^n\\
\end{equation} 
(usually $n=4,5$). Then we smooth the residual phase 
\begin{equation}
  \phi_\text{GW,res}(t) = \phi_\text{GW}(t) - \phi_\text{GW,fit}(t)  
\end{equation}
with a low-pass filter $\mathcal{F}_L$.  We have used either a `brick-wall' filter, where 
higher frequency modes are simply zeroed out in Fourier space, or a wavelet filter 
using Mathematica's discrete wavelet transform with an 8th-order Symlet 
wavelet~\cite{Daubechies92}. We then perform a nonlinear fit to a sinusoid,
\begin{equation}
  \label{eq:GW-phase-filtered-residual}
  \mathcal{F}_L [\phi_\text{GW,res}](t) = B \cos(\Omega_r t + \phi_0).
\end{equation}
Finally, we take an analytic time derivative of this fit, and reassemble a cleaned frequency 
quantity by adding the time derivative of $\phi_\text{GW,fit}(t)$ to obtain a cleaned GW 
frequency,
\begin{equation}
  \label{eq:omega_GW_cleaned}
  \omGWcl = \dot\phi_\text{GW,fit}(t) - B\Omega_r \sin(\Omega_r t + \phi_0).  
\end{equation}
The process is illustrated in Fig.~\ref{fig:phi_cleaning} for the NR GW phase from an equal-mass 
non-spinning simulation. Compared to orbital quantities, the GW signal is usable only after the 
junk radiation has passed, at around $500M$. This necessitates slightly longer simulations, 
so that 2-3 periods worth of residual oscillations are available. The resulting cleaned GW frequency
is shown along with the raw frequency in Fig.~\ref{fig:omega_cleaning}. 
It is worth noting that in some cases we can use the cleaned GW frequency at times earlier than the region used for the nonlinear fit, but in general this can lead to errors in matching to the phase of the oscillations.

\subsection{Measuring the eccentricity from the GW signal: \\ strain $h$ vs $\Psi_4$}
\label{sec:StrainVsPsi4}

As mentioned earlier, in our NR simulations we extract the GW signal from the Newman-Penrose
scalar $\Psi_4$. We then estimate the eccentricity from the phase of the $(\ell=2, m=2)$ mode
of $\Psi_4$ using Eqn.~(\ref{eq:e_phi_GW}). 

The measurable quantity in GW experiments is not $\Psi_4$, but the strain $h$, which is related
to $\Psi_4$ by two time derivatives. If we integrated $\Psi_4$ twice with respect to time, we
could instead calculate the eccentricity from the phase of $h$. 

We might naively expect that the ``eccentricity from the GW phase'' should be the same, irrespective
of how the GW signal is expressed. But this is not the case. Consider the complex GW strain as 
$h = A(t) e^{i \phi(t)}$. The eccentricity is related to the amplitude of oscillations in $\phi(t)$ 
with respect to the phase of a fiducial non-eccentric (quasi-circular) binary, 
$\phi(t) = \phi_{QC}(t) + 4 e_{\phi[h]}(t)$. If we now consider 
$\Psi_4 = \ddot{h}$ and express this also as $A'(t) e^{i \phi'(t)}$, then we see that after the 
application of two time derivatives we will not have $\phi(t) = \phi'(t)$. In practice the difference
between the two phases has been found to be very small, but the \emph{oscillations} in the
phase will increase in magnitude with each time derivative, and the difference between
$e_{\phi[h]}$ and $e_{\phi[\Psi_4]}$ will be significant. 

In Appendix~\ref{sub:eccentricity_estimators_for_gw} we consider the GW signal from an 
eccentric binary at the first-post-Newtonian (1PN) order, and find that the two eccentricities
are related by \begin{equation}
  \frac{e_{\phi[\psi_4]}}{e_{\phi[h]}} = 
  \frac{7}{4}-\frac{3 k \epsilon ^2}{2}, 
\end{equation} where $k$ is the periastron advance of the binary, and $\epsilon^2 = 1/c^2$ 
indicates the 1PN-order term. This relationship must be borne in mind when comparing
GW-phase eccentricities calculated from the strain and $\Psi_4$. 

To our knowledge this subtlety of GW-based eccentricity measures has not been noted in the literature to date.


\subsection{Numerical relativity examples} 
\label{sec:NR_example}

We are now ready to apply the eccentricity reduction procedure to NR simulations. We first
present an unequal-mass aligned-spin configuration with physical parameters $q=2$, $\chi_1 = 0$, and 
$\chi_2 = 0.25$. This is the example we will consider in the most detail, and in the 
remainder of the paper we will refer to this case as our ``reference example''.

As stated in the previous section, we need to skip the initial burst of junk radiation 
and evolve for about 3-4 orbits to get accurate estimates of the eccentricity based 
on the GW signal.

\begin{figure}[htbp]
  \centering
  \includegraphics[width=0.5\textwidth]{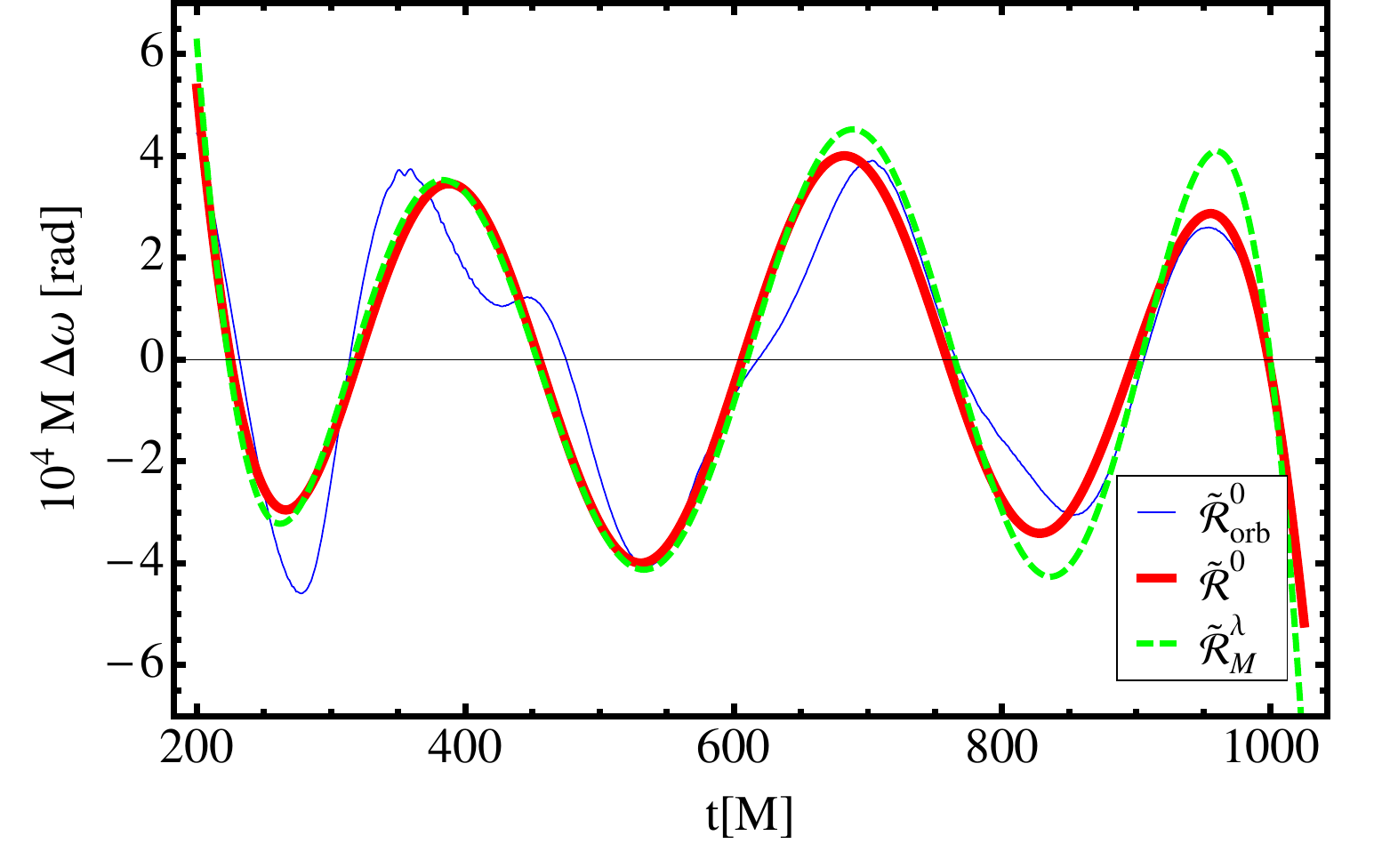}
  \includegraphics[width=0.5\textwidth]{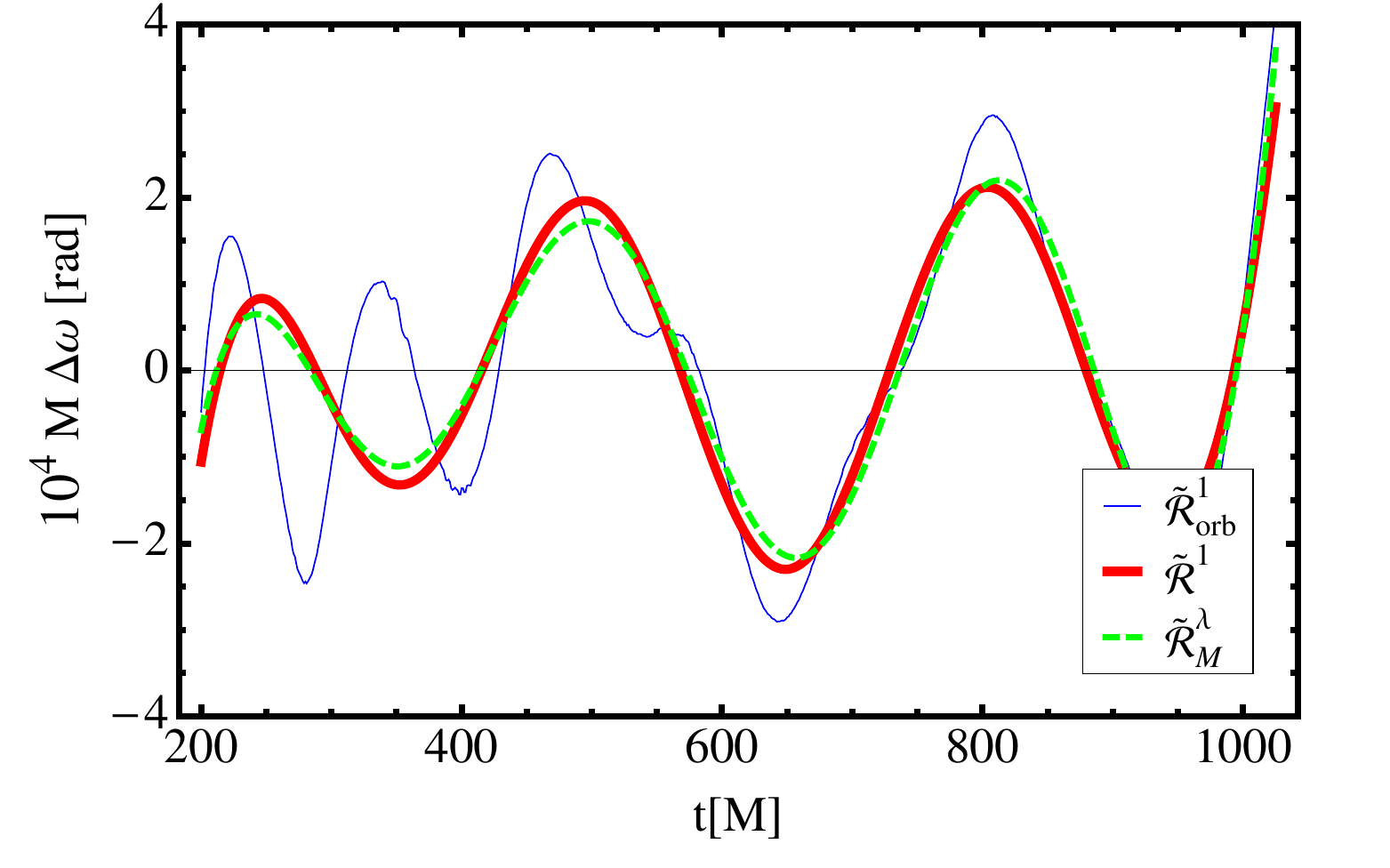}
  \includegraphics[width=0.5\textwidth]{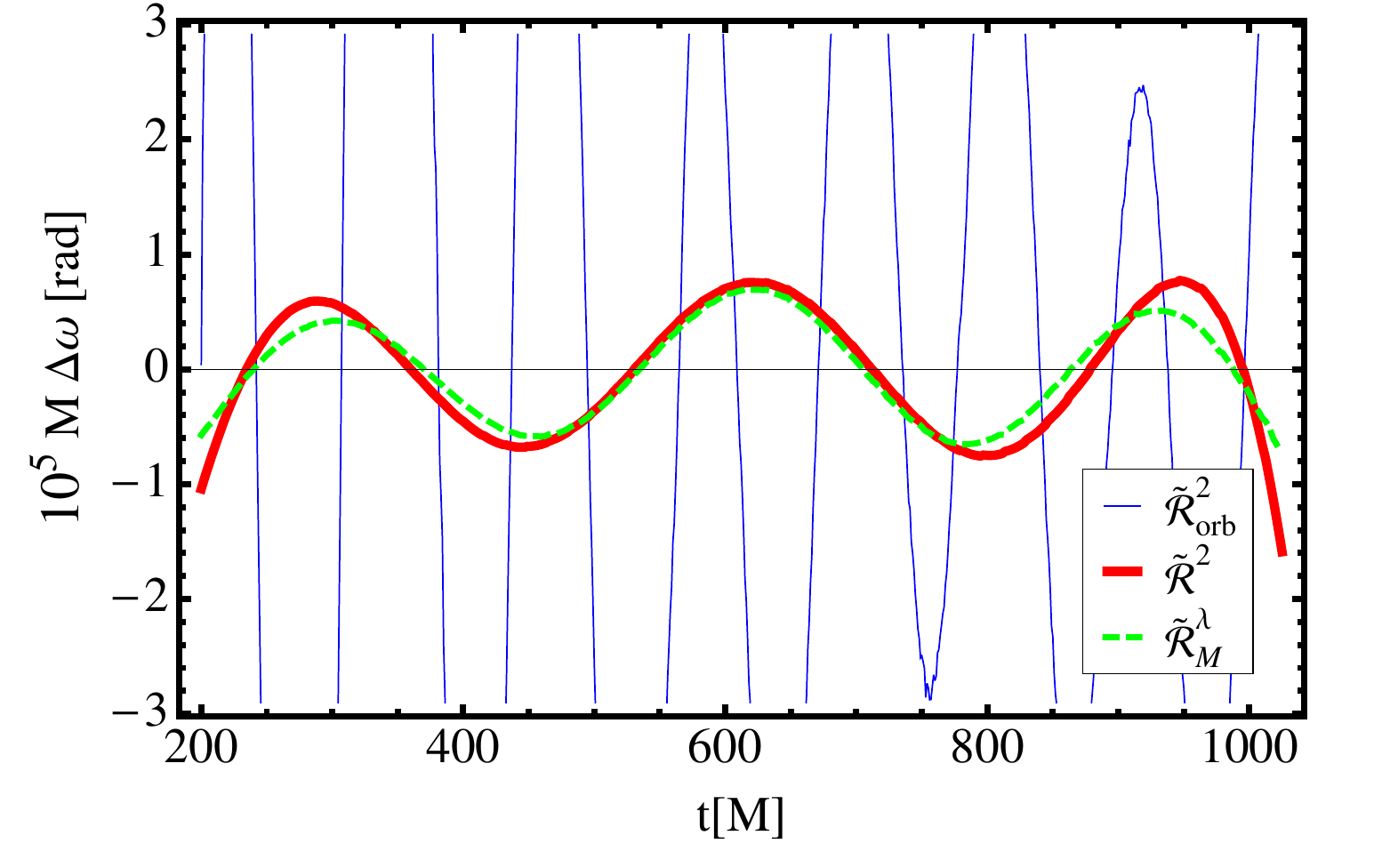}
  \caption{Eccentricity reduction steps for $q=2, \chi_1 = 0, \chi_2 = 0.25$. At each step the 
  NR residual $\R{}^i$ is calculated from the filtered GW signal $\omGW$ (red, thick line), 
  and for reference we also show $\R{orb}^i$ calculated from the orbital frequency $\omOrb$ 
  (thin blue line). The best-match perturbation parameters $\lambda^*$ lead to the residuals
  $\mathcal{R}^\lambda_M$ (green, dashed line), and are given in Tab.~\ref{tab:q2a0a025_results}, 
  which also provides the updated momenta. 
  The eccentricities are 0.006, 0.003, and $3 \times 10^{-4}$. 
  }
  \label{fig:q2a0a025_it}
\end{figure}

The PN/EOB QC initial parameters give rise to an initial eccentricity (measured from the GW phase) of 
$e_0 \approx 0.006$. The top panel of Fig.~\ref{fig:q2a0a025_it} plots the NR frequency residual 
calculated from both the GW phase and the orbital phase. For this level of eccentricity, 
we see that the two residuals agree well, and either could be used in the eccentricity reduction 
method. We find that a good match with residuals calculated from perturbing the reference EOB 
solution are obtained with $\lambda = (1, 1.0028)$. This match results in the adjustment 
$p_t \rightarrow p_t / 1.0028$. The best-match residual $\R{M}^\lambda$ is indicated by a 
green dashed line. 

The parameters that follow from the first iteration step lead to an eccentricity of 
$e_1 \approx 0.003$. The GW- and orbital-phase residuals from the subsequent simulation
are shown in the middle panel of Fig.~\ref{fig:q2a0a025_it}. Varying $\lambda$ to find a good match 
leads to the second adjustment of the initial momenta, $p_r \rightarrow p_r / 1.15$ and 
$p_t \rightarrow p_t / 0.999$. We see in this case that the orbital-phase residual now includes
some higher harmonics, and its amplitude is also different to that of the GW-phase residual;
it would now be difficult to obtain a reliable guess of the perturbation parameters based on the 
orbital-phase residual alone. 

The result of the second iteration step is shown in the lower panel of Fig.~\ref{fig:q2a0a025_it}.
The eccentricity is now $e_2 \approx 3 \times 10^{-4}$. In principle, we could take another step, 
and indeed the best-match perturbation residual has been calculated. 
However, the eccentricity is already very small, and as we discuss in Sec.~\ref{sec:errors_rex_fits}, 
for eccentricities on that 
order the uncertainty in estimating eccentricity and in calculating a cleaned GW phase is very 
high, around $50\%$.

\begin{figure}[htbp]
  \centering
  \includegraphics[width=0.5\textwidth]{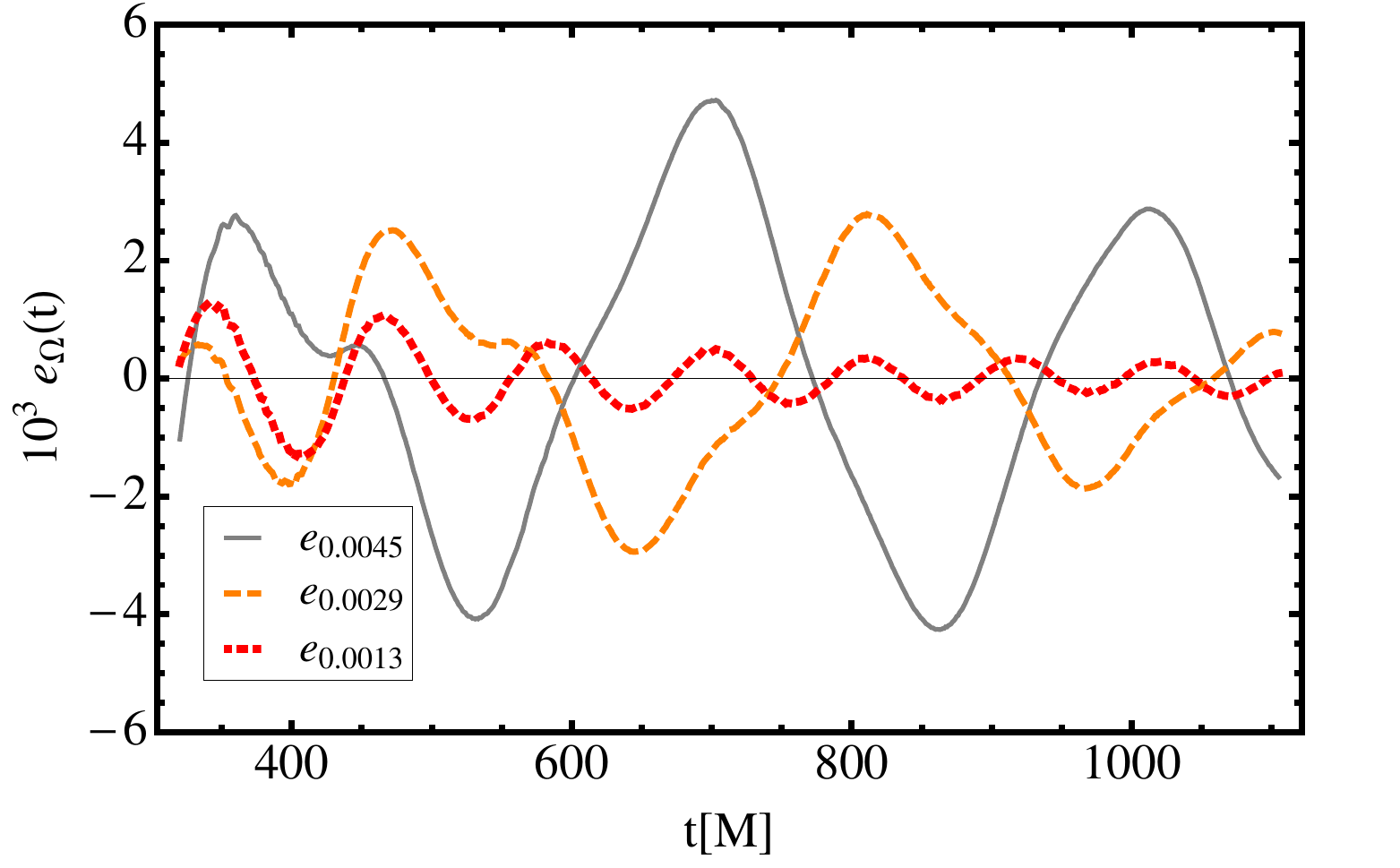}
  \caption{Eccentricity estimator based on the orbital motion, $\eomOrb$ for the example configuration.
  The oscillation period is about $330 M$ and the dominant feature in the original and first-iteration data, 
  shown in grey orange-dashed lines, respectively. At the second iteration step higher harmonics dominate 
  and result in an eccentricity (red, thick, small-dashed) that is considerably higher than the result of 
  $\ephiGW$. Refer to table \ref{tab:q2a0a025_results} for details. 
  }
  \label{fig:q2a0a025_e_omega_orb}
\end{figure}

The results for this example case are summarized in Tab.~\ref{tab:q2a0a025_results}. The progress of the 
eccentricity reduction method for two additional aligned-spin configurations is given in 
Tabs.~\ref{tab:q2aM075aM075_results} and \ref{tab:q1a05a05_results}. 
For all cases we provide eccentricity estimates from $\ephiGW$ along with the radial and tangential momenta 
and their scaling factors for each reduction step. 
To provide another example of the presence of higher harmonics in orbital quantities, we also give eccentricity 
estimates from $\eomOrb(t)$ for the example case in Tab.~\ref{tab:q2a0a025_results} and plot the orbital 
frequency estimator for the iteration steps in Fig.~\ref{fig:q2a0a025_e_omega_orb}. This plot again indicates
the unsuitability of the orbital phase for this method. We can see from 
both Fig.~\ref{fig:q2a0a025_e_omega_orb} and Tab.~\ref{tab:q2a0a025_results} that the orbital-motion-based
eccentricity estimator does not fall below 0.0013, which, if it were correct, would suggest a GW-phase eccentricity
of around 0.002. This is the basis of our claim that the orbital motion cannot be used to reduce to the 
eccentricity to below $e \sim 0.002$. We will study the behavior of the orbital motion further in 
Sec.~\ref{sub:gauge_dependence_of_the_orbital_motion}

We give two significant digits in $1-\lambda$. This choice is sensible in light of the discussion of errors  
in Sec.~\ref{sec:e_sensitivity}. Lastly, the uncertainties 
in the eccentricity values are about $25\%$ for $\ephiGW$ and, if we do not incorporate the contribution of 
gauge harmonics in the orbital frequency, about $5-10\%$ for $\eomOrb$. These uncertainties are discussed 
further in Sec.~\ref{sec:errors_rex_fits}. 

In summary, the data presented in this section demonstrate that starting from EOB QC initial parameters 
we can in general reach eccentricities well below $10^{-3}$ for aligned-spin NR configurations in two iteration 
steps. 


\section{Implementation issues} 
\label{sec:implementation}

\subsection{Determination of the scale factors}
\label{sec:full_er_algorithm}

In the previous PN and NR examples we determined the momenta scale factors $(\lpr,\lpt)$ by
trial and error. In this section we present a systematic procedure that can be automated.

\begin{table}[t]
  \catcode`#=\active\def#{\phantom0} 
  
  \subfloat[list of tables text 1]
  [
    Results for the example NR eccentricity reduction case discussed in the text, 
    $q=2, \chi_1 = 0, \chi_2 = 0.25$. 
    We give eccentricity estimates based on  both $\eomOrb$ and $\phi_\text{GW}$.
    ]
  {
    \begin{tabular}{ccccc@{\hspace{1.5mm}}||@{\hspace{1.5mm}}cc}
      \hline
      Iteration & $p_\text{r}$ & $p_\text{t}$ & $\ephiGW$ & $\eomOrb$
      & $\lpr$ & $\lpt$\\
      \hline
      0 & 0.000758 & 0.11710  & $0.006#$   & 0.0045  & 1    & 1.0028\\
      1 & 0.000758 & 0.11677  & $0.003#$   & 0.0029  & 1.15 & 0.999\\
      2 & 0.000660 & 0.11689  & $0.0003$   & 0.0013  &      &       \\ 
      \hline
    \end{tabular}
    \label{tab:q2a0a025_results}
  }

  \subfloat[list of tables text 2]
  [
    Results for the configuration $q=2, \chi_1 = -0.75, \chi_2 = -0.75$.
  ]
  {
    \begin{tabular}{cccc@{\hspace{1.5mm}}||@{\hspace{1.5mm}}cc}
      \hline
      \text{Iteration} & $p_r$ & $p_t$ & $\ephiGW$ & $\lpr$ & $\lpt$ \\
      \hline
      0 & 0.000677 & 0.11466 & ##0.0033      & 1.2 & 0.9985 \\
      1 & 0.000562 & 0.11483 & ##0.002#      & 0.9 & 1.0006 \\
      2 & 0.000623 & 0.11476 & $\sim 0.0008$ &   & \\      
      \hline   
    \end{tabular}
    \label{tab:q2aM075aM075_results}
  }

  \subfloat[list of tables text 3]
  [
    Results for the configuration $q=1, \chi_1 = 0.5, \chi_2 = 0.5$.
  ]
  {
    \begin{tabular}{cccc@{\hspace{1.5mm}}||@{\hspace{1.5mm}}cc}
      \hline 
      Iteration & $p_r$ & $p_t$ & $\ephiGW$ & $\lpr$ & $\lpt$ \\
      \hline
      0 & 0.000647 & 0.08764 & 0.006  & 1.2 & 1.003 \\
      1 & 0.000539 & 0.08737 & 0.001  & 1.3 & 1 \\
      2 & 0.000415 & 0.08737 & 0.0003 &   &   \\
      \hline
    \end{tabular}
    \label{tab:q1a05a05_results}
  }
  \caption{
    Results of the eccentricity reduction method for three aligned-spin BBH configurations.
    We give eccentricity estimates, the initial momenta $p_r, p_t$ and the obtained scaling factors $\lpr, \lpt$
    for EOB QC parameters and two iteration steps in each case.
  }
\end{table}

Recall that our goal is to ``match'' the NR and model frequency residuals modulo dephasing 
\begin{align}
  \label{eq:def_residuals_dephasing}
  \R{}^i &:= \mathcal{R}^i - {\mathcal{R}^i}_\text{fit}\\
  \RMd^{\lambda} &:= \RM^{\lambda} - {\RM^{\lambda}}_\text{fit},
\end{align}
and Eq.~\eqref{eq:match_residuals_vague} translates to the requirement that 
\begin{equation}
  \label{eq:match_residuals_vague_dephasing}
  \RMd^{\lambda} \approx \R{}^i.
  \end{equation}

In achieving this, the crucial observation is that the radial and tangential momenta perturbations make 
contributions to the (oscillatory) frequency residual $\RMd^\lambda$ that are \emph{out of phase}.
Consider a Newtonian binary with perfectly circular orbits. If we perturb the tangential momentum at 
$t=0$, then the resulting binary will now follow an elliptic orbit, and the location of the bodies at
$t=0$ will correspond to an extremum of the separation, and also an extremum of the instantaneous
orbital frequency. This point will therefore be an extremum in the frequency residual $\RMd$
calculated with respect to the circular orbit, and we can write the residual as $\RMd = 
A_t (\lpt) \cos(\Omega_r t)$, where $\Omega_r$ is the frequency of the eccentricity oscillations, and 
$A_t$ is the amplitude of the oscillations due to the perturbation of the tangential momenta. 
Similarly, a perturbation of the radial momentum will lead to an elliptic orbit in which $t=0$
corresponds to an extremum of the radial velocity, and therefore a zero in the frequency residual.
The frequency residual can then be written as $\RMd = A_r (\lpr) \sin(\Omega_r t)$. A more 
detailed and qualitative exposition of these points is given in Appendix~\ref{sub:conservative_dynamics}.

For a general perturbation of the momenta, we may then write the residual as 
\begin{eqnarray}
\RMd^\lambda & = & \Apr(\lpr) \sin(\Omega_r t) + \Apt(\lpt) \cos(\Omega_r t) \\
& = & A \cos(\Omega_r t + \Delta\Phi). \label{eq:Rmodel}
\end{eqnarray} We see in Appendix~\ref{sub:newtonian_eccentricity_estimators} that
the amplitudes of the PN/EOB residuals depend linearly on the momentum perturbations, i.e.,
$A_r = a_r (\lpr - 1)$ and $A_t = a_t (\lpt - 1)$. In principle, then, we could measure the amplitude
and phase of the numerical residual $\R{}^i$, and having determined $a_r$ and $a_t$, could
analytically calculate the appropriate 
scale factors $(\lpr, \lpt)$. 

This method can work well, but in practice we found that a more robust procedure consisted 
of simply performing two line searches: first vary the amplitude $A$ in 
Eq.~(\ref{eq:Rmodel}), but with the phase $\Delta \Phi$ fixed, and
then vary the phase with the amplitude fixed. The searches work as follows. 

Given the NR and model residuals, $\R{}^i(t)$ and $\RMd(t)$, we pick one extremum in $\R{}^i(t)$ 
in the middle of the available time window and record its location $t_i$ in time and its amplitude value 
$A_i = \R{}^i(t_i)$. 
Next, we locate the closest extremum with matching sign in $\RMd$ and again record its location 
$t_\text{M}$ and its amplitude $A_\text{M} = \RMd(t_\text{M})$. We typically store a list of data for 
one or two extrema to the left and right of the fiducial ones, and calculate the average. We then define 
the deviations $d_\text{A}$ and $d_\Phi$ as
\begin{eqnarray}
 d_\text{A} & = & 1 - \overline{A_\text{M}} / \overline{A_i},  \\
 d_\Phi & = & ( \overline{t_i - t_\text{M}} ) / T,
\end{eqnarray}
where bars denote averages over the chosen extrema and $T$ is the average time period of the
oscillations. It should be clear that $d_A$ will be zero when the amplitudes are equal, and $d_\Phi$ will
be zero when the residuals $\R{}^i(t)$ and $\RMd(t)$ are in phase. 

We first perform a line search for amplitude adjustment. We set $\lpr = 1$ and choose a symmetric interval in 
$\lpt$ around unity (with length on the order of $0.005$). 
We then choose $\lpt$ smaller or larger than unity depending on which choice gives better agreement of the phase of the residuals (i.e. where $|d_\Phi|$ is smaller).
Since $d_A$ is a continuous function of $\lambda$ and we have a bracketed root, we can then use a robust root-finding 
method that uses relatively few function evaluations, such as Brent's method~\cite{Brent73}, to quickly align 
the amplitude of the residuals.

The optimal amplitude from this first search then serves as input for the phase adjustment. By construction 
$\Delta_\Phi$ is periodic in the phase angle $\theta = Arg(\Apr + i\Apt) = \atan2(\Apt,\Apr)$ and is only continuous 
for comparison of a given set of extrema in both residuals, but not when passing to the next set of extrema 
in one residual due to adjusting the phase. If we are careful to select a bracket in a region of continuity of 
$\theta$, we can also use Brent's method for the phase adjustment and quickly find an optimal value for 
$\lambda$.
In practice some care must be taken to choose the size of the parameter intervals for the line searches. 
Otherwise the algorithm is automatic. 

\begin{figure}[htbp]
 \centering 
 \includegraphics[width=0.5\textwidth]{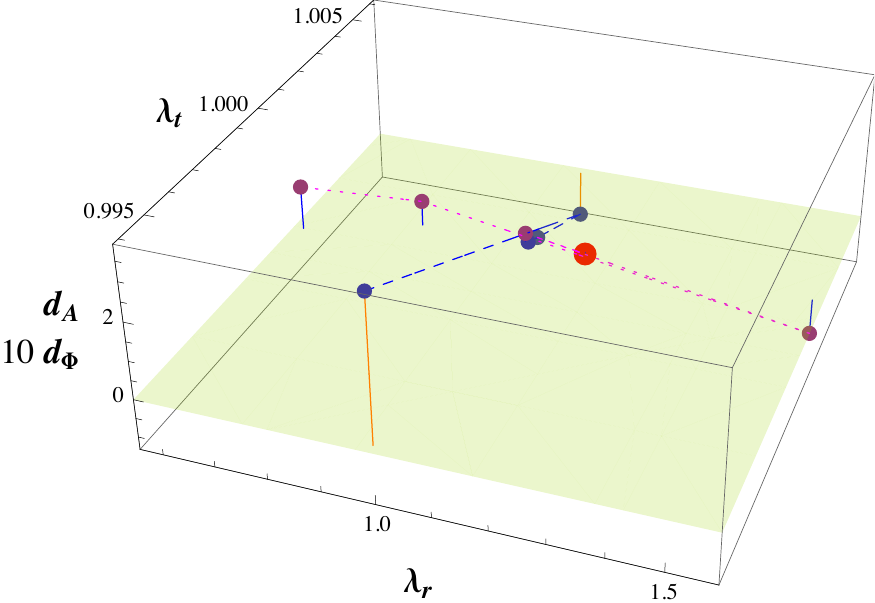}
 \caption{
 Line searches in the plane of perturbation scale factors $(\lpr, \lpt)$ for the first iteration of the $q=2, \chi_1 = 0, \chi_2 = 0.25$ configuration. An amplitude search to minimize $d_\text{A}(\lambda)$  (blue spheres with orange filling) and a phase search to minimize $d_{\Phi}(\lambda)$ along a line of constant amplitude (magenta spheres) are carried out. 
The final result of the two searches are the scaling factors $\lambda = \{1.12, 1.0024\}$ (larger red sphere).
}
 \label{fig:lineSearchEx3D}
\end{figure}

An example for the locations covered by the amplitude and phase adjustment line searches using Brent's 
method in the $\lambda$-plane is given in Fig.~\ref{fig:lineSearchEx3D}. In this example $\lpt$ dominates 
and $\lpr$ is close to unity, which is typical for eccentricities $ \sim 5 \times 10^{-3}$. This is for the same 
$q=2, \chi_1 = 0, \chi_2 = 0.25$ configuration as discussed in Sec.~\ref{sec:NR_example}. There, the adjustments 
were obtained manually, while the automatic search was developed later, and so the updated parameters are 
somewhat different to those given in Tab.~\ref{tab:q2a0a025_results}.

The eccentricity reduction algorithm described here has been implemented in Mathematica and will be made 
available at http://gw-models.org. The automated method to determine the optimal 
momentum scale factors $\{ \lpr^*, \lpt^* \}$ requires further optimization against a wider range of binary 
configurations, but will also be made available in due course.

\subsection{Computational cost of eccentricity reduction} 
\label{sub:computational_cost_of_eccentricity_reduction}
 
In this section we make a rough estimation of the relative cost of eccentricity reduction, 
within the overall production of high-quality numerical-relativity waveforms. 

Our experience suggests that we can measure the eccentricity with sufficient accuracy from 
the lowest resolution simulation of a convergence series. Furthermore, we assume 
that we will carry out two iteration steps, each 3-4 orbits long. Assuming that 
the full simulations will cover approximately 10 orbits~\cite{Hannam:2010ky,Ohme:2011zm}, the 
length of these two simulations 
together shall be comparable to the merger time for the configuration. 
For simplicity, let us choose the scale factor between the grid resolutions used in the convergence 
series about $\rho \sim 1.15$, although it usually varies slightly between different resolutions. 
The computational cost for each resolution is then proportional to $(1,\rho, \rho^2)^4$, 
about $(17\%, 30\%, 53\%)$ of the total cost. A conservative value for the relative cost overhead 
from eccentricity reduction is then about $20\%$.

For clustered configurations in a parameter study it can be argued that instead of a full convergence 
series a single high resolution simulation is sufficient for a subset of these configurations. In that 
case the overhead cost for eccentricity reduction is higher, about $35\%$. On the other hand, 
for longer waveforms or higher accuracy requirements on the final waveform, the overhead cost 
will be even lower. More importantly, in unexplored parts of the parameter space, it is usually necessary 
to perform test runs before launching production simulations. 
The cost of eccentricity reduction can then be partially absorbed into such test runs. This situation 
will be common, and therefore we estimate the overhead cost from eccentricity reduction in a large
parameter study as between a quarter and a third of the total computational cost.


\section{Gauge dependence of the orbital motion} 
\label{sub:gauge_dependence_of_the_orbital_motion}

In the NR examples in Sec.~\ref{sec:NR_example} we saw that the orbital motion 
cannot be used to make a reliable estimate of the binary's eccentricity. This is clear in 
Figs.~\ref{fig:q2a0a025_it} and \ref{fig:q2a0a025_e_omega_orb}.

In this section we explore in more detail the effects of the gauge choice on the puncture motion. 
The puncture motion is generated by the $\Gamma$-driver condition~\cite{Alcubierre:2002kk},
which, in its usual 
formulation, has one free parameter $\eta$. We study the accuracy of the
orbital frequency eccentricity estimator $e_\Omega$, as a function of standard choices of $\eta$. 

We first establish the theoretical relationship between the eccentricities calculated from orbital and GW quantities. 

The fundamental frequency of the oscillations due to residual eccentricity, $\Omega_r$, is related to the 
average orbital frequency $\Omega_\Phi$ via the fractional periastron advance per orbit $k = \Delta\Phi / (2\pi)$ 
by Eqn.~\eqref{eq:om_phi_om_r_1PN} (see also~\cite{Mroue:2010re}),
\begin{equation}
  \frac{\Omega_\phi}{\Omega_r}  = 1 + k(\Omega_r).
\end{equation}

As a consequence, eccentricity estimators for the orbital phase and frequency, or GW phase and frequency, 
are also related via this ratio,
\begin{equation}
  \frac{e_\phi}{e_\Omega} = 1 + k.
\end{equation}
However, the ratio between eccentricities calculated from the \emph{orbital} frequency  and \emph{GW} phase 
(i.e., one quantity is from the orbital motion and the other from $\Psi_4$) is different.
In appendix~\ref{sec:1pn_eccentricity_estimators} we show that to 1PN order and for low frequencies 
(or small $k$), the ratio is
\begin{equation}
  \kappa := \frac{e_{\phi[\psi_4]}}{e_\Omega} 
  \approx \frac{21}{16} +  \left(\frac{7 \nu}{32} - \frac{1}{4}\right) k \epsilon ^2,
\end{equation}
where $\nu$ is the symmetric mass-ratio and $\epsilon = 1/c$. For larger $k$ it is more accurate to form the ratio directly from Eqns.~\eqref{eq:e_om_orb_1PN} and \eqref{eq:e_phi_psi4}.

\begin{figure}[htbp]
  \centering
  \hspace*{-0.7cm}\includegraphics[width=0.53\textwidth]{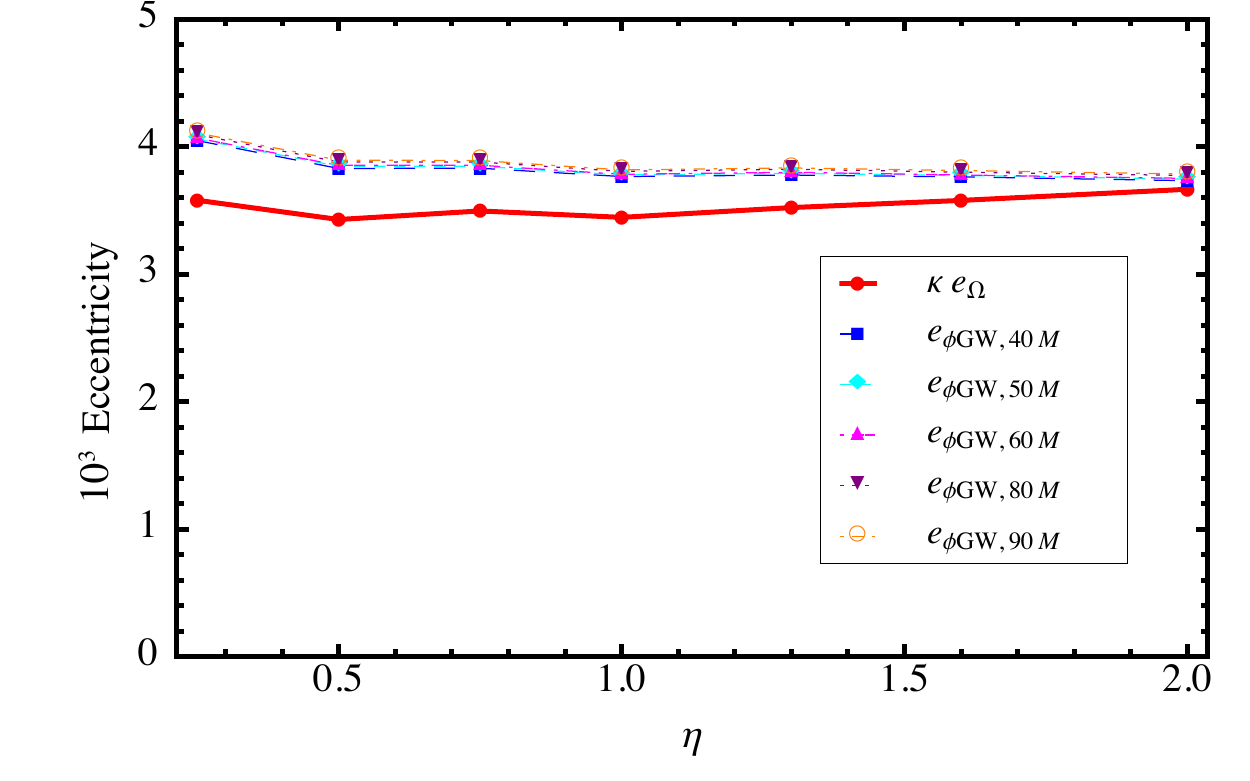}
  \caption{The eccentricity estimators $\eomOrb$ and $\ephiGW$ (for extraction radii $R_{ex} = 40, 50, 60, 80, 90M$) 
  as a function of the gauge parameter $\eta$ for equal-mass non-spinning evolutions with moderate 
  eccentricity initial data. $\eomOrb$ has been rescaled by a factor $\kappa = 1.25$ (see text). 
  }
  \label{fig:eccentricity_eta_moderate}
\end{figure}

\begin{figure}[htbp]
  \centering
  \hspace*{-0.7cm}\includegraphics[width=0.53\textwidth]{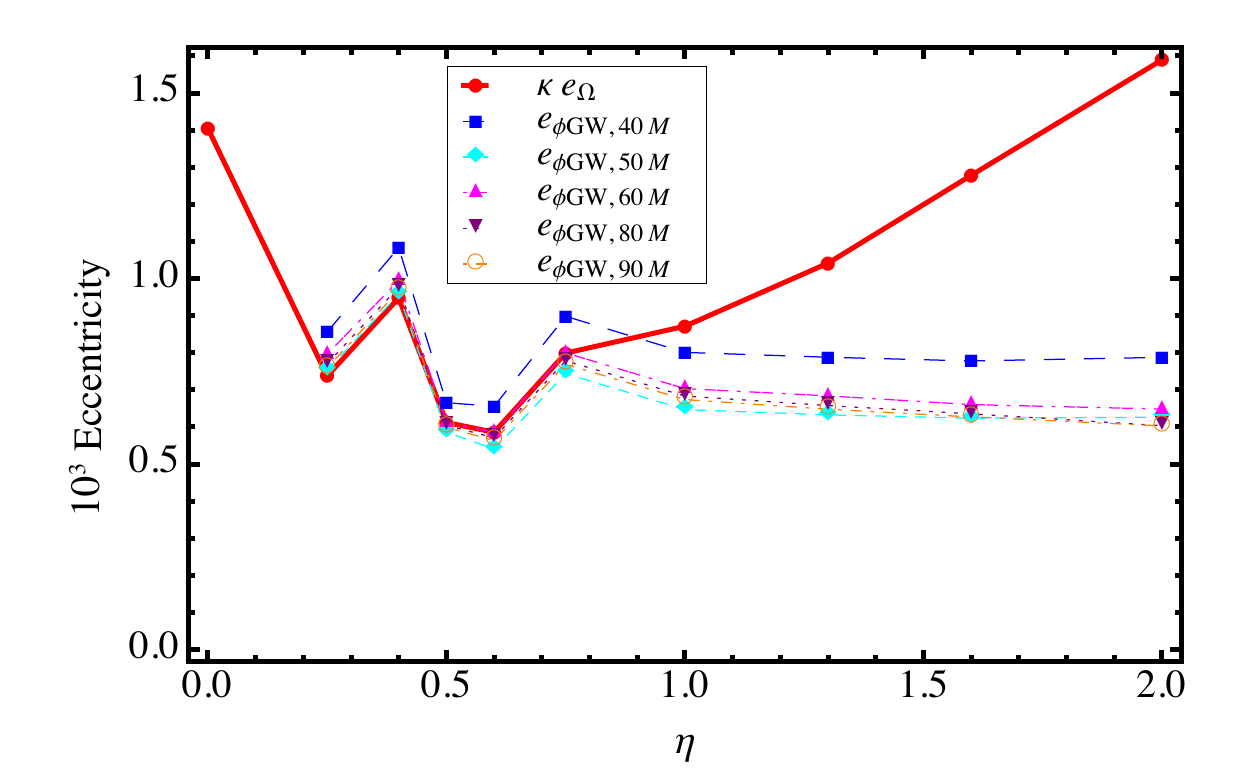}
  \caption{Same as Fig.~\ref{fig:eccentricity_eta_moderate}, but for an eccentricity almost one order of magnitude 
  lower.
}
  \label{fig:eccentricity_eta}
\end{figure}

We now consider a series of equal-mass non-spinning configurations with spatially constant $\eta$ ranging from 
zero to our standard value of $\eta = 2/M$. Simulations were performed for two different sets of initial parameters,
one which lead to moderate eccentricity $e \sim 0.004$ (Fig.~\ref{fig:eccentricity_eta_moderate}), and another for 
very low eccentricity $e \sim 0.0005$ (Fig.~\ref{fig:eccentricity_eta}). 

To reliably assess the residual eccentricity we have compared eccentricities computed from $\eomOrb$ and $\ephiGW$ 
for the same fitting window $t \in [400, 1200] M$. $\ephiGW$ has been computed for extraction radii 
$R_{ex} = \{40, 50, 60, 80, 90\}M$, while taking into account the retardation of the wave signal. 
We have computed the orbital eccentricities as the average between the minimum and the maximum of the estimators 
in the windows, and have made the required quasi-circular fits in Eqs.~(\ref{eq:e_phi_GW}) and (\ref{eq:e_omega_orb}) 
using a fifth order polynomial  for $\ephiGW$ and a fourth-order polynomial for $\eomOrb$. Additionally, $\ephiGW(t)$ 
was estimated using the smoothing procedures discussed in Sec.~\ref{sec:filtering}.

The average orbital frequency for the NR $q=1$ example in the fitting window is $M \Omega \approx 0.027$. 
Taking the fractional periastron advance, $k$, from the Fig.~7 of Ref.~\cite{Mroue:2010re}, a value $k \sim 0.38$ 
seems reasonable. This gives a factor $\kappa \approx 1.25$. (For comparison, the 1PN definition gives 
$k \approx 0.23$ which leads to $\kappa \approx 1.27$). 
This value of $\kappa$ has been used in Figs.~\ref{fig:eccentricity_eta_moderate} and \ref{fig:eccentricity_eta} to 
scale $\eomOrb$ so that a direct comparison with $\ephiGW$ is possible. 

Convergence in $\ephiGW$ as a function of the extraction radius for fixed $\eta$ is spoilt by uncertainties 
introduced by the level of noise in the waveform at these small eccentricities and by uncertainties from the 
sinusoidal fits to the data. 

\begin{figure}[htbp]
  \centering
  \hspace*{-0.3cm}\includegraphics[width=0.62\textwidth]{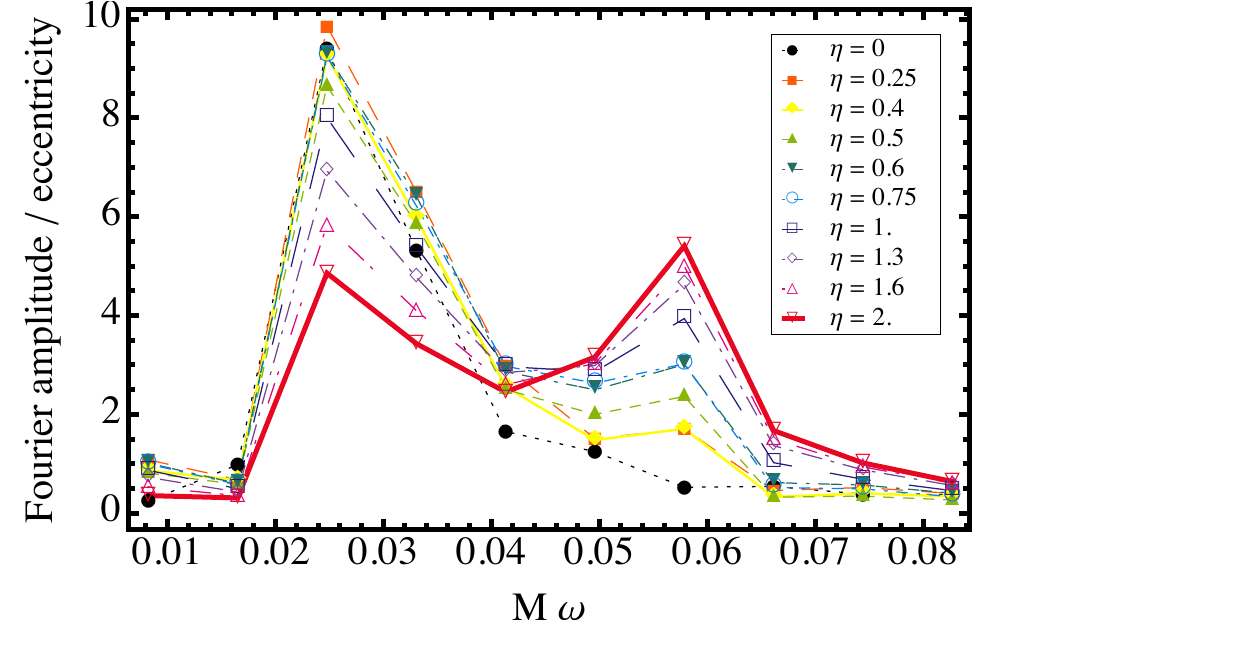}
  \caption{The amplitude of the lowest Fourier modes of the eccentricity residual $\eomOrb(t)$
  for equal-mass non-spinning evolutions ($M=1$) with varying gauge parameter $\eta$. 
  The average orbital frequency is at $M \omega = 0.027$. 
  }
  \label{fig:eta_harmonics}
\end{figure}

The resulting eccentricities in Fig.~\ref{fig:eccentricity_eta_moderate} are consistent within $10\%$ for both 
estimators and the dependence on $\eta$ is very weak at this moderate eccentricity. 
In Fig.~\ref{fig:eccentricity_eta} the rescaled orbital eccentricities align well with $\ephiGW$ near $\eta \sim 0.5$. 
For higher values of $\eta$ towards the standard value $2/M$ the amplitude of the second harmonic increases 
(see also Fig.~\ref{fig:eta_harmonics}). For $\eta = 0$ all quantities are extremely noisy, and $\eomOrb(t)$ 
had to be filtered and it was not possible to compute $\ephiGW$. 

We find that the absolute error $\eomOrb$ is roughly constant between the two cases; it is simply far more 
noticeable in Fig.~\ref{fig:eccentricity_eta}, where the eccentricity is roughly an order of magnitude lower than in
Fig.~\ref{fig:eccentricity_eta_moderate}. 

To see which frequency components contribute to the eccentricity values of $\eomOrb$, Fig.~\ref{fig:eta_harmonics} 
shows the ten lowest Fourier amplitudes in $\eomOrb(t)$ for the simulations shown in Fig.~\ref{fig:eccentricity_eta} 
as a function of $\eta$. The amplitude of the harmonic at $\sim 2 \Omega_\Phi$ increases with $\eta$ and for 
$\eta=2/M$ is higher than the fundamental frequency.

One implication of the above analysis is that eccentricity estimators computed from the orbital motion become 
unreliable below a certain eccentricity, $e \approx 0.002$, and will lead to estimates higher than the physical 
eccentricity present in the evolution. 
Having shown consistency between $\eomOrb$ and $\ephiGW$, we rely in the following on the estimator 
$\ephiGW(t)$. 
A study of Figs.~\ref{fig:eccentricity_eta_moderate} and \ref{fig:eccentricity_eta}, along with 
Fig.~\ref{fig:eta_harmonics}, might suggest that the orbital-frequency eccentricity estimator \emph{could} be used if we 
used, for example, $\eta = 1/M$ instead of $\eta = 2/M$. However, we only know that this value of $\eta$ gave a 
reasonable estimate of the eccentricity after performing a detailed study the $\eta$ dependence of the estimator 
for this case. The optimal choice of $\eta$ for another configuration could be quite different, and may change yet
again when a spatially varying $\eta$ is used. The GW-frequency-based eccentricity estimator is clearly far less
dependent on the choice of gauge (as we would expect), and we choose to use that in all subsequent work. 


\section{Errors} 
\label{sub:errors}

In the following sections we discuss sources of error in the determination of the scale 
factors, the measurement of the eccentricity, and in the final waveforms. 

\subsection{Sensitivity of eccentricity on scale factors}
\label{sec:e_sensitivity}

We first consider how the eccentricity depends on the uncertainties in the scale factors.

From Newtonian considerations one can see explicitly that the variation in the eccentricity
depends linearly on the scale factors, i.e., $\pd{e}{\lambda_\bullet}$ is constant for $e \sim 0$,
where $\lambda_\bullet$ denotes either of $\lpr$ or $\lpt$.  
The absolute error in the eccentricity is then proportional to the error in $\lambda_\bullet$: 
$\Delta e = \pd{e}{\lambda_\bullet} \Delta\lambda_\bullet \propto \Delta\lambda_\bullet$.
For example, if $\lpr = 1$, the Newtonian eccentricity formula~\eqref{ecc:Newtonian} gives 
\begin{equation}
  \pd{e}{\lpt} = 2 \lpt \approx 2.
\end{equation}
Let $p_\bullet^\circ$ denote the QC value of the momentum $p_\bullet$, so that 
$e(p_r^\circ, p_t^\circ) = 0$.
In NR cases it is easier to estimate $\Delta e / \Delta p_t \approx 16$. Using the chain 
rule and the relation $p_\bullet = \lambda_\bullet p_\bullet^\circ$ we have
\begin{equation}
\pd{e}{\lambda_\bullet} = \pd{e}{p_\bullet} \pd{p_\bullet}{\lambda_\bullet} = \pd{e}{p_\bullet} p_\bullet^\circ,
\end{equation}
and find $\pd{e}{\lpt} \approx 1.9$, which is very close to the Newtonian sensitivity. (Note that
this agreement is the basis of our of statement that $\nabla_p e_M \approx \nabla_p e_\text{NR}$ 
in Sec.~\ref{sub:requirements_on_the_model}.)
Due to the lack of radiation reaction, the Newtonian model fails to give a useful estimate 
of the sensitivity due to perturbations of the radial momentum. In order to get a theoretical
estimate of this value, we instead performed a sample of EOB/PN perturbations and found 
 $\pd{e}{\lpt} \approx 2$ and $\pd{e}{\lpr} \approx 0.005$ for our reference example.

We can now estimate a lower bound on the achievable eccentricity with our method, using 
the EOB/PN sensitivities and the fact that the perturbation factors are usually 
very close to unity (typically, $1 - \lpr \sim 0.1$ and $1 - \lpt \sim 0.001$). 
Assuming an error of $x\%$ in $\lpt$, 
the error in the eccentricity will be $\Delta e = \pd{e}{\lpt} \Delta\lpt \sim x/50$. 
For $\Delta e \lesssim 0.001$, we must therefore have $x \lesssim 0.05\%$, i.e., $\lpt$ 
has to be accurate to $5 \times 10^{-4}$, and by a similar argument 
$\lpr$ has to be accurate to within $15\%$.
Clearly, eccentricity is a lot less sensitive to changes in $\lpr$ than to $\lpt$, but on the 
other hand, the adjustments are larger for $\lpr$. While our automatic procedure adjusts 
both momenta in each step, we have found that $\lpt$ is far more important, especially 
in the first iteration step. This is the reason why $\lpr$ is sometimes equal to unity in the 
example configurations shown in Tabs.~\ref{tab:q2a0a025_results}, \ref{tab:q2aM075aM075_results},
and \ref{tab:q1a05a05_results}, where the adjustments have been checked by eye. 

This raises the obvious question: what is the minimum eccentricity we can obtain with 
$p_r = 0$? If it were sufficiently low (for example, less than $10^{-3}$), then our eccentricity
reduction procedure could be dramatically simplified by locating only the appropriate
tangential momentum. 
This can easily be tested on the PN level or estimated using the computed sensitivities 
and a required $\Delta p_r = p_r$. We calculate $\pd{e}{p_r} \approx 8$ and have 
$\Delta e \approx \pd{e}{p_r} \Delta p_r \approx 0.005$ for the reference 
$(q = 2, \chi_1 = 0, \chi_2 = 0.25)$ configuration. This is lower than the eccentricity that often 
results from the first-guess EOB/PN parameters, but is higher than the $10^{-3}$ threshold
that we have been aiming for. We also expect this value to be higher for configurations with
high anti-aligned spins, where the inspiral is more rapid, and therefore the appropriate
$p_r$ is higher. 


\subsection{Finite-extraction-radius and fitting errors}
\label{sec:errors_rex_fits}

We extract the gravitational wave signal at a number of finite radii. For each radius we can 
compute an approximate match and scale factors ($\lpr, \lpt$). How large is the related
error in the scale factors, compared to their required accuracy in order to attain a given 
eccentricity? 
An analysis of iteration 1 of the reference example shows that this error is about 
$2.5 \times 10^{-4}$ in $\lpt$  
and $0.07$ 
in $\lpr$. 
Comparing with the results from the sensitivity analysis above, this level of error will allow us
to attain an eccentricity on the order of $e \sim 5 \times 10^{-4}$, although in practice we seem
to do a little better (see table \ref{tab:q2a0a025_results}). At this point it is difficult to say 
whether numerical noise, errors due to fitting artifacts, or errors due to finite extraction radius 
are the dominant source that limits the accuracy of the iteration steps.

Apart from the errors due to gauge oscillations in $\eomOrb$, which we have pointed out above, 
eccentricity estimators introduce an additional source of error through the polynomial fits involved.
Two relevant parameters are the order of the fitting polynomial and the duration of the fitting window. 
We choose the order of the polynomial based on the signal length (2 - 3 periods) in order to avoid 
also picking up the eccentricity oscillations. For our preferred estimator $\ephiGW$ we also apply 
a low-pass filter to the residual, perform a nonlinear fit to a sinusoid, and take its amplitude as 
the eccentricity value. From tests with a NR signal with an artificially injected eccentricity (i.e., a 
sinusoidal modification to the GW phase), we find that $\ephiGW$ tends to underestimate the 
eccentricity by $5 - 20 \%$. Since real NR data will have less than perfect sinusoidal oscillations, 
we take a conservative error estimate of $25 \%$ relative error in $\ephiGW$. For very low 
eccentricities $e \sim 5 \times 10^{-4}$, the amount of noise in the NR phase degrades the 
accuracy further and we estimate the error to be about $50 \%$. 



\subsection{Phase errors and mismatches between eccentric hybrids} 
\label{sub:phase_errors_mismatches_between_eccentric_hybrids}

The motivation for this work is waveform modeling for gravitational-wave 
astronomy. Most black-hole-binaries visible to the second-generation ground-based
detectors Advanced LIGO and Virgo will follow non-eccentric orbits. We therefore
want to simulate non-eccentric binaries. Since we cannot construct simulations of
binaries with precisely zero eccentricity, we are then faced with the question: what
level of eccentricity in NR waveforms is \emph{tolerable}? More precisely, if NR 
waveforms of a given eccentricity are used to produce waveform models, which 
are in turn used for GW searches and parameter estimation, how many signals 
will the eccentricity cause us to lose, and of those signals that we observe, 
how much will our parameter measurement be skewed? (The effect on detection 
of the opposite problem --- using non-eccentric models when the real signals are
from eccentric binaries --- is considered in~\cite{Brown:2009ng}.)

The standard tool with which to address such questions is the \emph{mismatch}. We choose
one waveform, $h_1$, as the true signal, and another $h_2$ as the model that will be used
in GW searches and parameter estimation. The most basic mismatch is defined as
\begin{equation}  
{\cal M} = 1 -  \max_{\tau,\Phi} \frac{\langle h_1 | h_2 \rangle }{\sqrt{
     \langle h_1|h_1\rangle \langle h_2|h_2\rangle }},  \label{eqn:mismatch}
\end{equation} where the inner product between the two waveforms is further defined 
as~\cite{Cutler94},
\begin{equation}
 \label{eq:scalar_prod}
 \langle h_1|h_2\rangle := 4 \, {\rm Re} \left[ \int_{f_{\rm
         min}}^{f_{\rm max}} 
     \frac{ \tilde h_1(f) 
     \tilde h_2^\ast (f)}{S_n(f)} \, df \right] \, .
 \end{equation} The inner product is calculated in terms of the frequency-domain waveforms
 $\tilde{h}(f)$, and is weighted by the power spectral density $S_n(f)$ of a given detector. 
 The frequency range in which the detector is deemed sensitive is $[f_{\rm min},f_{\rm max}]$.
 In calculating the mismatch, the inner product is maximized over time and phase offsets of 
 the waveforms, so that two waveforms produced by exactly the same source (even at
 different times and with different initial phases) would lead to a mismatch of zero. In our
 calculations we use for $S_n(f)$ the Advanced LIGO zero-detuned, high-power~\cite{T0900288} 
 noise curve, and choose $f_{\rm min} = 20$\,Hz, and $f_{\rm max} = 8$\,kHz.
 
 In assessing the usefulness of a given model in a GW \emph{search}, we should also maximize
 over the physical parameters of the model. By not doing that, we produce an upper limit on
 the mismatch between the signal and the model. We do not perform the additional
 maximization because (a) we cannot, because our NR waveforms only model discrete 
 configurations, and (b) we can also use the mismatch~(\ref{eqn:mismatch}) to estimate
 the \emph{indistinguishability} of the waveforms, which characterizes the 
 effect of the waveform's error on parameter estimation. If $h_1$ is observed at a given 
 signal-to-noise ratio (SNR), then it cannot be distinguished from $h_2$ if 
 $| \delta h| = | h_1 - h_2 |^2 < 1$~\cite{Lindblom:2008cm}. If this is the case, then the parameter estimation 
 uncertainties will not be effected by any errors in the model $h_2$. 
 The indistinguishability can be related to 
 the mismatch by~\cite{McWilliams:2010eq} $|\delta h| / \rho^2 = 2 {\cal M}$, where $\rho$ is
 the SNR.  Note that for any $h_2$ there will be an SNR sufficiently low that 
 we cannot distinguish it from $h_1$, and an SNR sufficiently high that we \emph{can}
 distinguish them (unless of course they are in fact identical). 
 
 The primary application of our waveforms will be in producing phenomenological waveform
 models~\cite{Ajith:2007qp,Ajith:2007kx,Ajith:2007xh,Ajith:2009bn,Santamaria:2010yb}. 
 These are based on hybrids between PN early-inspiral waveforms and the NR
 late-inspiral-and-merger waveforms. In calculating mismatches, we will therefore consider
 PN-NR hybrids of our waveforms. And since we are aiming for a non-eccentric model, the
 PN portion of the waveform will always have zero eccentricity. We produce hybrid waveforms
 using the standard technique where we 
 align the PN and NR 
 waveforms over a small time interval around the frequency $M\omega = 0.055$ 
 and then smoothly blend them together. We consider the reference example
 ($q=2, \chi_1 = 0, \chi_2 = 0.25$)
 configuration, and use TaylorT1 as the PN approximant. 
 
\begin{figure}[htbp]
  \centering
  \includegraphics[width=0.5\textwidth]{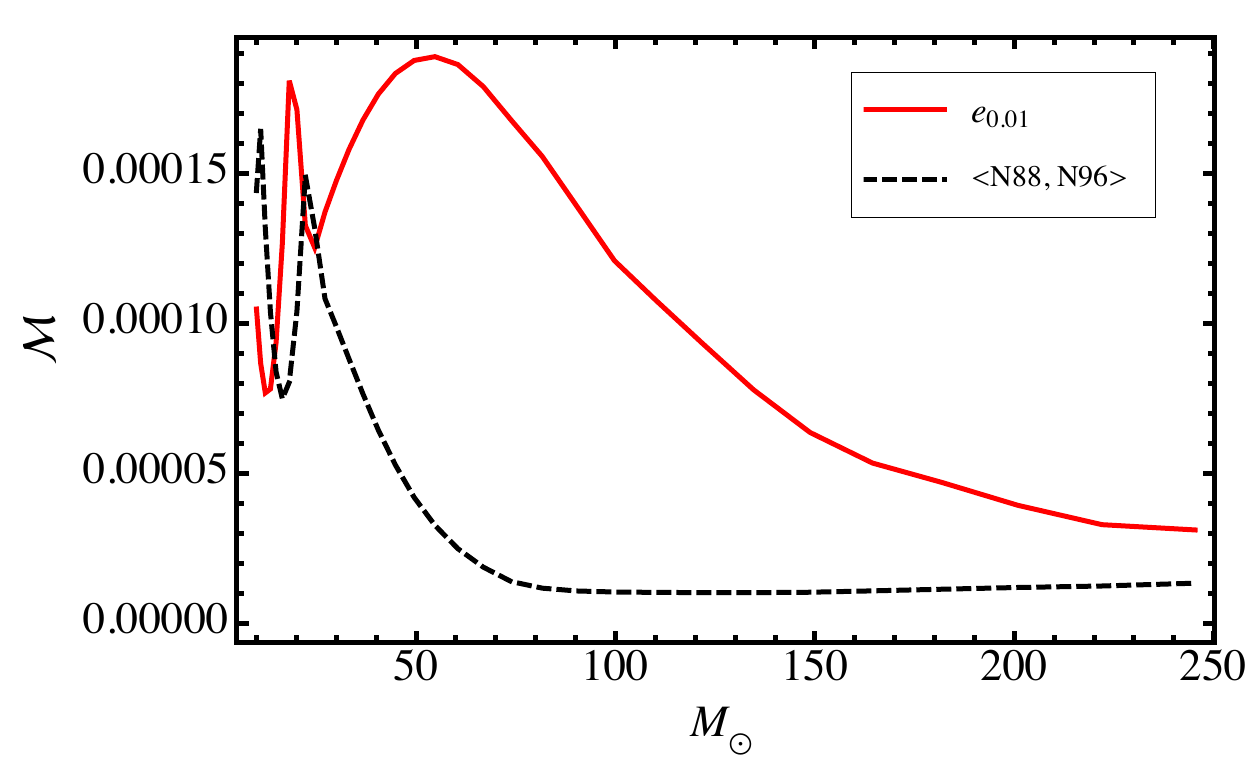} %
  \caption{
  We show a mismatch between PN-NR hybrids of eccentricity $e = 0.01$ and a reference hybrid ($e \sim 0.0003$) 
  for the $q=2, \chi_1 = 0, \chi_2 = 0.25$ configuration (red curve). 
  The hybrids use the same numerical resolution ($N = 88$ grid) and extraction radius.
  For the sake of comparison, a second mismatch is computed between different resolutions $N=88$ and $N=96$ 
  for eccentricity $e \sim 0.0003$.
  }
  \label{fig:matches}
\end{figure}

 Fig.~\ref{fig:matches} shows the mismatch results between a ``model'' waveform with NR eccentricity 
 $e = 0.01$, and a ``signal'' waveform with NR eccentricity $e = 3 \times 10^{-4}$.
 For comparison we also show the mismatch between simulations with different levels of 
 numerical resolution.
 The precise mismatch values in the figure should not be taken too seriously; 
 these are sensitive to the details of the hybrid construction and the mismatch calculation. 
 The main point is that the mismatches are remarkably low. For reference: 
 (a) a commonly used criterion for detection requires mismatches between the signal and 
 model of less than 0.03; (b) the mismatches between a series of equal-mass nonspinning-binary 
 NR waveforms, each produced with a different NR code, were found to be typically five times 
 larger than the $10^{-4}$ that we see here~\cite{Hannam:2009hh}, and mismatches for larger
 mass ratios and spinning black holes tend to be higher~\cite{Santamaria:2010yb,Ajith:2012tt}; 
 and 
 (c) these waveforms would be indistinguishable for SNRs 
 below 100 in Advanced LIGO. An observation with such a high SNR in Advanced LIGO 
 would be truly exceptional.
 In third-generation detectors, such as the Einstein Telescope (ET)~\cite{Punturo:2010zz}, and 
 space-based detectors (for example~\cite{AmaroSeoane:2012km}),
 however, far larger SNRs would be likely, and these differences will be distinguishable. 
 
 This plot suggests, then, that the use of NR waveforms with eccentricities as high as even
 $e \sim 0.01$ will have negligible impact on GW detection and parameter estimation over the
 next decade. So long as the raw PN/EOB momenta parameters lead to eccentricities below 
 this value, we could conclude that we do not need to apply any eccentricity reduction at all.
 
 The weakness of mismatch plots is that they are effectively a measure of how well the phases
 of two waveforms can be aligned, and no more. In particular, they do not tell us how 
 the errors in a set of waveforms will affect the calibration of coefficients in a phenomenological or
 EOB model. This question cannot be answered until we attempt to calibrate those
 models. What we do know is that the eccentricity reduction process is relatively computationally
 cheap to perform (see Sec.~\ref{sub:computational_cost_of_eccentricity_reduction}), 
 but repeating a large family of simulations if they turn out to be inadequate
 is extremely computationally expensive. For this reason, we take the conservative view that
 we want the phase errors due to eccentricity to be no larger (and preferably smaller) than those due 
 to other numerical errors. 
 
\begin{figure}
  \centering
  \includegraphics[width=0.5\textwidth]{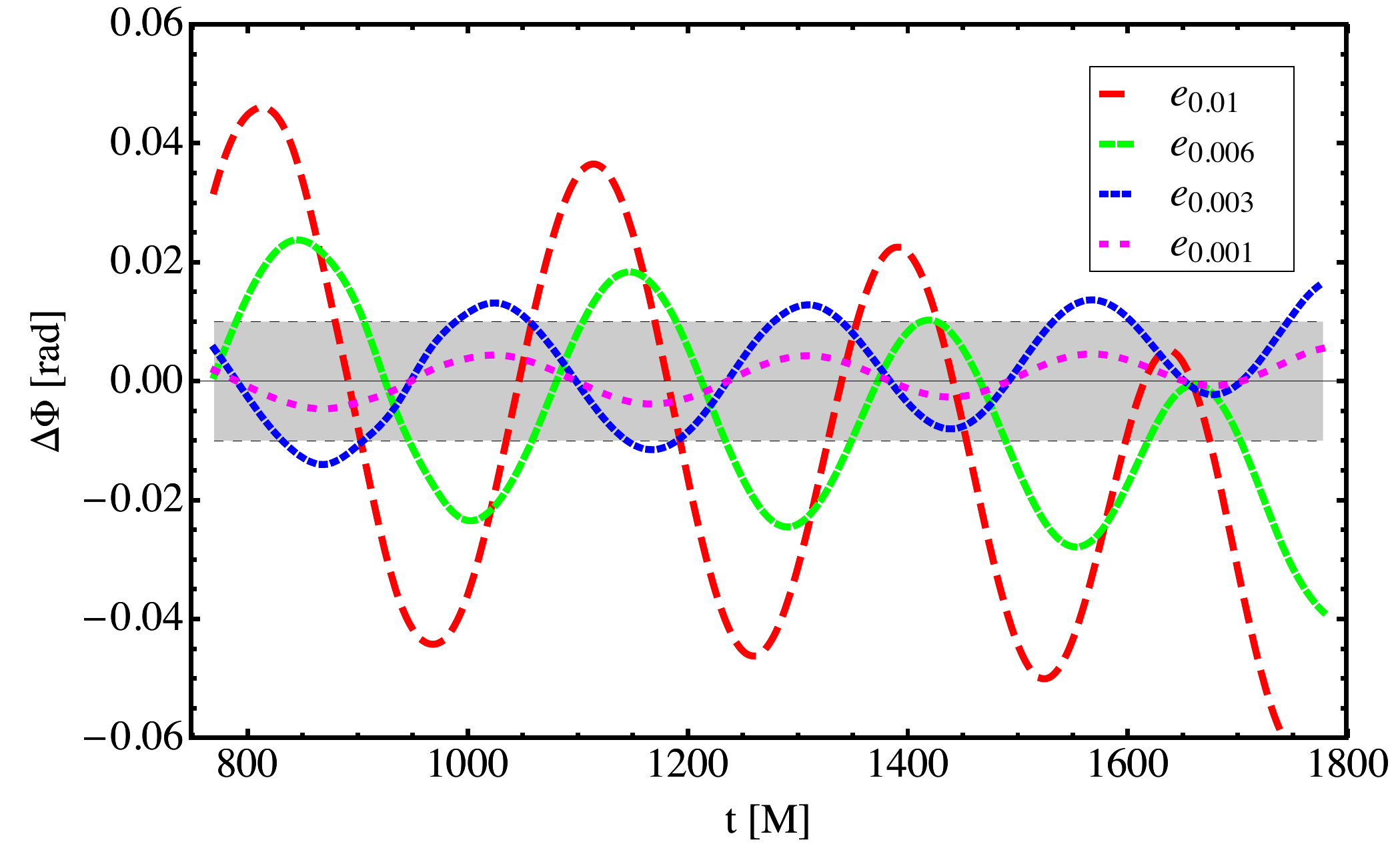}
  \caption{Phase differences 
  from $\Psi_4$ for $q=2, \chi_1=0, \chi_2=0.25$ simulations of varying 
  eccentricity with $N=88$ numerical grids. 
  All differences are with respect to the $\ephiGW=0.0003$ simulation. 
  We also estimate the curve for a simulation with eccentricity  
  $\ephiGW=0.001$ (magenta, short-dashed); see text. 
  A stringent NR phase error requirement of $\pm 0.01$\,rad is indicated by the shaded region.
  }
  \label{fig:phase_errors}
\end{figure}

This requirement motivates Fig.~\ref{fig:phase_errors}, in which we compare the $\Psi_4$ phase 
differences between inspiral NR waveforms of varying eccentricity. 
All curves are with reference to the lowest eccentricity phase $(e \sim 3\times 10^{-4})$. The phases 
are aligned at $M\omega = 0.055$ using a non-eccentric fit through the data. We see that the 
phase differences have a secular and an oscillatory part. Since the amplitude of the phase oscillations
is linearly related to the eccentricity, we were also able to estimate the curve for a
simulation with our threshold eccentricity requirement of $e = 10^{-3}$ (the short-gap-dashed 
magenta line).

Note that simple alignment at a fixed frequency can lead to large secular phase errors 
that depend sensitively on the alignment frequency. 
The worst-case phase errors due to eccentricity for this set of 
simulations agree well with a Caltech-Cornell estimate in Eq.~61 in~\cite{Boyle:2007ft}. 
As an example, for phases aligned at $M\omega = 0.1$ the backwards-in-time phase error for 
$e = 0.003$ reaches 
$0.8 \, \textrm{rad}$ at $M\omega = 0.055$.
But it is important to realize that this is a pessimistic estimate. By using non-eccentric fits to the phase
evolution, we have effectively aligned to an underlying ``mean quasi-circular frequency'', and now
the secular drift is far less. This also mimics what is effectively done in producing hybrid
waveforms, where the root-mean-square integrated phase disagreement is minimized over a time
interval that includes at least one GW cycle. 

Based on our previous experience with moving-puncture simulations, we expect the accumulated
numerical phase error through inspiral to be less than 0.01\,rad, and this is shown by the shaded
region in the figure.
We see that the oscillations in the phase disagreement are greater than 0.01\,rad for 
the simulations with $e = 0.003,0.006, 0.01$, but are well within this tolerance for $e = 10^{-3}$.

This figure does not include dephasing through merger and ringdown. We expect that the eccentricity
will have negligible effect on the merger/ringdown waveform, but the phase oscillations through 
inspiral that are visible in Fig.~\ref{fig:phase_errors} may cause the eccentric binary to merge slightly
early or later than its non-eccentric counterpart. We can estimate this dephasing effect as follows. 
Based on the figure, we see that near merger (we choose $M\omega = 0.1$), the non-eccentric 
and $e = 10^{-3}$ binaries may be
as much as 0.01\,rad out of phase, or $\sim$1/500th of a cycle. The orbital period of the last full orbit is
roughly $T \sim 100M$, and so this dephasing corresponds to a relative time lag of 
$\delta t \approx 0.2M$. If we take the GW frequency function for a non-eccentric waveform, 
$\omega_{e=0}(t)$, and integrate $(\omega_{e=0}(t-\delta t) - \omega(t))$ from $M\omega = 0.1$
through merger and ringdown, we can estimate the additional accumulated dephasing. This
effect will be largest when the ringdown frequency is highest, i.e., for an equal-mass binary with
large aligned spins. Using the results from a previous simulation of a $q = 1$, $\chi_i = 0.85$ 
configuration~\cite{Hannam:2010ec}, we find an additional accumulated dephasing of 0.2\,rad. 
This is well 
below the lowest accumulated \emph{numerical} phase error that we have recorded to date
(1.0\,rad for the $q = 1, \chi_i = 0.5$ configuration, in Tab.~III of Ref.~\cite{Hannam:2010ec}) 
and suggests that
our threshold of $e = 10^{-3}$ limits the phase uncertainty due to eccentricity to a level lower or
(at worst) comparable to the numerical phase error. 

There are two points that should be made about this estimate. We should recall that this is a
conservative requirement on the eccentricity; the mismatches shown in Fig.~\ref{fig:matches}
suggest that a requirement of $e < 10^{-3}$ is much lower than what is required for GW
astronomy in the advanced detector era. On the other hand, if one does wish the 
eccentricity-induced phase effects to be below numerical uncertainty, then the eccentricity
threshold will be different for simulations at different levels of accuracy. If the numerical
phase errors were an order of magnitude lower (as in the simulation discussed in 
Ref.~\cite{Scheel:2008rj}) then the same argument would demand $e < 10^{-4}$, which 
is what was used in that case. 
 
\section{Comparison between eccentricity reduction methods} 
\label{sub:comparison_between_eccentricity_reduction_methods}

We now compare our new techniques with the iterative eccentricity-reduction method used by the 
Caltech-Cornell-CITA group in
Refs.~\cite{Pfeiffer:2007yz,Boyle:2007ft,Buonanno:2010yk}. We will refer to this as the ``CCC'' method.
The CCC-method was able  to efficiently 
reduce eccentricity down to the order of $10^{-5}$ for generalized-harmonic formulations of Einstein's 
equations~\cite{Scheel:2006gg} using conformal-thin-sandwich (CTS) excision initial
data~\cite{Cook:2004kt}. Since this method has proved so successful, it is important to explain 
why we have gone to the trouble of devising an entirely new method, and why we have not simply
adopted their method. The key reasons are associated with the gauge choice, and we
explain this further below. First we provide a brief description of the CCC method.

We discuss here only the latest version of the method~\cite{Buonanno:2010yk}. In a nutshell, the 
time-derivatives of the coordinate separation and the orbital frequency are fit against model functions 
consisting of a non-oscillatory part that approximates the inspiral, plus a sinusoidal term to capture 
the residual eccentricity oscillations. The fit results are then used to construct updating formulas 
for the initial velocity and the initial orbital frequency at a starting separation and partly make use of 
information from the Newtonian Hamiltonian. 

In principle we could use the CCC method with the moving-puncture orbital quantities. 
However, there are three difficulties in doing this, all of which relate to the gauge choice.
First, the $\Gamma$-driver shift condition, which has proven very robust and is routinely employed in 
moving-puncture simulations, is usually initialized with a vanishing shift vector. Consequently, it takes 
typically at least one orbit until initial transients have decayed in the orbital tracks. It is thus not possible, 
for example, to compute the orbital frequency at $t=0$ and the required fits can be performed only 
after the system has reached quasi-equilibrium. 
Second, depending on the choice of a free parameter $\eta$ in the $\Gamma$-driver shift, gauge 
dependent oscillations at about twice the orbital frequency contaminate the orbital tracks throughout the 
inspiral for eccentricities on the order of $10^{-3}$ and below, which prevents us from accurately performing 
the required fits. 
Third, while fitting to time-derivatives simplifies the model functions, it also amplifies numerical noise and 
increases the need for filtering even at eccentricities on the order of $5 \times 10^{-3}$. 

To be specific, we consider one variant of the latest version of the method~\cite{Buonanno:2010yk}, 
which only makes use of the frequency and therefore could in principle be used with a filtered GW frequency 
instead of the orbital frequency to avoid the problems mentioned above. This variant requires the following fit
\begin{equation}
  \label{CC:omega_model}
  \dot \Omega_\text{NR}(t) = S_\Omega(t) + B_\Omega\cos\left( \omega_\Omega t + \phi_\Omega \right),
\end{equation}
with fitting parameters $B_\Omega$, $\omega_\Omega$, and $\phi_\Omega$ and the following non-oscillatory 
model
\begin{equation}
  S_\Omega^k = \sum_{n=0}^{k-1} A_k \left( T_c - t \right)^{-11/8 - n/4}, 
\end{equation}
with $k=1$ or $k=2$ and free parameters $A_k$ and $T_c$.
Various choices for the updating formulas for $\dot r$ and $\Omega$ at $t=0$ can be made, but, 
as far as we are aware, only the following choice allows us to rely exclusively on frequency information
\begin{align}
  \Delta\dot r &= \frac{r_0 B_\Omega}{2\Omega_0} \cos\phi_\Omega,\\
  \Delta\Omega &= - \frac{B_\Omega}{4\Omega_0} \sin\phi_\Omega.
\end{align}

When using the time-derivative of the cleaned GW signal, one problem with this method is that the 
GW signal for moving-puncture evolutions is extremely noisy for the first one to two orbital periods. 
This implies that the cleaned GW signal and the residual eccentricity oscillations will only be accurate 
in the actual fitting window, starting after roughly two orbits. The phase offset $\phi_\Omega$ of the 
eccentricity oscillations, which is the crucial quantity for this variant of the CCC-method, and is 
determined by the fit at $t=0$, may therefore not be accurate enough to efficiently reduce the 
residual eccentricity. Moreover the frequency $\omega_\Omega$ is not constant over time, which 
exacerbates the error in the phase offset further. This can be partially addressed by adding a term 
proportional to $t^2$ in the argument of the cosine in~\eqref{CC:omega_model} as mentioned 
in~\cite{Buonanno:2010yk}. 

We briefly compare the essence of the variant of the CCC-method discussed above to our PN-based 
eccentricity reduction method.
The objective of both methods is to compute approximate corrections to initial parameters at a 
starting time $t=0$ and a radial separation $r_0$. To do that, residual eccentricity information 
from one NR frequency quantity is extracted. In the CCC-method this is done by fitting the frequency 
to a nonlinear model function \eqref{CC:omega_model}, while in our method the initial parameters 
of a PN model are adjusted such that the residual of the time-evolution of this PN-model matches with 
the NR frequency residual. As a second step, updating formulas for the initial parameters are used in 
the CCC-method that rely on Newtonian information. In our method, we simply apply the matching 
perturbation in the opposite direction to recover our updated initial parameters. While the computation 
of the match in our method is more involved that the fit in the CCC-method, we can avoid a nonlinear 
fit in the method proper and obtain a robust method. 

While there are technical problems in applying the CCC method directly to moving-puncture simulations,
on the other hand it should be possible to apply the method that we have presented to SpEC 
CTS-excision simulations.  It would be interesting in the future to compare the two methods in that setting.



  
\section{Summary and discussion} 
\label{sec:conclusion}

We have developed a robust iterative procedure to reduce the eccentricity in moving-puncture simulations
of black-hole binaries. In this method the eccentricity is measured from the phase of the 
GW signal, reducing the gauge dependence of previous methods, which used 
the orbital motion. 

The method relies on calculating differences (residuals) between the GW frequency of an NR simulation,
and that of an analytic model of the same physical system. The key requirements of the model are 
that it (a) accurately capture the phase evolution of the binary; (b) be parametrized by the same initial 
parameters as the NR simulation, which in our case are the radial and tangential momenta $(p_r, p_t)$; 
and (c) we can produce a model solution with zero eccentricity. We use solutions 
of a set of PN/EOB equations as the model (Appendix~\ref{sec:eob_pn_equations_of_motion}). 
We calculate two residuals with respect to the zero-eccentricity model frequency, 
$\omega_\text{M}(p_r^\circ,p_t^\circ)$. In one we use the NR GW frequency, and in the other we use 
a perturbed model frequency, $\omega_\text{M}(\lpr p_r^\circ, \lpt p_t^\circ)$. The essence
of the method is in finding the scale factors $(\lpr, \lpt)$ such that the two residuals agree in both 
amplitude and phase. The amplitude and phase can be adjusted independently, making it possible
to semi-automate this procedure. We then update the NR initial parameters with the inverse parameters, 
$p_r \rightarrow p_r / \lpr$, $p_t \rightarrow p_t / \lpt$. We find that with this procedure the eccentricity can
typically be reduced from $e \sim 0.01$ to $e < 10^{-3}$ in two iteration steps. 

For the method to work, we must filter the NR GW signal, and account for dephasing effects in the 
frequency residuals. Even then, noise in the NR waveform prevents us from reducing the eccentricity to lower
than about $2 \times 10^{-4}$. However, in studies of the mismatch between hybrid PN+NR waveforms, 
 we have 
seen no evidence that eccentricities as high as 0.01 in the final $\sim$10 orbits will have any
 noticeable influence on GW searches or 
parameter estimation in the Advanced detector era.
This is somewhat surprising, since at this level the 
eccentricity is visible by eye in the waveform. To be conservative we prefer to lower the eccentricity 
to a level where the eccentricity-induced oscillations and secular drifts in the GW phase are well below 
the numerical phase errors in our simulations. We choose a tolerance of $e \sim 10^{-3}$, 
which produces oscillations in the GW phase with an amplitude of $\Delta \phi \sim 0.01$\,rad during
inspiral, and an accumulated phase offset through merger and ringdown of less than 0.2\,rad. This
is well within our numerical phase errors. 
 We note that our analysis considered
only one configuration, but we do not expect the relationship between phase oscillations 
and eccentricity to change very much across the parameter space.

We have shown that, for typical gauge choices in moving-puncture simulations, the frequency of the 
orbital motion cannot be used to accurately measure the eccentricity below $e \sim 0.002$, and
even at this level the eccentricity estimator is contaminated by gauge effects; 
see, for example, Fig.~\ref{fig:q2a0a025_e_omega_orb}.

In large studies of the black-hole-binary parameter space, we estimate that the computational 
overhead in performing this eccentricity reduction is between 25\% and 35\%. 

In our implementation of the method the GW signal is given by the Newman-Penrose scalar 
$\Psi_4$. We make the important observation that the eccentricity measured from the phase
of $\Psi_4$ will be different to that from the GW strain $h$. In 
Appendix~\ref{sub:eccentricity_estimators_for_gw} we find a simple 
expression for the relative scale factor between the two measures. 

So far the method has been applied to non-precessing binaries. Precession will introduce
additional oscillations into the GW amplitude and frequency, 
which will make it difficult to isolate the effects of eccentricity and of precession, and hence
complicate our method. However, techniques that simplify the phase evolution of the 
precessing-binary waveform, like that suggested 
in~\cite{Schmidt:2010it,O'Shaughnessy:2011fx,Boyle:2011gg}, may alleviate this 
problem. We will consider this further in future work. 

Because our method is applied to the GW signal, it can be adapted to any evolution method,
and is not limited to moving-puncture simulations. It could also be adapted to other compact binary 
simulations, for example neutron-star (NS-NS) binaries, and black-hole--neutron-star (BH-NS) 
binaries.

  
  
\section{Acknowledgements}

It is a pleasure to thank Ian Hinder, Abdul Mrou{\'e}, Harald Pfeiffer, Frank Ohme and Barry Wardell
for discussions.

MP and MH were supported by the Science and Technology Facilities Council grants
ST/H008438/1 and ST/I001085/1. MP was also supported in part by the FWF (Fonds zur 
F\"orderung der wissenschaftlichen Forschung)  grant  P22498. 
SH was supported by the Spanish Ministry of Economy and Competitiveness 
(projects FPA2010-16495, CSD2009-00064), the Conselleria d'Economia Hisenda i Innovacio 
of the Govern de les Illes Balears, and by European Union FEDER funds.
Part of this work was carried out during the ``Dynamics of General Relativity'' workshop at the  
Erwin Schr\"odinger Institute, Vienna, 2011. MP also thanks the Universitat de les Illes Balears 
(UIB) for hospitality. 
Numerical simulations were performed on the Vienna Scientific Cluster (VSC), the 
Leibniz-Rechenzentrum (LRZ), Advanced Research Computing Cardiff (ARCCA), 
MareNostrum at BSC-CNS and the PRACE clusters Hermit and Curie.


\appendix

\section{EOB/PN equations of motion} 
\label{sec:eob_pn_equations_of_motion}

For our post-Newtonian evolutions we are using the Hamiltonian equations of motion in the standard 
Taylor-expanded form, as we have done previously \cite{Husa:2007rh,Hannam:2010ec}, and in the 
EOB form \cite{Buonanno:1998gg,Damour:2001tu}. 
For the Taylor-expanded version, we use the non-spinning 3PN accurate Hamiltonian
\cite{Jaranowski:1997ky,Damour:2001bu,Damour:2000kk} (see also 
\cite{Blanchet:2000ub,deAndrade:2000gf,Blanchet:2002mb}) and 3.5PN 
accurate radiation flux \cite{Blanchet:1997jj,Blanchet:2001aw,Blanchet:2004ek}. We add both leading-order
\cite{Barker1970,Barker1974,Barker1979,Kidder:1995zr,Damour:2001tu,Poisson:1997ha}
and next-to-leading order \cite{Blanchet:2006gy,Faye:2006gx,Damour:2007nc}
contributions to the spin-orbit and spin-spin Hamiltonians, and
spin-induced radiation flux terms as described in \cite{Buonanno:2005xu} 
(see also \cite{Kidder:1995zr,Poisson:1997ha}). In addition we include the flux contribution due to the energy flowing in to 
the black holes, which appears at the relative 2.5PN order, as derived in 
Ref.~\cite{Alvi:2001mx}.

For the EOB equations, we use the 3PN accurate resummation \cite{Damour:2000we} of the above 
non-spinning Hamiltonian, and add the Taylor-expanded spinning terms. We evolve the resulting 
EOB Hamiltonian evolution equations in the same ADM-TT coordinate system used for the 
Taylor-expanded Hamiltonian approach, i.e.,  representing the EOB momenta and generalized 
coordinates in terms of the ADM-TT expressions. For simplicity, we perform this canonical coordinate 
transformation from the EOB to the ADM-TT phase space only to 2PN order, 
as described in~\cite{Buonanno:1998gg}. This is sufficient for our purposes, as demonstrated by the 
successful of our eccentricity reduction procedure. In the future we do however plan to use the 
3PN-order transformation \cite{Damour:2000we}.


\section{Eccentricity perturbations in Newtonian dynamics} 
\label{sec:newtonian_dynamics_perturbing_initial_parameters}

We consider the effect of momenta perturbations on eccentricity in the simple setting of 
Newtonian binaries. We consider both conservative motion, and the inclusion of quadrupole radiation reaction.
This allows us to derive some basic results that were useful in developing and implementing our general
eccentricity-reduction procedure. 

In Sec.~\ref{sub:newtonian_eccentricity_estimators} we first define the customary Newtonian eccentricity 
estimators that are based on the analytical solution to the Kepler problem linearized in eccentricity. We consider
the effects of momenta perturbations to conservative dynamics in Sec.~\ref{sub:conservative_dynamics}, and show
that the effects of radial and tangential momenta perturbations are out of phase. We include quadrupole radiation 
reaction in Sec.~\ref{sub:newtonian_evolutions_with_quadrupole_flux}, which allows us to study the dephasing
due to eccentricity. 

An understanding of the behavior of GW-signal-based eccentricity estimators
requires us to go beyond Newtonian order, and in Appendix~\ref{sec:1pn_eccentricity_estimators} we consider
1PN effects. This calculation allows us to derive the leading-order ratio between the eccentricity measured from
the GW strain $h$ and the Newman-Penrose scalar $\Psi_4$.

As a simple model for an inspiraling binary we consider Newtonian dynamics with quadrupole 
radiation reaction as discussed in~\cite{Buonanno:2005xu}.
The Hamiltonian for the Kepler problem in polar coordinates $q_i = (r,\phi)$ in the center-of-mass frame with total
mass $M = m_1 + m_2$ and symmetric mass-ratio $\nu = m_1 m_2 / M^2$ is
\begin{equation}
  H_N = \frac{p_r^2}{2\nu M} + \frac{p_\phi^2}{2r^2\nu M} - \frac{\nu M^2}{r},
\end{equation}
where we have set $G=1$.
The equations of motion with an added nonconservative radiation reaction force are
\begin{align}
  \label{eq:Newtonian_RR_eqns}
  \dot q_i &=   \pd{H_N}{p_i},\\
  \dot p_i &= - \pd{H_N}{q_i} + F_i,
\end{align} 
where the ``radiation reaction force'' is proportional to the momentum vector and the rate of energy loss,
\begin{equation}
  \label{eq:RR-force}
  F_i =  p_i  \frac{1}{\Omega L}\frac{dE}{dt}.  
\end{equation}
The quadrupole flux is given as 
\begin{equation}
  \label{eq:RR-quadrupole}
  \frac{dE}{dt} = -\frac{32}{5}\nu^2 (M \Omega)^{10/3},  
\end{equation}
where $\Omega = \dot\phi$ is the orbital frequency. 
The radial momentum is $p_r = \nu M \dot r$, and the angular momentum $L \equiv p_\phi$ is conserved 
for vanishing energy flux. Instead of $L$ we use the tangential momentum $p_t = L/r$ in the following sections.

 
\subsection{Newtonian eccentricity estimators} 
\label{sub:newtonian_eccentricity_estimators}

In this section we collect the customary definitions of eccentricity estimators used in the context of BH binary 
evolutions. These estimators are based on Taylor expansions of the Kepler solution to linear order in eccentricity. 
This conservative Newtonian setting is the \emph{only} one in which eccentricity can be uniquely defined.

We first summarize the explicit solutions (see e.g.,~\cite{Maggiore2008-GW-vol1}) for to the Kepler problem. 
Here we drop the subscript $_\text{orb}$ for orbital quantities and only give it for the eccentricity estimators.

The radial separation as a function of the phase is given by
\begin{equation}
  \label{eq:r_of_phi_Newtonian}
  r(\phi) = \frac{a (1-e^2)}{1 + e \cos\phi},
\end{equation}
where $a = L^2 / (M^3 \nu^2 (1-e^2))$
is the semi-major axis, and the orbital frequency is
\begin{equation}
  \Omega(\phi) = \frac{L}{\nu M r(\phi)^2}.
\end{equation}
The Newtonian orbital phase
\begin{equation}
  \phi = \phi_0 + A_e(u),
\end{equation}
is defined in terms of the \emph{true anomaly}
\begin{equation} 
  A_e(u) = 2 \arctan\left[ \left(\frac{1+e}{1-e}\right)^{1/2} \tan\frac{u}{2} \right],
\end{equation}
and \emph{eccentric anomaly} $u$, which is related to $t$ by the Kepler equation
\begin{equation}
  \beta := \Omega_0 \, t = u - e \sin u,
\end{equation}
where $\beta$ is the \emph{mean anomaly} and the average orbital frequency is $\Omega_0^2 = M / a^3$.

Expanding to linear order in the eccentricity, we find
\begin{align}
  \label{eq:Newtonian-r-om-linearized}
  r(\phi)      &= \frac{L^2}{M^3 \nu^2} (1 - e \cos\phi) + \bigO(e^2),\\
  \Omega(\phi) &= \frac{M^5 \nu^3}{L^3} (1 + 2 e \cos\phi) + \bigO(e^2),\\
  \phi(t)      &= \phi_0 + \Omega_0 t + 2 e \sin(\Omega_0 t) + \bigO(e^2),
\end{align}
and can define eccentricity estimators from the orbital separation $r$, orbital frequency $\Omega$ 
and orbital phase $\phi$ as follows,
\begin{align}
  e_{r,\text{orb}}    &= \frac{r(t) - r_\text{fit}(t)}{r_\text{fit}(t)},\\
  \eomOrb             &= \frac{\Omega(t) - \Omega_\text{fit}(t)}{2 \Omega_\text{fit}(t)},\\
  e_{\phi,\text{orb}} &= \frac{\phi(t) - \phi_\text{fit}(t)}{2}.
\end{align}

 
\subsection{Conservative dynamics} 
\label{sub:conservative_dynamics}

In this section we study at an analytic level the effect on eccentricity of momenta perturbations in a Newtonian binary. 
We start with conservative dynamics.

For the Kepler problem the orbital eccentricity can be written explicitly as a function of the separation $r$ and the radial and tangential momenta $p_r$ and $p_t$
\begin{equation}
  \label{ecc:Newtonian}
  e(p_t, p_r, r) = \sqrt{1 + \frac{2 E L^2}{\nu^3 M^5}}.
\end{equation}
Note that for a bound system the energy $E$ is negative.

Circular orbits satisfy $p_r = 0$ and $\dot p_r = 0$ at all times, and this requirement leads to the solution $p_t^\circ
= \nu M \sqrt{M / r_0}$. We now perturb either $p_r$ or $p_t$, which will give rise to eccentricity. The eccentricity
$e(p_r,p_t)$ has a single minimum at the circular-orbit momenta values, which implies that given generic elliptical data
both momenta need to be adjusted to find the circular initial values.

\begin{figure}[htbp]
  \centering
    \includegraphics[width=1.5in]{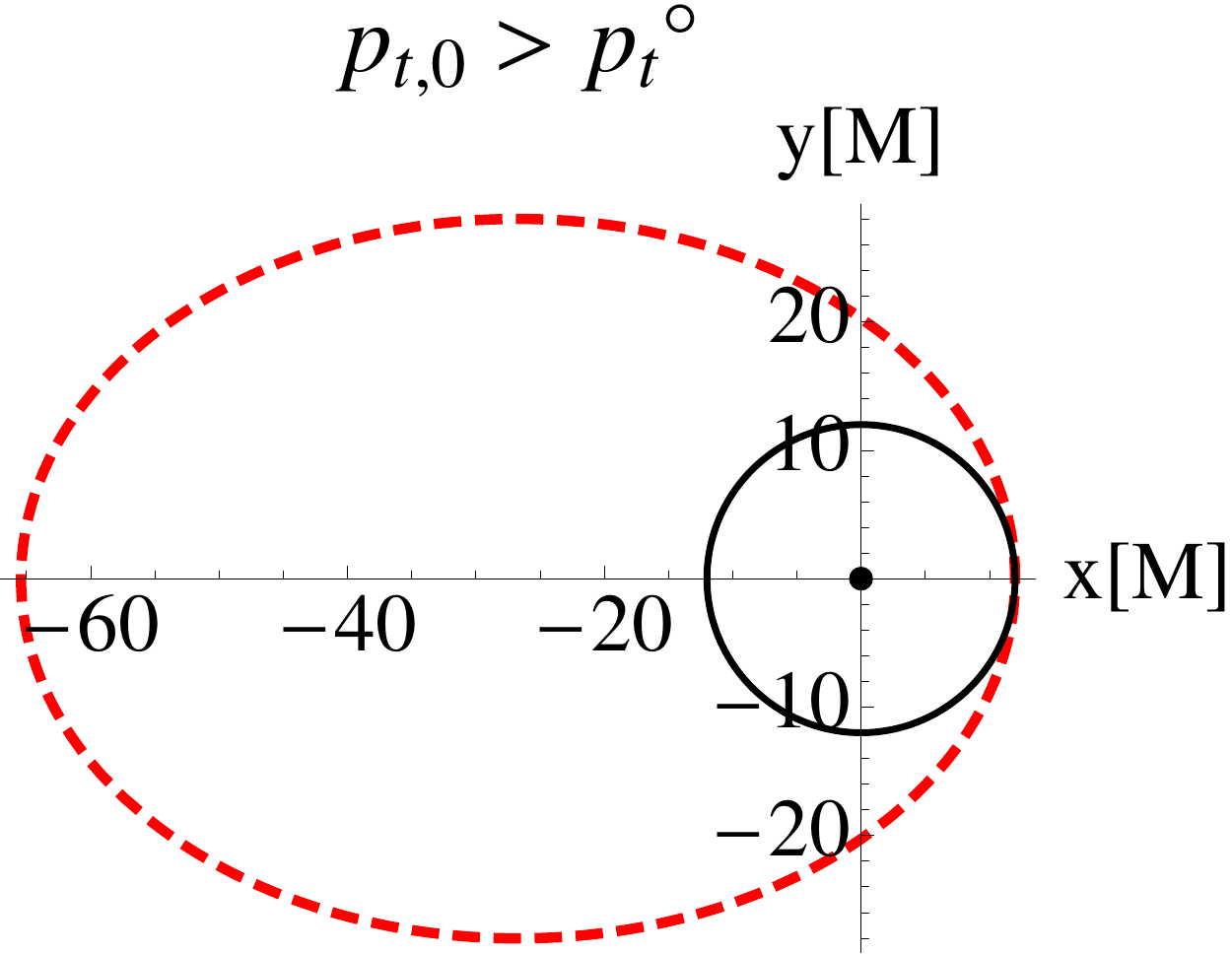}
    \includegraphics[width=1.5in]{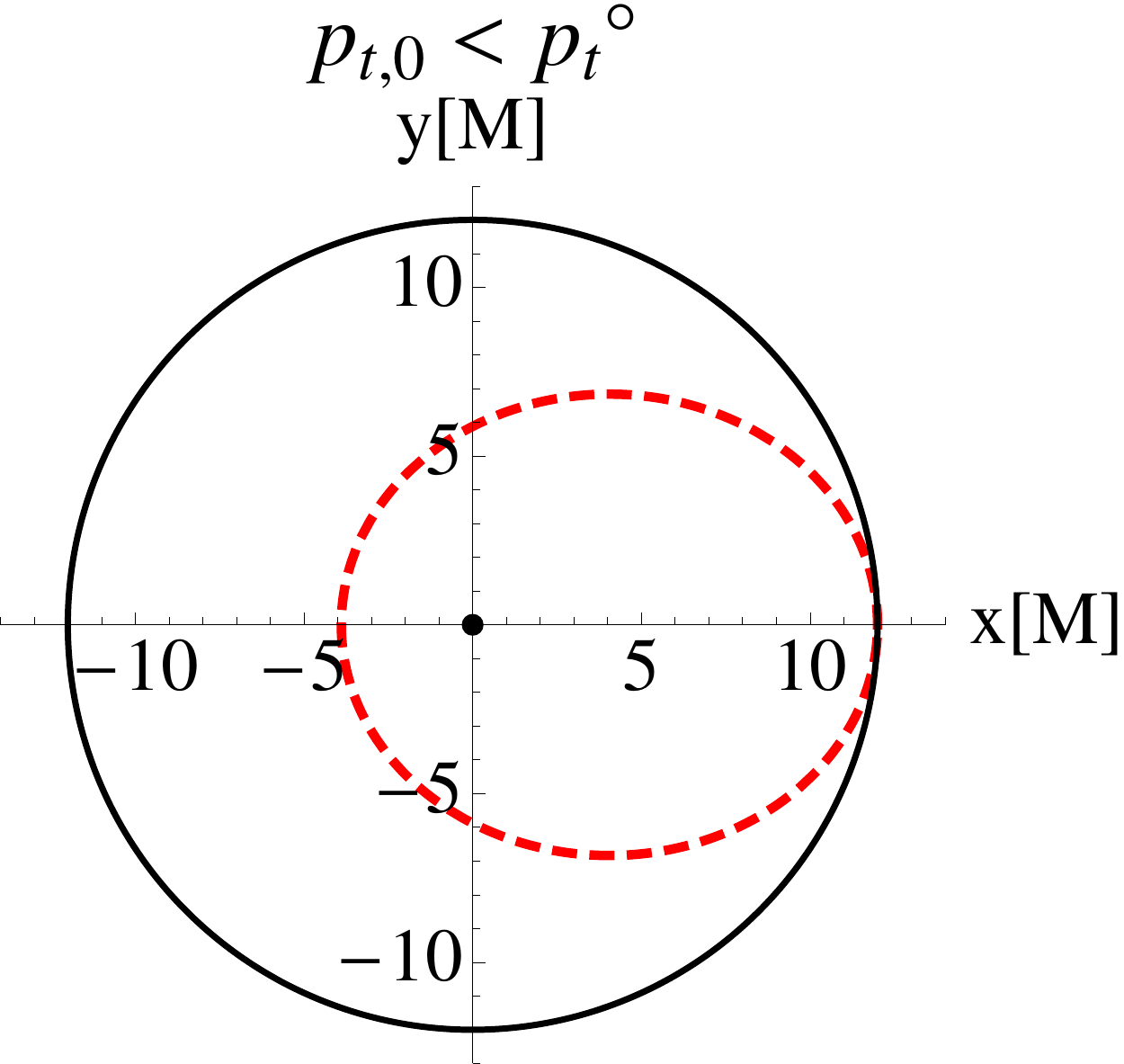}
  \caption{Newtonian orbits for perturbation of initial tangential momentum from circular initial parameters for $q=2,  r_0=12 M$. 
  The circular orbit is shown in blue. For the left plot $p_{t,0} = 1.3 p_t^\circ$ and for the right plot $p_{t,0} = 0.7 p_t^\circ$.
  }
  \label{fig:NewtonianBinaryZeroFlux_pert_pt}
\end{figure}

\begin{figure}[htbp]
  \centering
    \includegraphics[width=1.5in]{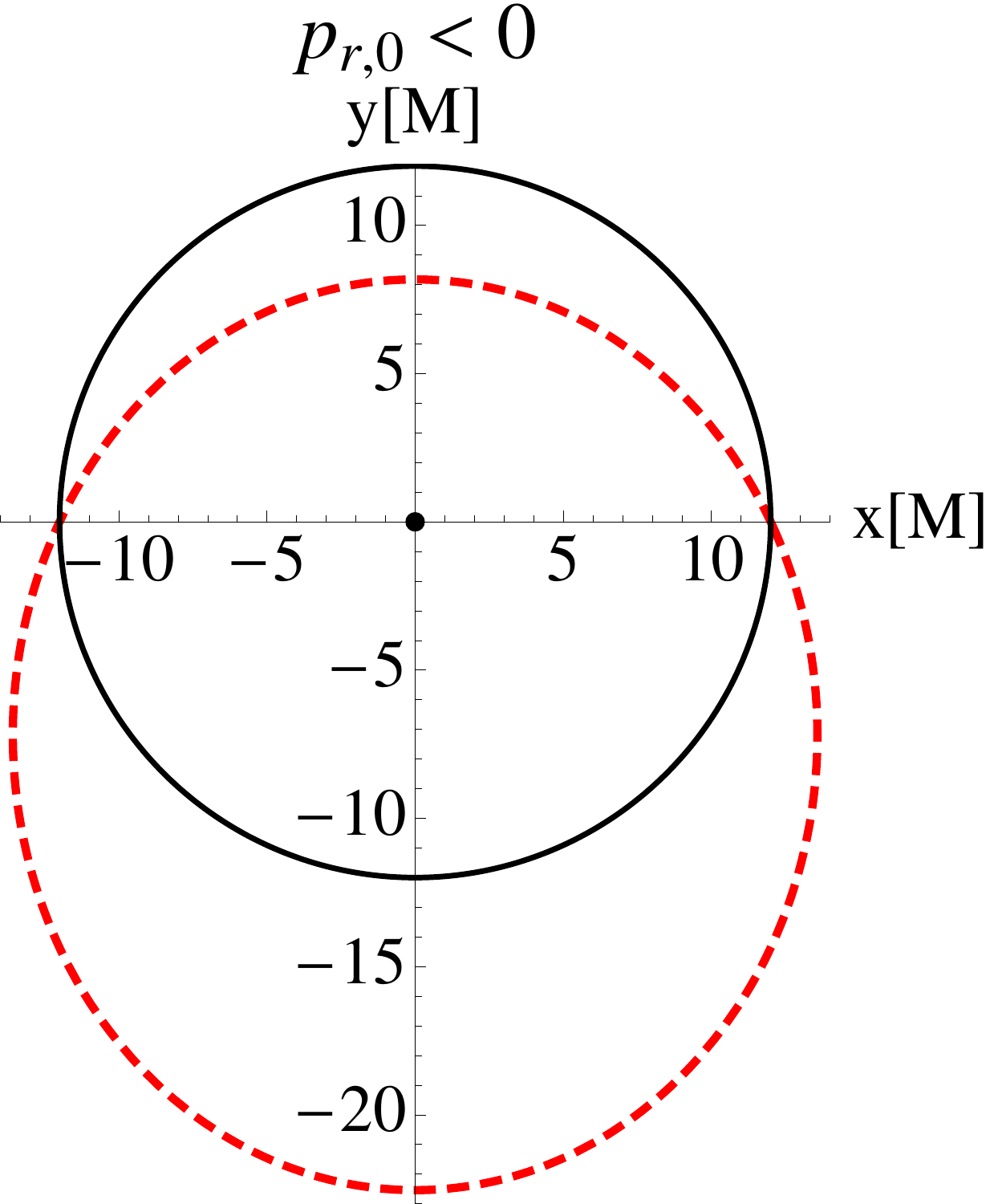}
    \includegraphics[width=1.5in]{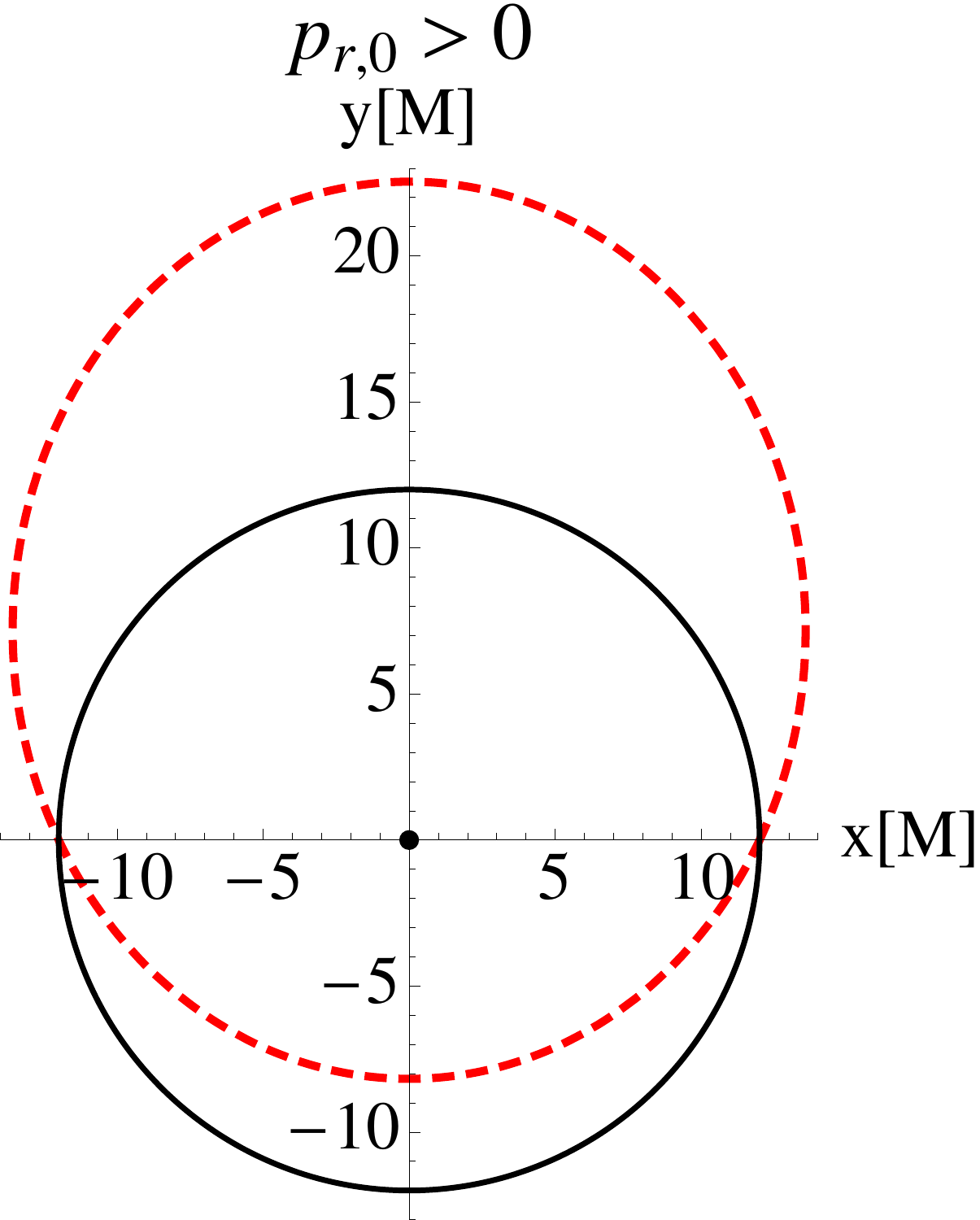}
  \caption{Newtonian orbits for perturbation of initial radial momentum from circular initial parameters for $q=2, r_0=12 M$. 
  The circular orbit is shown in blue. For the left plot $p_{r,0} = -0.03 M$ and for the right plot $p_{r,0} = 0.03 M$.
  }
  \label{fig:NewtonianBinaryZeroFlux_pert_pr}
\end{figure}

Fig.~\ref{fig:NewtonianBinaryZeroFlux_pert_pt} illustrates the effect of perturbing the initial tangential momentum
$p_{t,0}$ in a $q=2$ binary, where either $p_{t,0} > p_t^\circ$ (left panel), or $p_{t,0} < p_t^\circ$ (right panel).
Large perturbations were chosen to exaggerate the effect. We see that an increase in the initial tangential momentum
leads to a larger radial separation, with a maximum at the phase $\phi = \pi$ (apastron), while the radius keeps the
same value at the periastron. In contrast, the perturbation of the orbital frequency starts at a maximum and reaches a
minimum at the apastron. The frequency is also shifted by a constant offset.
Choosing the initial radial momentum smaller than its circular value zero leads to the perturbation of the radial
separation that vanish at the periastron and apastron and take extremal values half-way in between. The perturbation of
the orbital frequency behaves similarly, but with the opposite sign; see Fig.~\ref{fig:NewtonianBinaryZeroFlux_pert_pr}.

\begin{figure}[htbp]
  \centering
    \includegraphics[width=1.5in]{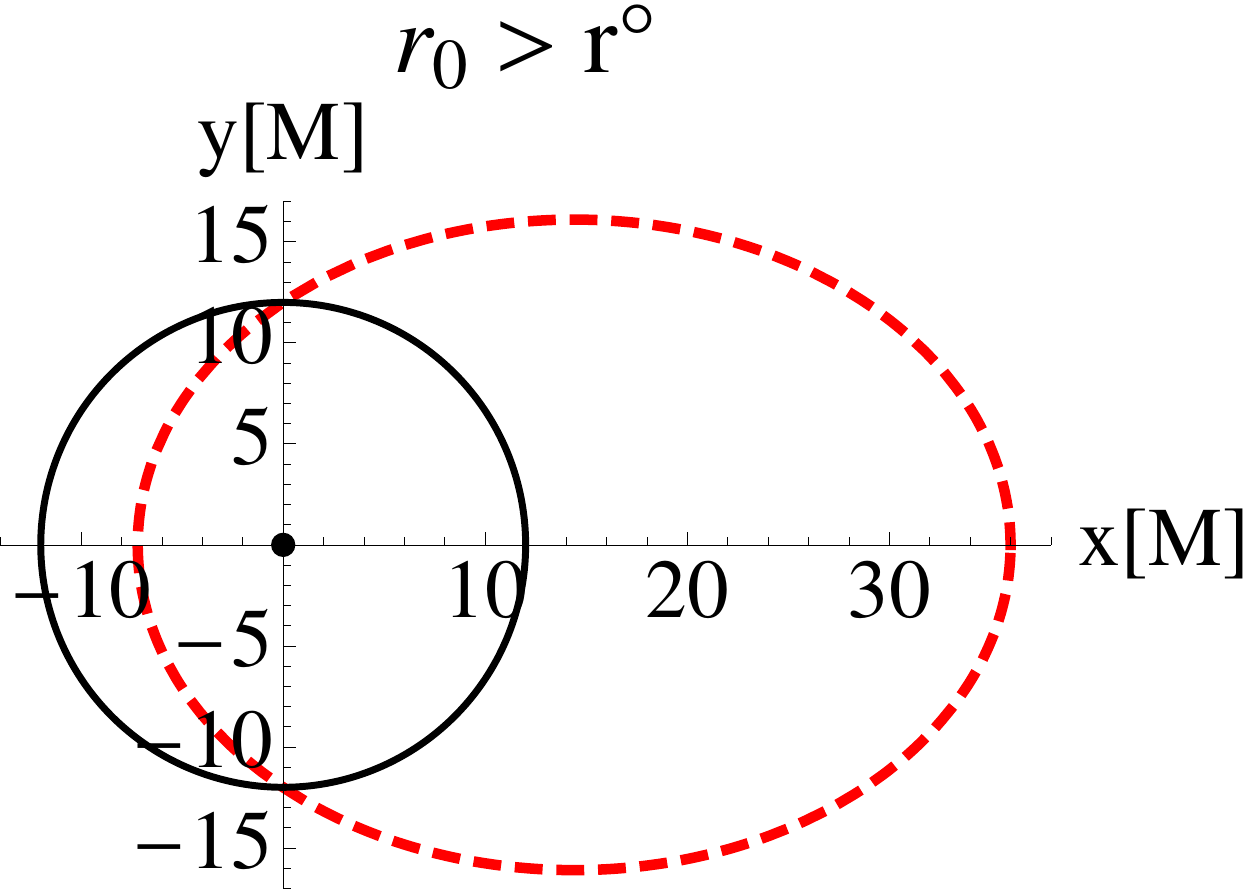}
    \includegraphics[width=1.5in]{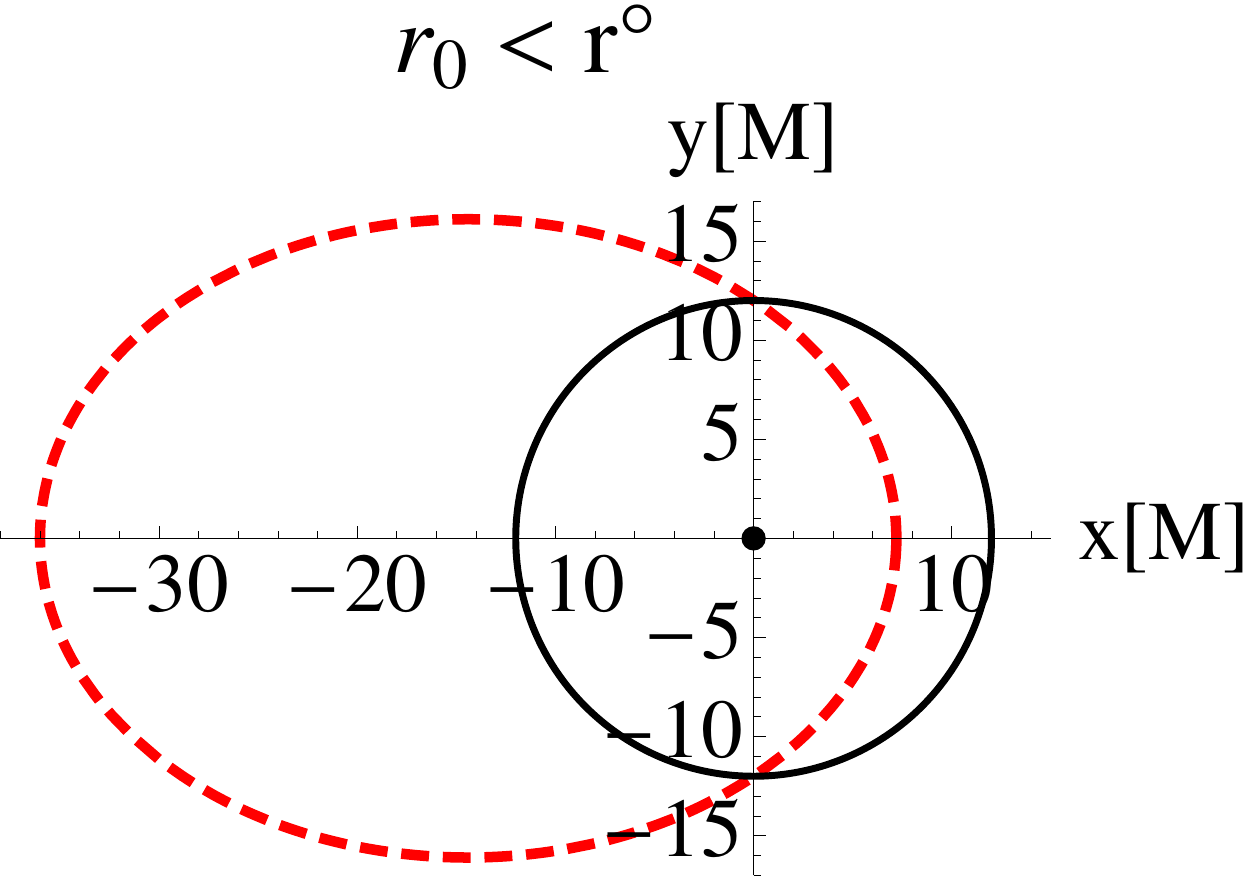}
  \caption{Newtonian orbits for a perturbation of the initial radial separation from its circular-orbit value for $q=2, r^\circ=12 M$. 
  The circular orbit is shown in with a solid line. We refer to the radial separation of the reference circular
  parameters as $r^\circ$. For the left plot $r_0 = 3 r^\circ$ and for the right plot $r_0 = 0.6 r^\circ$.
  }
  \label{fig:NewtonianBinaryZeroFlux_pert_r}
\end{figure}

For completeness, we also consider the effect of perturbing the initial separation away from its circular orbit value
$r^\circ$, which will also introduce eccentricity. Choosing $r_0 < r^\circ$ results in an elliptic orbit that starts at
the periastron (see Fig~\ref{fig:NewtonianBinaryZeroFlux_pert_r}). The perturbation is therefore qualitatively similar
to increasing the tangential momentum from its circular value and the associated mode for the orbital frequency is
proportional to a cosine without the offset that is present in the latter case. This is not very surprising, as we still
have $p_{r,0} = 0$ and only $p_{t,0}$ does not have its correct circular value for the new initial radial separation.
Choosing $r_0 > r^\circ$ again flips the sign of the mode.

We can analyze the analytical solution for the Kepler problem given in section~\ref{sub:newtonian_eccentricity_estimators} 
to gain more insight into the behavior of these momenta perturbations. 
For our purposes it is sufficient to do this up to linear order in eccentricity.
To simplify the solutions we consider the circular initial momenta for a given initial separation $r_0$
\begin{align}
  p_r^\circ &= 0,                  \label{eq:Neqtonian-p_r_circ}\\
  p_t^\circ &= \nu M \sqrt{M/r_0}, \label{eq:Neqtonian-p_t_circ}
\end{align}
as the reference values and write the reference frequency as $\Omega^\circ := \Omega^{p_t^\circ, p_r^\circ} = M^5 \nu^3 / {L^\circ}^3$.
Perturbing the initial tangential momentum $p_{t,0}$ at $t=0$ by a factor $\lpt$ is equivalent to perturbing the angular momentum 
$L$ by the same factor, as long as the initial separation stays fixed. 
If we choose $\lpt > 1$ and $\phi_0 = 0$, the orbital frequency perturbation is
\begin{align}
  \begin{split}
  \mathcal{R}_\Omega^{\lpt p_t^\circ} &:=  \Omega^{\lpt p_t^\circ} - \Omega^\circ\\
  &= \Omega^\circ \left[ \left( \frac{1}{\lpt^3} - 1\right) 
  + \frac{2 e(p_r^\circ, \lpt p_t^\circ,r_0)}{\lpt^3} \cos\phi(t)  \right] + \bigO(e^2)\\
  &= \Omega^\circ \left(\lpt - 1\right) \left(-3 + 4\cos\phi(t)\right) + \bigO((\lpt - 1)^2).
  \end{split}
\end{align}
Similarly, we can perturb the radial momentum $p_r$ from its circular value zero. 
Since  $p_r^\circ = 0$, we cannot scale it to obtain an eccentric solution. Therefore, we choose $p_{r,0}<0$ and find
\begin{align}
  \begin{split}
    \mathcal{R}_\Omega^{p_{r,0}} &:=  \Omega^{p_{r,0}} - \Omega^\circ\\ 
    &= 2 \Omega^\circ e(p_{r,0},p_t^\circ,r_0) \sin\phi(t) + \bigO(e^2)\\
    &= 2 \Omega^\circ |p_r| \sqrt\frac{{r_0}}{\nu^2 M^3} \sin\phi(t) + \bigO(p_r^2).\\
  \end{split}
\end{align}
Note that Eqn.~\eqref{eq:r_of_phi_Newtonian} assumes that the motion starts at the periastron. Choosing $p_{r,0}<0$ and $\phi_0 = 0$ the periastron is shifted by $\pi/2$, hence the sine.

\begin{figure}[htbp]
  \centering
    \includegraphics[width=.5\textwidth]{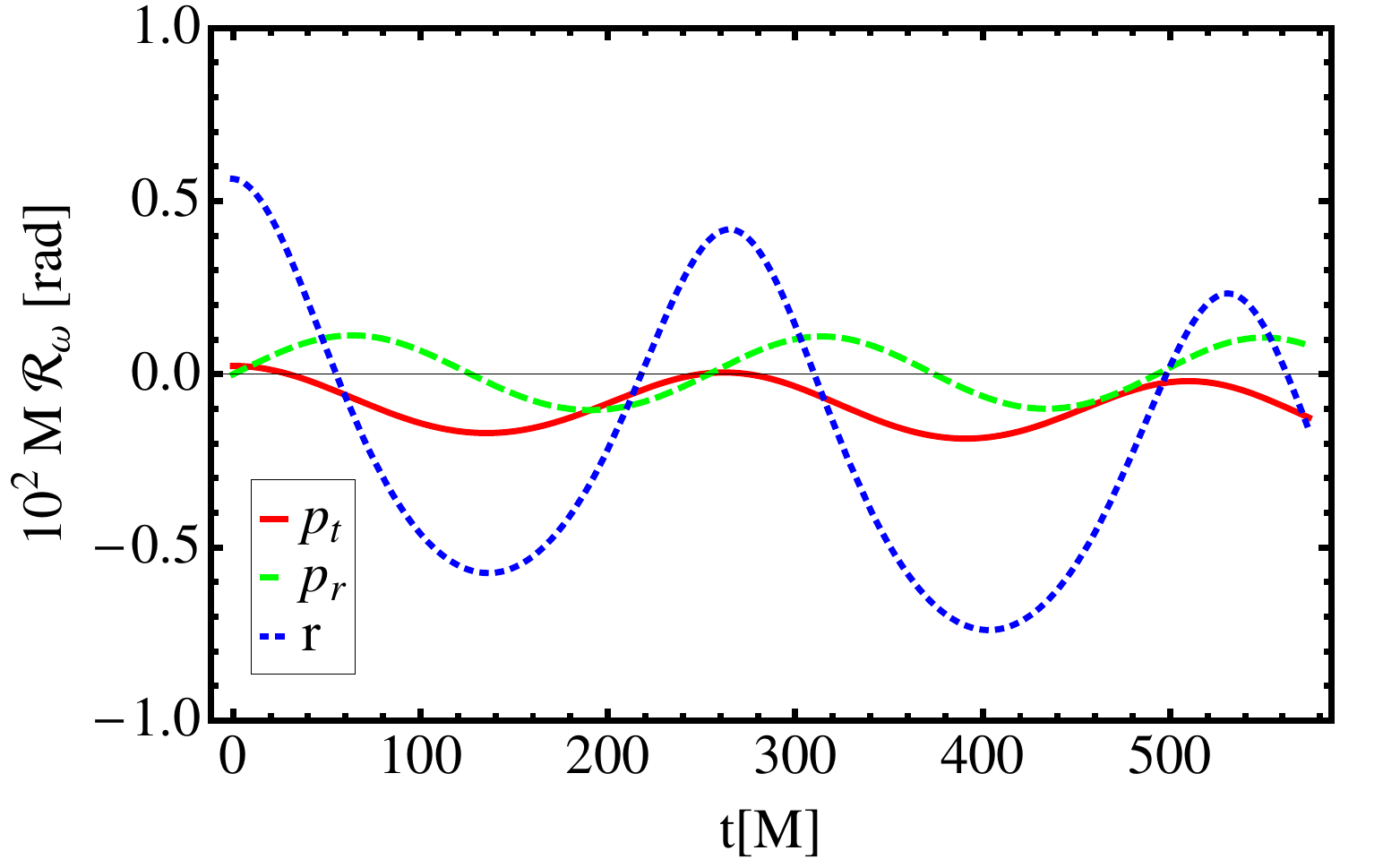}
  \caption{Frequency residuals for perturbations of the tangential momentum, radial momentum, and radial separation 
  for a conservative Newtonian model.
  The residuals are shown for a $q=2, r_0 = 12 M$ binary and perturbations $p_{t,0} = 1.03 p_t^\circ$ (red curve), 
  $p_{r,0} = -0.01$ (green long-dashed curve) and $r_0 = 0.9 r^\circ$ (blue short-dashed curve).
  }
  \label{fig:NewtonianBinaryZeroFlux_residuals}
\end{figure}

The perturbation of $p_t$ is associated with a cosine mode with an offset, while the perturbation of 
$p_r$ gives rise to a sine mode. In Fig.~\ref{fig:NewtonianBinaryZeroFlux_residuals} we plot frequency residuals for 
perturbations of $p_t$, $p_r$, and $r$, which show this mode behavior. This is consistent with the heuristic argument 
presented at the beginning of Sec.~\ref{sec:full_er_algorithm}. This fact allows us to extract the oscillatory information 
of a single frequency quantity to adjust two physical parameters (i.e., the radial and tangential momenta), and is a key 
feature of our eccentricity reduction method. One could also develop a simpler method, based on the variation of only
one parameter (for example the initial binary separation), but such a method would not be able to reduce the eccentricity
between some bound. See Sec.~\ref{sec:e_sensitivity}, where we estimate that this bound is around $e \sim 0.005$.

\subsection{Newtonian evolutions with quadrupole flux} 
\label{sub:newtonian_evolutions_with_quadrupole_flux}

In this section we include quadrupole radiation-reaction. This leads to dephasing between the quasi-circular and 
eccentric configurations. 

The addition of a radiation reaction through Eqns.~\eqref{eq:RR-force} and \eqref{eq:RR-quadrupole} to the Newtonian
equations of motion~\eqref{eq:Newtonian_RR_eqns} leads to inspiral, and the radial separation for a quasi-circular (QC)
configurations is
\begin{equation}
  \label{eq:rNewtonian_QC}
  r(t) = 4\left(\frac{\nu M^3}{5}(T_c - t)\right)^{1/4},
\end{equation}
(see e.g.,~\cite{Baumgarte2010-NR-book}), where $T_c$ is the coalescence time and can be expressed in terms of the 
initial radial separation as 
\begin{equation}
  \label{eq:T_c}
  T_c(r_0) = \frac{5 r_0^4}{256 \nu M^3}.  
\end{equation}
We choose the initial tangential momentum as in the conservative case, by Eqn.~\eqref{eq:Neqtonian-p_t_circ}.
The initial radial momentum that will result in QC inspiral is now nonzero and can be found by
combining equations \eqref{eq:rNewtonian_QC} and \eqref{eq:T_c} and taking a time derivative
\begin{equation}
  p_{r,\text{QC}} =  \nu M \dot r\left(t=0;T=T_c(r_0)\right) = -\frac{64 M^4 \nu^2}{5 r_0^3}.
\end{equation}

\begin{figure}[htbp]
  \centering
    \includegraphics[width=.5\textwidth]{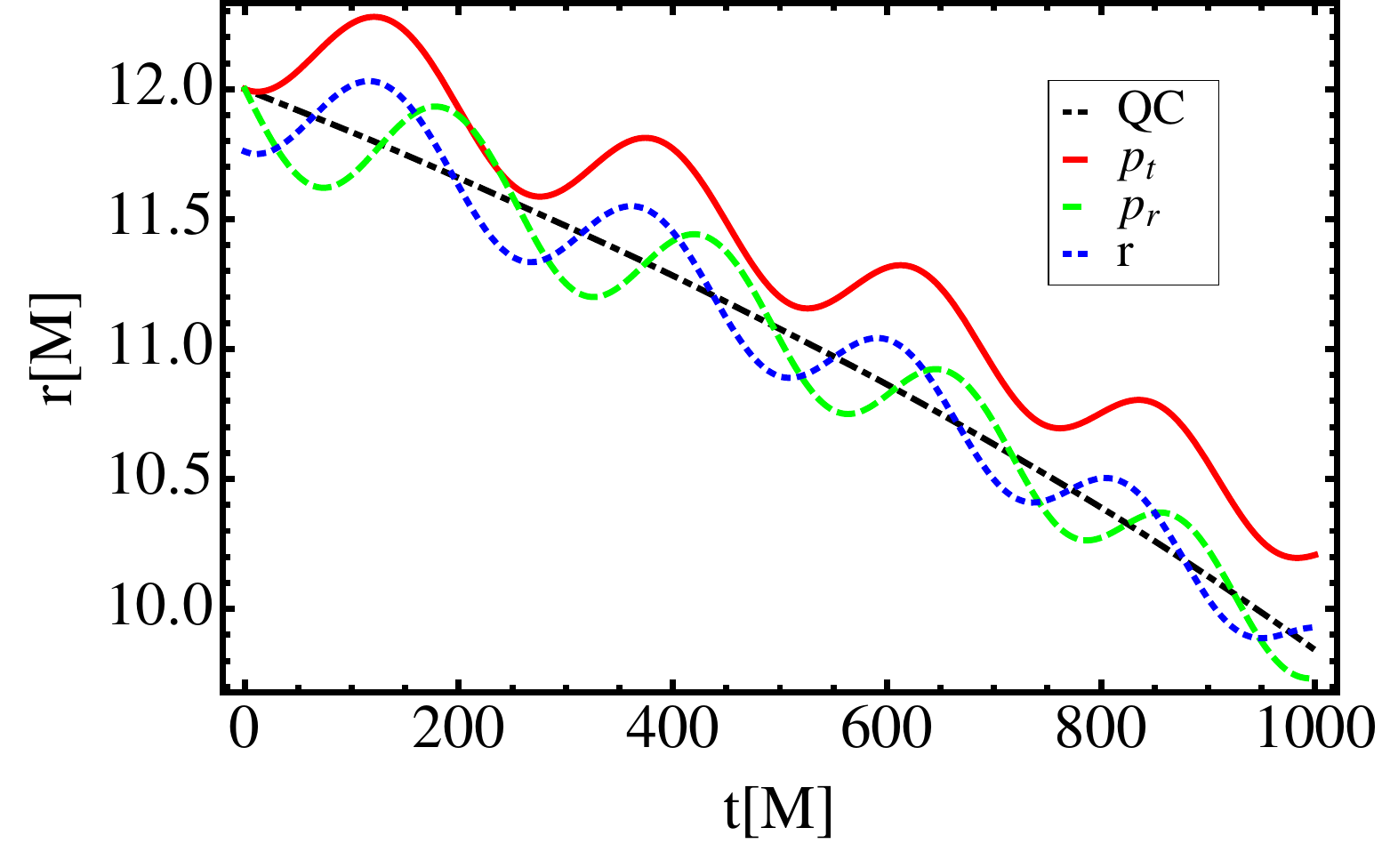}
    \includegraphics[width=.5\textwidth]{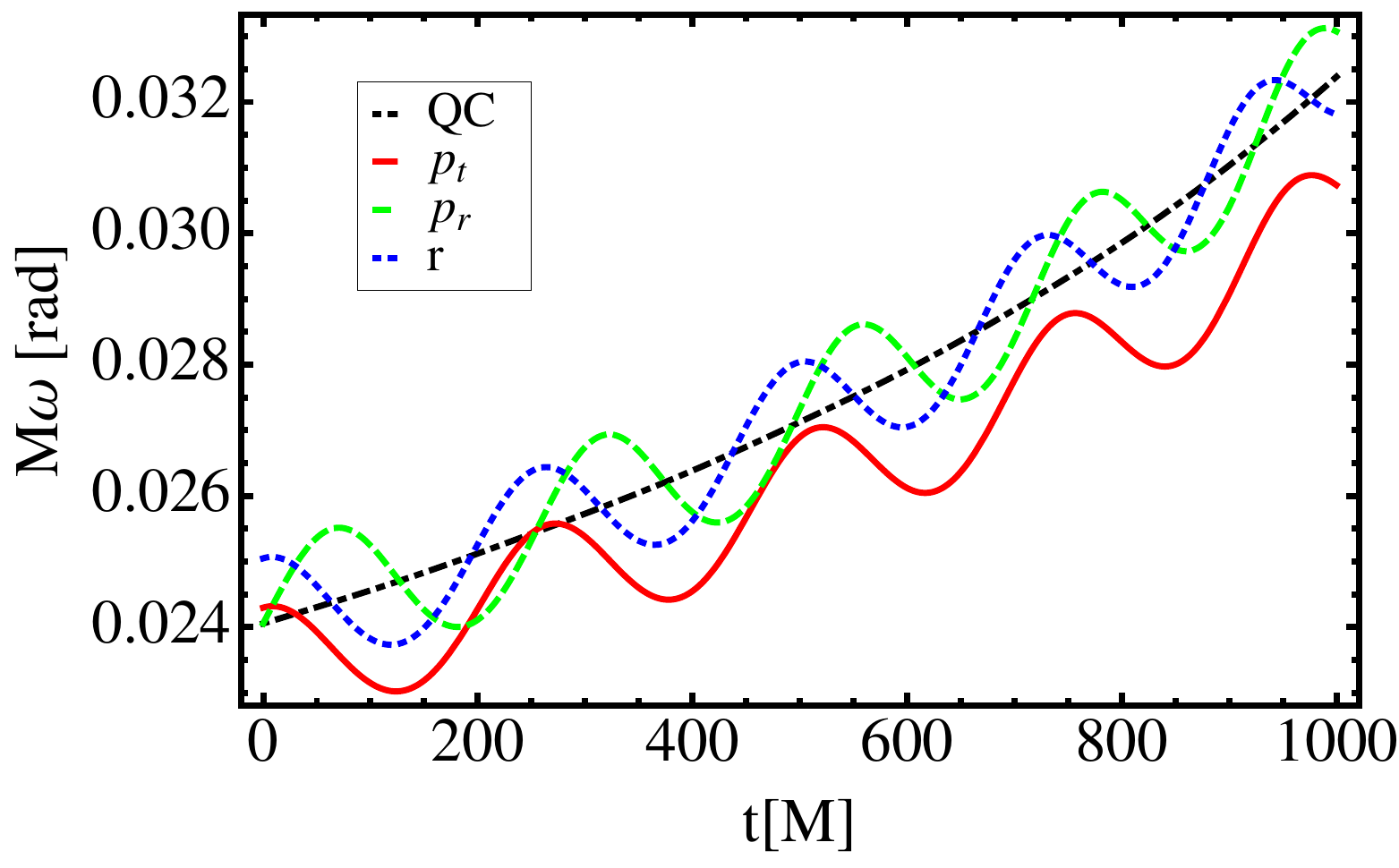}
  \caption{Radial separation (top) and orbital frequency (bottom) for Newtonian plus quadrupole flux inspiral for 
  QC data ($q=2, r_0 = 12 M$) and perturbations $p_{t,0} = 1.01 p_{t,\text{QC}}$, $p_{r,0} = 5 p_{r,\text{QC}}$ 
  and $r_0 = 0.98 r_\text{QC}$ from the QC values.
  }
  \label{fig:NewtonianBinaryFlux_inspiral}
\end{figure}

\begin{figure}[htbp]
  \centering
   \hspace*{0.15cm}\includegraphics[width=.497\textwidth]{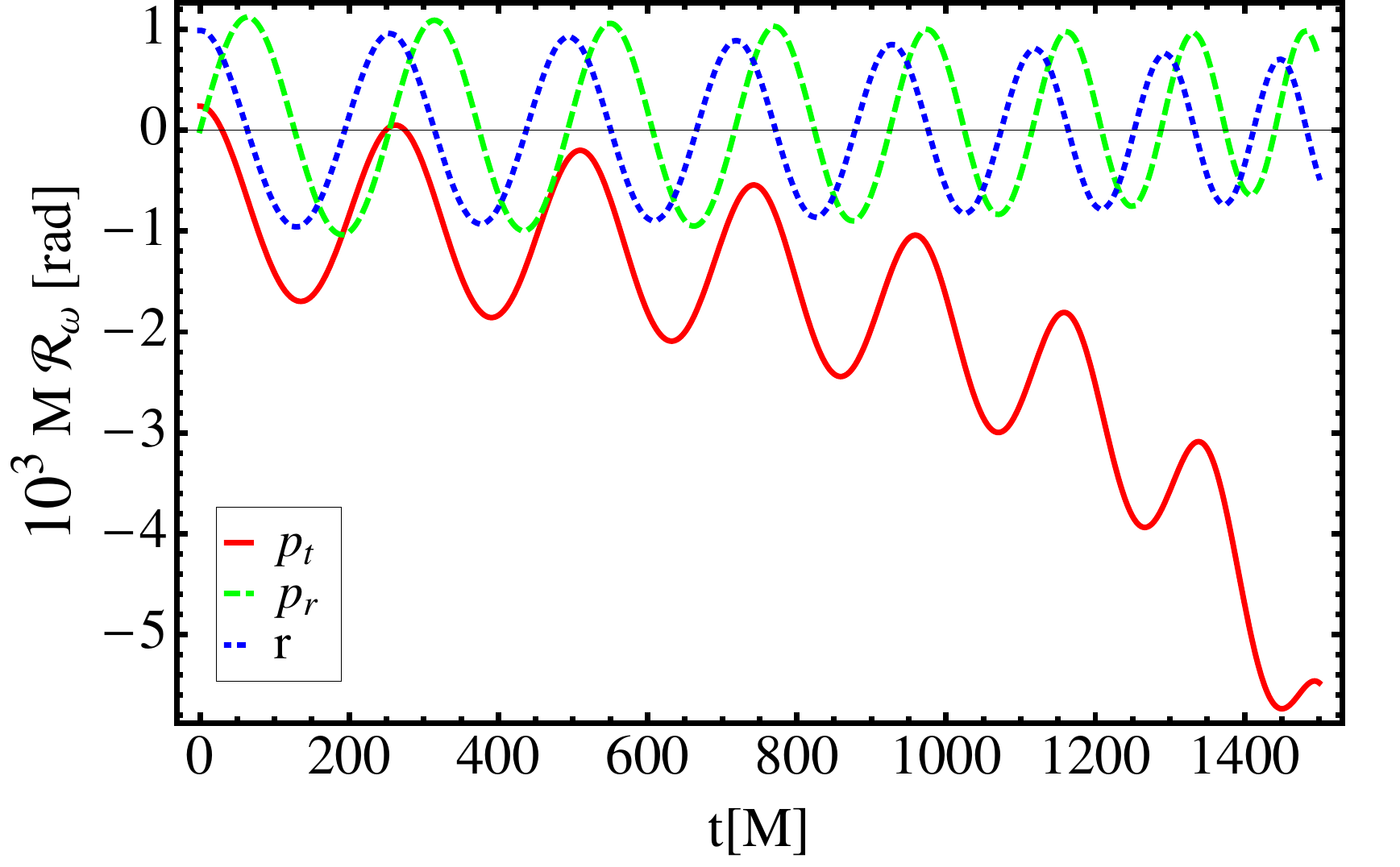}
 \hspace*{-0.15cm}\includegraphics[width=.5\textwidth]{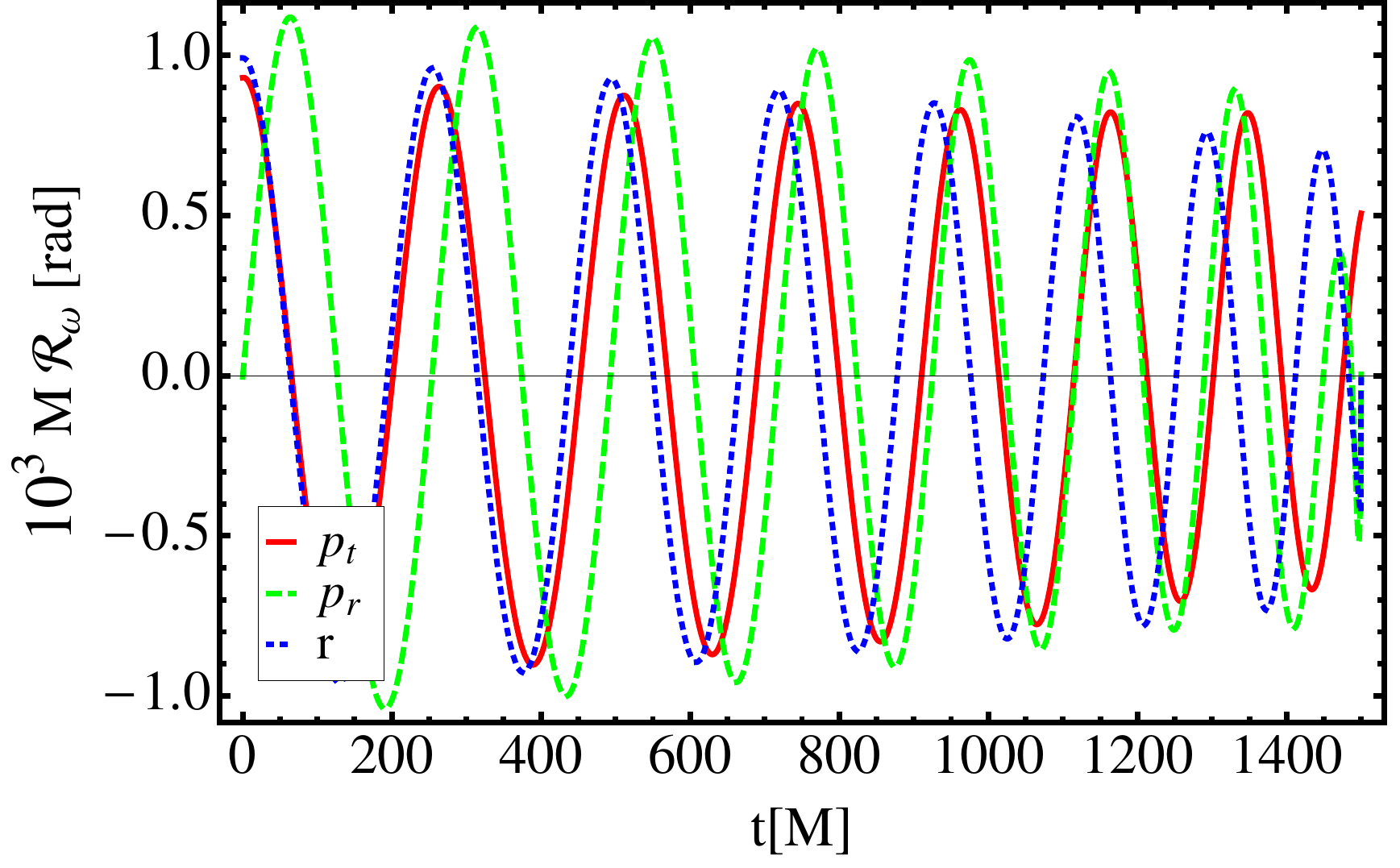}
  \caption{Frequency residuals for perturbations of tangential momentum, radial momentum, and radial separation for 
  the Newtonian-plus-quadrupole-flux model. The residuals are computed from perturbations 
  $p_{t,0} = 1.01 p_{t,\text{QC}}$, $p_{r,0} = 5 p_{r,\text{QC}}$ and $r_0 = 0.98 r_\text{QC}$ from QC data for a 
  $q=2, r_0 = 12 M$ binary. The upper panel illustrates the dephasing effect. In the lower panel the dephasing has
  been removed by a fit to \eqref{eq:model_dephasing}.
  }
  \label{fig:NewtonianBinaryFlux_inspiral_residuals}
\end{figure}

In Fig.~\ref{fig:NewtonianBinaryFlux_inspiral} we show the radial separation and the orbital frequency for  
quasi-circular (QC) inspiral and perturbations thereof, which lead to eccentric inspiral.
We show the frequency residuals for these cases in Fig.~\ref{fig:NewtonianBinaryFlux_inspiral_residuals} (top panel). 
It is apparent that it is only the perturbation of the tangential momentum that causes a significant `dephasing' (red line). 
For the simple Newtonian-plus-quadrupole-flux model, a perturbation of $p_r$ causes no dephasing in the frequency 
and a perturbation of $r$ only a very slight dephasing. For perturbations of PN/EOB approxmiants, however, significant 
dephasing is generic and therefore we need to correct for it. The simple model considered here serves mainly as an illustration 
of the dephasing effect. 

We can calculate the dephasing effect in our Newtonian model in the following way. First we find the orbital frequency 
for QC inspiral by combining Kepler's law $\Omega^2 = M/r^3$ with the evolution of the radial 
separation \eqref{eq:rNewtonian_QC} to obtain
\begin{equation}
  \label{eq:omega-Newtonian_QC}
  \Omega(t) = \frac{1}{8} \left( \frac{5}{\left( \mu M^{2/3} (T_c-t) \right)} \right)^{3/8}.  
\end{equation}
Writing the coalescence time as in~\eqref{eq:T_c}, it is apparent that the single initial parameter $r_0$ in the orbital 
frequency~\eqref{eq:omega-Newtonian_QC} can be perturbed. (Alternatively, $T_c$ could we written as a function of 
initial frequency, so that $\Omega_0$ can be perturbed.)
We expand the dephasing $\mathcal{D}_\Omega$ and  
$\mathcal{D}_\Omega = \Omega(t; T_c(r_0 \delta)) - \Omega(t; T_c(r_0))$ around $\delta = 1$ and obtain
\begin{multline}
  \mathcal{D}_\Omega = 
  - C \frac{3 T_c}{2 (T_c - t)^{11/8}}  (\delta - 1)\\
  + C \frac{3  \left(6 t T_c + 5 T_c^2\right)}{8 (T_c - t)^{19/8}} (\delta -1)^2 + \bigO((\delta - 1)^3),
\end{multline}
where $C = \frac{1}{8} 5^{3/8} \left[1 / (\nu  M^{5/3})\right]^{3/8}$.

The functional form of the dephasing can be used as a fitting model beyond its Newtonian setting by letting the 
prefactors of the rational terms be free parameters. We then arrive at the following model 
\begin{multline}
  \label{eq:model_dephasing}
  \mathcal{D}_{\Omega,\text{model}}(t;\delta, T_c, A, B) = 
  -\frac{A (\delta -1)T_c}{\left(T_c-t\right){}^{11/8}}\\
  + \frac{B (\delta -1)^2 \left(6 t T_c+5 T_c^2\right)}{\left(T_c-t\right){}^{19/8}},
\end{multline}
with parameters $\delta, T_c, A, B$.

We can use this rational model to remove the dephasing, as shown in the bottom panel of 
Fig.~\ref{fig:NewtonianBinaryFlux_inspiral_residuals}. However, when applying our eccentricity reduction 
method to full NR simulations, we find that nonlinear fits to this model are difficult to tune, and in practice
we have found that the polynomial fit (\ref{eq:Rfit}) to capture the dephasing behavior just as well, while 
also being more robust. The only drawback of the polynomial fit is that it results in artifacts at the beginning 
and end of the fitting window, where the polynomial is prone to picking up parts of oscillations in the residuals. 
This problem can be alleviated by discarding part of the fitting window near the boundary.

\section{Eccentricity for 1PN compact binaries} 

This discussion is based on the 1PN equations of motion in a quasi-Keplerian parametrization as given 
in~\cite{Tessmer:2010sh}, with the notation in same places modified for consistency with the rest of this paper. 
After defining the relevant quantities we linearize in the ``time-eccentricity'' $e_t$
and evaluate the usual Newtonian eccentricity estimators with the orbital phase and frequency. Instead of $e_t$ we could have chosen the ``radial eccentricity'' $e_r$ or the ``phase eccentricity'' $e_\phi$ defined below.

In section ~\ref{sub:eccentricity_estimators_for_gw}, we use the quadrupole formula to obtain the strain polarizations 
$h_+$ and $h_\times$ from the 1PN accurate orbital phase. Furthermore, we calculate the phases of $h$ and 
$\Psi_4$ to linear order in $e_t$, and evaluate the usual eccentricity estimator for the GW phase. 
Finally, we relate the results of the orbital and GW eccentricity estimators for the 1PN compact binary as functions 
of the periastron advance parameter $k$. This result is surprising at first glance: if we measure the eccentricity from 
the GW phase, we obtain \emph{different answers} if we use the strain $h$ and the Newman-Penrose scalar
$\Psi_4$. It is important to bear this result in mind whenever the GW signal is used to measure the eccentricity
(as for example in~\cite{Mroue:2010re}); fortunately to leading order the two can be related.

\subsection{Definitions} 
\label{sec:1pn_eccentricity_estimators}

The following is based on the treatment in~\cite{Tessmer:2010sh}.

We define the orbital phase
\begin{equation}
  \phi_{orb} = \phi_0 + (1+k) A_e,
\end{equation}
the true anomaly 
$A_e$
in terms of the eccentric anomaly $u$
\begin{equation} 
  A_e(u)
  = 2 \arctan\left[ \left(\frac{1+e_\phi}{1-e_\phi}\right)^{1/2} \tan\frac{u}{2} \right],
\end{equation}
and the mean anomaly
\begin{equation}
  \beta := n (t-t_0) = u - e_t \sin u.
\end{equation}
The radial separation is given by
\begin{equation}
  r = a_r \left( 1 - e_r \cos u\right),
\end{equation}
where
\begin{equation}
  a_r = \frac{3}{k} \left(1 + \frac{k}{6}\frac{\nu - 7}{4} \right).
\end{equation}
The eccentricities $e_\phi$, $e_r$ and $e_t$ can be related in terms of the periastron advance parameter $k$
\begin{align}
  \frac{e_\phi}{e_t} &= 1 + \frac{k}{6}(1-e_t^2) (-2(\nu - 4)),  \label{eq:ratioeper} \\
  \frac{e_r}{e_t}    &= 1 + \frac{k}{6}(1-e_t^2) (8 - 3\nu).
\end{align}

The frequency of radial eccentricity oscillations is 
\begin{equation}
  n\equiv \Omega_r = 2\pi / P_r, \label{eq:ndefn}
\end{equation} 
where $P_r$ measures the time between two consecutive periastron passages, while the average orbital frequency is
\begin{equation}
  \label{eq:om_phi_om_r_1PN}
  \Omega_\phi = (1+k) \Omega_r.
\end{equation} 
The fractional periastron advance per orbit, $k$, depends on the frequency of eccentricity oscillations
\begin{equation}
  \label{eq:k_of_frequency}
  k = 3 \epsilon^2 \frac{n^{2/3}}{1-e_t^2},
\end{equation}
where $\epsilon$ counts the order of the inverse speed of light $c$, $\epsilon^2 = 1/c^2$.

Expanding up to linear order in eccentricity (the different eccentricities can be related to each other) we find for the orbital phase
\begin{equation}
  \phi_{orb} = \phi_0 + \Omega_\phi t + e_t (1+k) \left(1 + \frac{e_\phi}{e_t}\right) \sin(\Omega_r t).
\end{equation}
The Newtonian definition of the orbital phase eccentricity then yields
\begin{equation}
  e_{\phi,orb} = \frac{\phi_{orb} - \phi_{orb}(e=0)}{2}
  = e_t (1+k) \frac{1 + e_\phi/e_t}{2}.
\end{equation}
The orbital frequency is simply the time-derivative of the orbital phase
\begin{equation}
  \Omega = \Omega_\phi + e_t \Omega_\phi \left(1 + \frac{e_\phi}{e_t}\right) \cos(\Omega_r t)
\end{equation}
and the associated eccentricity is
\begin{equation}
  \label{eq:e_om_orb_1PN}
  e_{\Omega} = \frac{\Omega - \Omega(e=0)}{2\Omega}
  = e_t \frac{1 + e_\phi/e_t}{2}.
\end{equation}
The ratio between the orbital phase and frequency eccentricities is
\begin{equation}
  \frac{e_{\phi,orb}}{e_\Omega} = 1+k,
\end{equation}
as given in \cite{Mroue:2010re}. At Newtonian order there is no periastron advance, and $k=0$, 
and, as we have already emphasized, all eccentricity estimators agree. 


\subsection{Eccentricity estimators for the GW signal} 
\label{sub:eccentricity_estimators_for_gw}

In this section we calculate the ratio between the eccentricities measured from the phase of
the GW strain $h$ and the Newman-Penrose scalar $\Psi_4$. 

We first calculate expressions for $\Psi_4$ and the strain $h$ by combining the 1PN phase given in 
Sec.~\ref{sec:1pn_eccentricity_estimators} with the quadrupole formula
\begin{equation}
  h_{ij}^{TT} = \frac{2}{r} \ddot{\mathcalstd{I}}_{ij}^{TT}(t-r),
\end{equation}
where $\mathcalstd{I}_{ij}$ is the reduced quadrupole moment and $TT$ projects out the transverse traceless 
part of a tensor. The strain can be decomposed as
\begin{equation}
  h_{ij}^{TT} = h_+ e^+_{ij} + h_\times e^\times_{ij},
\end{equation}
in terms of two polarization tensors $e^+_{ij}$ and $e^\times_{ij}$~\cite{Baumgarte2010-NR-book}.

For an observer in the wave zone at a distance $r$ and inclination angle $\theta$ relative to 
the plane in which the binary orbits, the two polarization modes of the strain are
\begin{multline}
  h_+ =
  -\frac{G \mu}{c^4 r}
  \Bigg\{
    (1 + \cos\theta^2)
    \bigg[ 
        2 \, \dot r \,r \,\dot\phi \sin(2 \phi)\\
        + \left( 
             \frac{G M}{r} + r^2 \dot \phi^2 - \dot r^2 
          \right) \cos(2 \phi) 
      \bigg]\\
    + \sin\theta^2 \left( 
            \frac{G M}{r} - r^2 \dot\phi^2 - \dot r^2
          \right)    
  \Bigg\}
\end{multline} 

\begin{multline}
  h_\times =
  -\frac{2G \mu}{c^4 r}
  \Bigg\{
    \cos\theta
    \bigg[\sin (2 \phi ) 
      \left(
        \frac{G M}{r} + r^2 \dot{\phi}^2 - \dot{r}^2
      \right)\\
      -2\, r \, \dot{r} \, \dot{\phi } \cos (2 \phi )
    \bigg]
  \Bigg\}      
\end{multline}

The only spherical-harmonic modes of interest are the $(\ell, m) = (2,\pm 2)$ modes. 
The $(\ell, m) = (2,\pm 1)$ modes vanish irrespective of the choice of $\theta$ at 1PN order, 
and the $(\ell, m) = (2,0)$ mode is real and 
does not contribute to the phase.
To make the computation of the phases tractable we assume that the binary is optimally oriented to the 
observer and set $\theta = 0$, which implies that only the  $(\ell, m) = (2,2)$ mode remains. This is 
sufficient, because out eccentricity reduction method considers only the frequency from $\Psi_{4,22}$.

The wave strain
\begin{equation}
  h = h_+ - i h_\times
\end{equation}
is related to the Newman-Penrose scalar $\Psi_4$ by
\begin{equation}
  \Psi_4 = \ddot{h}_+ - i \ddot{h}_\times.
\end{equation}
The phase of $\Psi_4$ and $h$ can be obtained locally in time by computing the complex argument 
of either $\Psi_4$ or $h$. 

Applying the customary definition of an eccentricity estimator for the GW phase
\begin{equation}
  e_{\phi,GW} = \frac{\phi_\text{GW} - \phi_{fit}}{4},
\end{equation}
we find
\begin{equation}
  \label{eq:e_phi_h}
  e_{\phi[h]} = 
  e_t \frac{3 + 4 k \epsilon^2}{4}
  ,
\end{equation}
and
\begin{equation}
  \label{eq:e_phi_psi4}
  e_{\phi[\psi_4]} = 
  e_t \frac{21 + 10 k \epsilon^2}{16}.
\end{equation}

We may Taylor expand the factor between $e_{\phi[\psi_4]}$ and $e_{\phi[h]}$ in $\epsilon$ up to 1PN-order 
($\epsilon^2 = 1/c^2$) to obtain an approximation in the low $k$ or low frequency limit
\begin{equation}
  \label{eq:e_phi_psi4_vs_e_phi_h_ratio}
  \frac{e_{\phi[\psi_4]}}{e_{\phi[h]}} = 
  \frac{7}{4}-\frac{3 k \epsilon ^2}{2}.
\end{equation}

It is clear that in general we must specify whether the GW-phase eccentricity was measured from
$h$ or $\Psi_4$. 
In the main text we refer only to the `GW phase', $\phi_\text{GW}$, by which we mean
the phase computed from $\Psi_4$. 

\begin{figure}
\vspace{0.3cm}
\centering
\hspace*{-0.2cm}
  \includegraphics[width=0.5\textwidth]{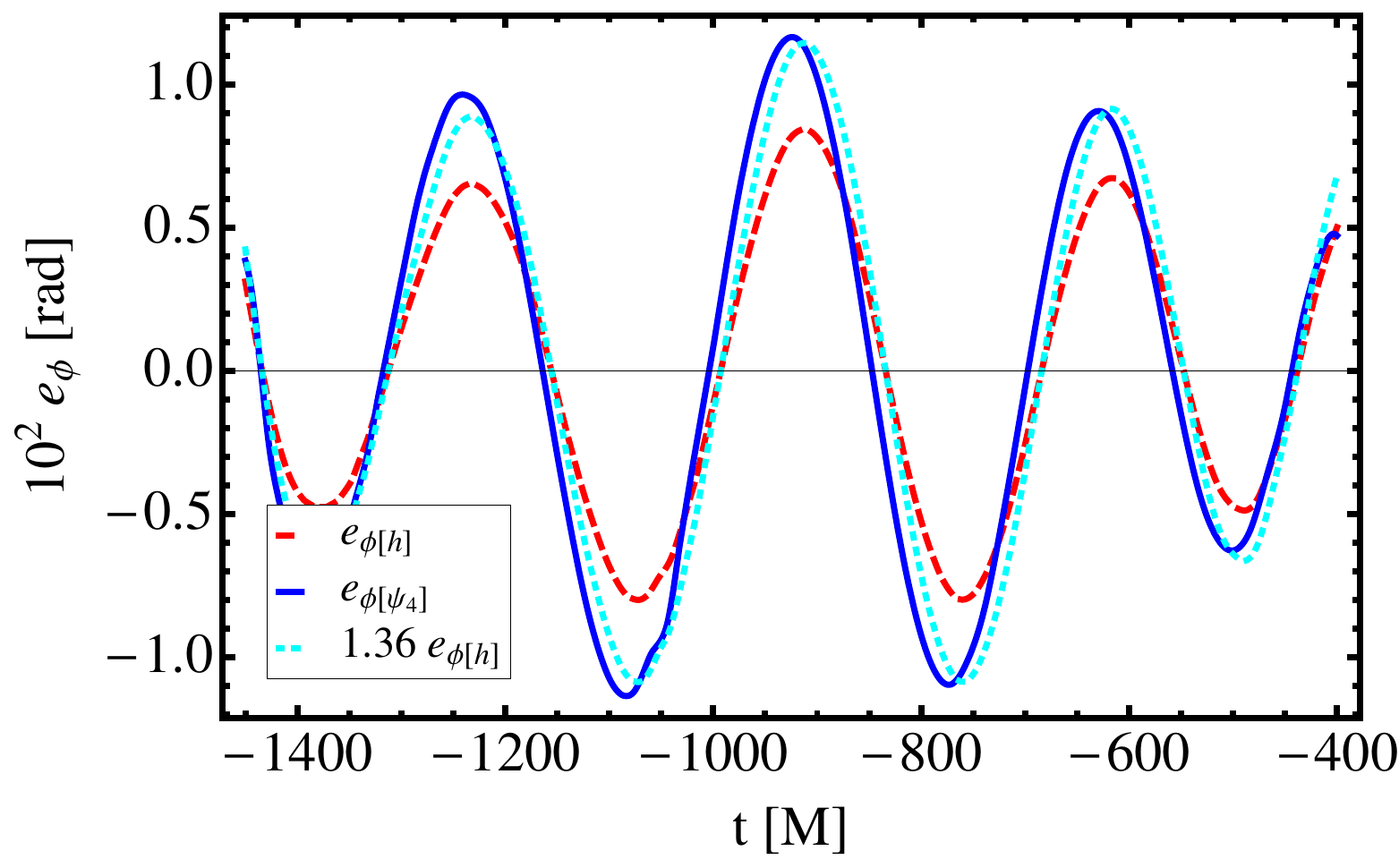}
  \caption{Comparison of eccentricity estimators $e_{\phi[\psi_4]}$ and $e_{\phi[h]}$ for the reference 
  ($q=2, \chi_1=0, \chi_2=0.25$) configuration with $e=0.01$. The theoretical scale factor between them is 1.36 
  (see text).
  }
  \label{fig:e_phi_h_vs_psi4}
\end{figure}

Figure~\ref{fig:e_phi_h_vs_psi4} compares the eccentricity estimators $e_{\phi[\psi_4]}$ and $e_{\phi[h]}$ for our
reference example $q=2, \chi_1=0, \chi_2=0$ configuration, with high eccentricity $e=0.01$. 
We take $k = 0.4$ from the NR results in Fig.~7 of~\cite{Mroue:2010re} (assuming it still holds for $q=2$) 
for an average orbital frequency in the early evolution of $M \omega \sim 0.027$.  We then calculate the ratio 
$e_{\phi[\psi_4]} / e_{\phi[h]} \approx 1.36$. This factor agrees with the numerical data to about $15\%$ accuracy 
in the amplitude. We have repeated the analysis for the same configuration with eccentricity $e = 0.006$, and
find similar agreement.  For smaller eccentricities the noise in the $\Psi_4$ residual makes the comparison 
less conclusive. 

For the sake of completeness, we compute the eccentricity estimator for the frequency of $\Psi_4$,
\begin{equation}
  e_{\omega[\psi_4]} = e_t \frac{1}{16} \left(21-11 k \epsilon ^2\right).
\end{equation}
Up to 1PN order the ratio between $e_{\phi[\psi_4]} $ and $e_{\omega[\psi_4]}$ equals $1 + k$, as expected.

In Sec.~\ref{sub:gauge_dependence_of_the_orbital_motion} we compare the eccentricity measured from 
both $\Psi_4$ and the orbital-motion frequency $\Omega$. The ratio between the phase- and frequency-based
eccentricities is $1+k$, as discussed earlier in Appendix~\ref{sec:1pn_eccentricity_estimators}. 
But we are now considering the phase and frequency from different physical quantities, 
respectively $\Psi_4$ and the orbital motion, and so we do not expect that the same relationship will hold. 

To obtain an approximation for the low-$k$ or low-frequency limit, we Taylor expand the factor 
between $e_{\phi[\psi_4]}$ and $e_\Omega$ up to 1PN-order in $\epsilon$. We combine
Eqns.~(\ref{eq:ratioeper}), (\ref{eq:e_om_orb_1PN}), and (\ref{eq:e_phi_psi4}),
\begin{equation}
  \label{eq:e_phi_psi4_vs_e_om_orb_ratio}
  \kappa := \frac{e_{\phi[\psi_4]}}{e_\Omega} 
  \approx \frac{21}{16} +  \left(\frac{7 \nu}{32} - \frac{1}{4}\right) k \epsilon ^2.
\end{equation}

The average orbital frequency for the NR $q=1$ example is $M \omega \approx 0.027$. We can recover 
$k$ at the 1PN level by combining Eqns.~(\ref{eq:ndefn})--(\ref{eq:k_of_frequency}) to give
\begin{equation}
  k = 3 \left(\frac{0.027}{1+k}\right)^{2/3},
\end{equation}
which yields $k \approx 0.23$.
Then, $\kappa \approx 1.27$.
Taking $k$ from the Fig.~7 of Ref.~\cite{Mroue:2010re}, a value $k \sim 0.38$ seems reasonable. 
This gives a factor $\kappa \approx 1.25$, a bit smaller than the factor $1+k$.

We find good agreement with a factor $1.25$ between eccentricities in Figs.\ref{fig:eccentricity_eta_moderate} 
and \ref{fig:eccentricity_eta}. The disagreement is at most $10\%$, which is inside the error bars of the 
eccentricity estimators discussed in section~\ref{sec:errors_rex_fits}. 
Moreover, we expect that the 1PN expressions also contribute a significant error.




\bibliography{PN_eccentricity_reduction}

\end{document}